\documentclass[11pt]{article}
\usepackage[a4paper,left=17mm,top=20mm,right=17mm,bottom=25mm]{geometry}

\usepackage{authblk}
\usepackage[numbers, sort&compress]{natbib}
\usepackage[margin=30pt,font=small,labelfont=bf]{caption}
\usepackage{amsmath,amsopn,amsthm,amssymb}

\usepackage{graphicx}
\usepackage{mathptmx} 
\usepackage[mathscr]{euscript}
\usepackage{natbib}
\usepackage{mathtools}
\usepackage{enumitem}
\usepackage{xspace}
\usepackage{scalerel}
\usepackage{bold-extra}

\usepackage[
 	colorlinks=true,
	urlcolor=black,
	linkcolor=blue
]{hyperref}

\usepackage{algorithm}
\usepackage[noend]{algpseudocode}
\algrenewcommand\algorithmicrequire{\textbf{Input:}}
\algrenewcommand\algorithmicensure{\textbf{Output:}}

\newtheorem{theorem}{Theorem}[section]
\newtheorem{lemma}[theorem]{Lemma} 
\newtheorem{corollary}[theorem]{Corollary}
\newtheorem{proposition}[theorem]{Proposition}

\newtheorem{definition}[theorem]{Definition}
\newtheorem{observation}[theorem]{Observation}

\newtheorem*{problem}{Problem}

\newcommand{\Hasse}[1][]{\mathscr{H}\ifthenelse{\equal{#1}{}}{}{(#1)}}
\newcommand{\hasse}[1][]{\mathscr{H}\ifthenelse{\equal{#1}{}}{}{(#1)}}
\DeclareMathOperator{\CC}{\mathtt{C}}
\newcommand{\LCA}{\ensuremath{\operatorname{LCA}}}
\newcommand{\lca}{\ensuremath{\operatorname{lca}}}
\DeclareMathOperator{\child}{child}
\DeclareMathOperator{\indeg}{indeg}
\DeclareMathOperator{\outdeg}{outdeg}
\DeclareMathOperator{\klca}{\textit{k}-lca}

\DeclareMathOperator{\2LCA}{\textit{2}-LCA}
\DeclareMathOperator{\kLCA}{\textit{k}-LCA}

\newcommand{\One}{\mathscr{I}^{\scaleto{1}{3pt}}}

\newcommand{\rel}{\textsc{Rel}\xspace}

\newcommand{\Olca}{$\One$-$\lca$\xspace}
\newcommand{\OLCA}{$\One$-$\LCA$\xspace}
\newcommand{\Olcarel}{$\One$-$\lca$-$\rel$\xspace}
\newcommand{\OLCArel}{$\One$-$\LCA$-$\rel$\xspace}
\newcommand{\OlcaTMP}{$\mathscr{I}$-$\lca$\xspace}
\newcommand{\OLCATMP}{$\mathscr{I}$-$\LCA$\xspace}
\newcommand{\OlcarelTMP}{$\mathscr{I}$-$\lca$-$\rel$\xspace}
\newcommand{\OLCArelTMP}{$\mathscr{I}$-$\LCA$-$\rel$\xspace}

\newcommand{\PROBLEM}[1]{\texttt{\textit{#1}}}
\newcommand{\PROBLEMK}[2]{${#1}$-\texttt{\textit{#2}}}
\newcommand{\PROBLEMONE}[2]{${#1}$-\texttt{\textit{#2}}-\texttt{\textit{Rel}}}

\makeatletter
	\newcommand{\hatC}{%
		\m@th\widehat{\raisebox{0pt}[0.8\height]{$\mathfrak{C}^*$}}%
	}
\makeatother

\usepackage{xcolor}

\providecommand{\keywords}[1]{\textbf{\textit{Keywords: }} #1}

\title{Simplifying and Characterizing DAGs and Phylogenetic Networks via Least Common Ancestor Constraints}

\author[ ]{Anna Lindeberg}
\author[ ]{Marc Hellmuth}

\affil[ ]{Department of Mathematics, Faculty of Science,
  Stockholm University, SE-10691 Stockholm, Sweden} 

\date{\ }

\begin{document}
\sloppy

\maketitle

\abstract{ \noindent
	Rooted phylogenetic networks, or more generally, directed acyclic graphs (DAGs), are widely used
	to model species or gene relationships that traditional rooted trees cannot fully capture,
	especially in the presence of reticulate processes or horizontal gene transfers. 
	Such networks or DAGs are typically inferred from observable data (e.g., genomic sequences of extant species), 
	providing only an estimate of the true evolutionary history.
	However, these inferred DAGs are often complex and
	difficult to interpret. In particular, many contain vertices that do not serve as least common
	ancestors (LCAs) for any subset of the underlying genes or species, thus may lack direct support
	from the observable data. In contrast, LCA vertices are witnessed by historical traces 
	justifying their existence and thus 
	represent ancestral states substantiated by the data.
	To reduce unnecessary complexity and eliminate unsupported vertices, we aim to simplify a DAG to
	retain only LCA vertices while preserving essential evolutionary information.

	In this paper, we characterize $\mathrm{LCA}$-relevant and $\mathrm{lca}$-relevant DAGs, defined as those in
	which every vertex serves as an LCA (or unique LCA) for some subset of taxa. We introduce methods
	to identify LCAs in DAGs and efficiently transform any DAG into an $\mathrm{LCA}$-relevant or
	$\mathrm{lca}$-relevant one while preserving key structural properties of the original DAG or network.
	This transformation is achieved using a simple operator ``$\ominus$'' that mimics vertex
	suppression. 
}

\smallskip
\noindent

\section{Introduction}

Rooted networks and, more generally, directed acyclic graphs (DAGs), are essential in mathematical
phylogenetics for modeling complex evolutionary relationships that traditional rooted trees cannot
fully represent \cite{Huber2022-gu,HRS:10,Huson:11}. In a DAG $G$, the leaf set $L(G)$ represents
extant taxa, such as genes or species, while internal vertices $v \in V(G)\setminus L(G)$ correspond
to ancestral states and are associated with sets $\CC_G(v)$ of descendant leaves known as
``hardwired clusters'' \cite{Huson:11,vIKRH:10,HR:08}, or \emph{clusters} \cite{Hellmuth2023} for
short. Typically, only the leaf set $L(G)$ is available and a primary task is to reconstruct the
evolutionary history - i.e., phylogenetic networks or DAGs - using information provided solely by
the taxa in $L(G)$. This information, often derived from genomic sequence data and sequence
similarities, can reveal clusters within the unknown DAG $G$ that are used to reconstruct $G$
\cite{vIKRH:10,HR:08,Huson:11,Gambette2017}.

However, DAGs and networks inferred from genomic data can be highly complex and tangled, often
containing redundant information \cite{DM:09,FRANCIS2021107215}. In particular, unlike phylogenetic
trees, the number of vertices in a DAG $G$ is in general not asymptotically bounded by a function
depending solely on the number of leaves. As a result, various methods have been developed to
simplify DAGs while preserving their most significant features
\cite{Heiss2024,FRANCIS2021107215,HUBER201630}. Our research builds on this line of work and focuses
on eliminating vertices from a DAG $G$ that are ``less relevant'' in the sense that they are not
least common ancestors of certain subsets of $L(G)$. 

A least common ancestor (LCA) of a subset $A \subseteq L(G)$ is a vertex $v$ that is an ancestor of
all $x \in A$ and that has no descendant which also satisfies this property. LCAs are essential for
understanding and interpreting evolutionary relationships in phylogenetics
\cite{Schaller2021,Hellmuth:15a,Lafond2016,NEDWH:18}. In evolutionary biology, there is a general
consensus that inferred networks and DAGs should be phylogenetic, that is, they should not contain
vertices with in- and out-degree one. The reason is simple: such vertices cannot be observed from
any biological data since there is no historical trace left justifying their existence
\cite{Hellmuth2017}. By similar reasoning, LCA vertices should represent ancestral relationships
evidenced by a clear phylogenetic signal in the observable data, i.e., the leaves in $G$. Vertices
that are not LCAs of any subset of taxa of the underlying data may lack direct relevance to the
observed ancestral relationships.

\begin{figure}
	\centering
		\includegraphics[width=0.9\textwidth]{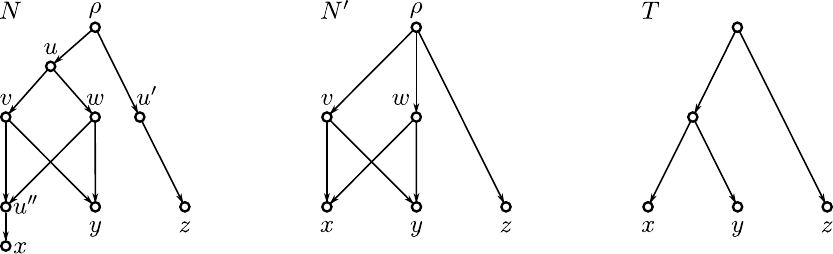}	
	\caption{
	Shown are three networks $N$, $N'$ and $T$. All have the same clustering system $\mathfrak{C} =
	\{\{x\},\{y\},\{z\},\{x,y\},\{x,y,z\}\}$ and leaf set $X = \{x,y,z\}$. Here, only $N'$ and $T$
	are phylogenetic. The network $N$ is not phylogenetic, since $N$ contains the vertex $u'$ with
	in- and out-degree one. Moreover, vertices $u$, $u'$ and $u''$ in $N$ are not LCAs of any subset
	of leaves. ``Removing'' $u$, $u'$ and $u''$ from $N$ via the ``$\ominus$''-operator -- as
	explained in detail in Section~\ref{sec:ominus-lcaRel} -- yields the simplified network $N' =
	N\ominus \{u,u',u''\}$ in which all vertices are LCAs of some subset of $X$. Hence, $N'$ is
	$\LCA$-\rel. In particular, $N'$ is precisely the simplification $\varphi_{\LCA}(N)$ as explained
	in Section~\ref{sec:simplify}. However, $N'$ is not $\lca$-\rel as the vertices $v$ and $w$ are
	not unique LCAs in $N'$ for any subset of $X$. If desired, $N'$ can now be further simplified by
	``removing'' one of $v$ or $w$ and resulting shortcuts which yields the phylogenetic and
	$\lca$-\rel tree $T$. The tree $T$ is the unique phylogenetic tree whose clustering system
	$\mathfrak{C}$ is identical to those in $N$ and $N'$.
	}		
\label{fig:simple-example}		
\end{figure}

In the network $N$ shown in Figure~\ref{fig:simple-example}, the three vertices $u$, $u'$, and $u''$
do not serve as the LCA for any subset of $L(N) = \{x, y, z\}$. Removing or suppressing these
vertices using the ``$\ominus$-operator'' defined in Section~\ref{sec:ominus-lcaRel} yields the
simplified network $N'$. While the term ``simplification'' may be subjective and influenced by
specific interests, the process described here satisfies fundamental axioms for network
simplification as proposed by Heiss et al.\ \cite{Heiss2024} (see Section~\ref{sec:simplify} for
more details). All vertices in $N'$ serve as an LCA for some subset of leaves and can thus be
considered to be supported by historical traces observable in the data $L(N) = X$, a property not
satisfied by the vertices $u$, $u'$, and $u''$ in $N$. For instance, the vertex $u'$ in $N$ has in-
and out-degree one, with no justification for its existence provided by the set of leaves.
Similarly, since $u$ and $u''$ are not LCAs for any subset of $X$, they can be considered less
supported by the data. We are aware that $u$ and $u''$ are ``binary'' vertices, i.e., they have
in-degree one and out-degree two, and in-degree two and out-degree one, respectively. However,
binary (fully resolved) networks often result from artificially resolving non-binary vertices and,
thus, may impose assumptions about evolutionary history that extend beyond what is provided by the
data \cite{Hellmuth:15a,vIKRH:10}. This provides justification for removing some binary vertices as
well. Again, the primary reason to suppress vertices $u$ and $u''$ here is that they do not serve as
the LCA of any subset of leaves. Note that $u''$ can also be considered redundant because the
reticulation event leading to taxa $x$ it represents is still captured in $N'$ after suppressing
$u''$. If one is additionally interested in networks that include only vertices representing unique
LCAs of subsets of leaves, one can further simplify $N'$ into the phylogenetic tree $T$ by removing
one of $v$ or $w$. Notably, $T$ is the unique phylogenetic tree whose clustering system is identical
to those in $N$ and $N'$. In general, the networks $N'$ and $T$ can always be considered as
simplifications of $N$ showing a ``trend of evolution''. 

As argued above, non-LCA vertices may lack clear interpretation and could be considered less
significant or redundant in an evolutionary context. Therefore, simplifying a DAG by ``removal'' of
non-LCA vertices resulting in a DAG in which each vertex is a (unique) LCA of at least some leaves
is a natural next step. We demonstrate that this transformation can be performed efficiently while
preserving the structural integrity of the original DAG. The central questions considered here are
as follows:

\smallskip
\begin{enumerate}
    \item Is a given vertex a (unique) LCA of a specific, known subset $A \subseteq L(G)$?\smallskip
    \item Is a given vertex a (unique) LCA of some unknown subset $A \subseteq L(G)$, possibly with a prescribed size $|A|$?\smallskip
    \item Can one characterize and recognize DAGs $G$ in which every vertex is a (unique) LCA of some subset of $L(G)$?\smallskip
    \item Is it possible to efficiently remove  all vertices from a DAG $G$ that do not satisfy (1) or (2) 
    		and thus, to simplify $G$ to a DAG in which each vertex is a (unique) LCA of some subset of $L(G)$
    		while preserving as many structural features of $G$ as possible?
\end{enumerate}
\smallskip

We will address these problems from different perspectives. Numerous results have been established
for Question~1 and 2 for the case $|A|=2$
\cite{KL:07,NU:94,GFP+05,MVX:22,CKL:07,GILPU:21,Bender:01,KL:05,HT:84}, or when assuming that $A =
\CC_G(v)$ for a given vertex $v$ \cite{Nakhleh:05}.

This paper is organized as follows. We start with introducing the basic definitions needed in
Section~\ref{sec:basics}. In Section~\ref{sec:LCA-rel}, we define the notions of $\lca$-\rel and
$\LCA$-\rel DAGs, as well as $k$-$\lca$ and $k$-\LCA\ vertices. In short, a vertex $v$ is a $k$-$\LCA$
(resp., $k$-$\lca$) in $G$ if there exists a subset $A\subseteq L(G)$ of size $|A|=k$ such that $v$
is a LCA (resp., unique LCA) of the vertices in $A$. As we will see, a vertex $v$ is a $k$-$\LCA$
vertex (resp., $k$-$\lca$ vertex) for some $k$ precisely if $v$ is a LCA (resp., unique LCA) of the
vertices in $\CC_G(v)$ (cf.\ Corollary~\ref{cor:klca<=>v=lca(C(v))}). A DAG is $\LCA$-\rel (resp.,
$\lca$-\rel) if each of its vertices is a LCA (resp., unique LCA) for some subset $A$. We then show
that the set of least common ancestors of a set $A \subseteq L(G)$ can be determined in linear time
when $|A| \in O(1)$ is constant. Additionally, we demonstrate that recognizing $\lca$-\rel and
$\LCA$-\rel DAGs can be done in polynomial time.

In Section~\ref{sec:regular-CL}, we continue by characterizing $\lca$-\rel and $\LCA$-\rel DAGs. As
shown in Theorem~\ref{thm:LCArel-char}, $\LCA$-\rel DAGs are precisely those DAGs that do not
contain adjacent vertices $u$ and $v$ with the same cluster, i.e., $\CC_G(v)=\CC_G(u)$. We then
provide several characterizations of $\lca$-\rel DAGs in Theorem~\ref{thm:strongCLChar}. Among other
results, $\lca$-\rel DAGs are exactly those $\LCA$-\rel DAGs with the path-cluster-comparability
(PCC) property, meaning there is a directed path between two vertices $v$ and $u$ if and only if
their clusters are comparable with respect to inclusion. Moreover, we show a close connection
between $\lca$-\rel\ DAGs and so-called regular DAGs where the latter, loosely speaking, are DAGs
that are completely determined by their set of clusters. In particular, $\lca$-\rel DAGs with all
shortcuts removed are regular (cf.\ Corollary~\ref{cor:Onerel=>regular}). Novel characterizations of
regular DAGs are presented in Theorem~\ref{thm:regular-Char}. Similar to phylogenetic trees, the
number of vertices in $\lca$-\rel DAGs $G$ is asymptotically bounded above by a function on the
number of leaves, see Lemma~\ref{lem:bound-V-by-X}.

Not all DAGs are $\lca$-\rel or $\LCA$-\rel. Hence, the question arises whether one can transform a
given DAG $G$ into an $\lca$-\rel or $\LCA$-\rel DAG $H$ while preserving as much of the structure
of $G$ as possible. In Section~\ref{sec:ominus-lcaRel}, we provide an axiomatic framework for the
phrase ``\emph{preserving as much structure of $G$ as possible}'' resulting in five structural
properties (S1) -- (S5). These properties are informally stated here and will be defined more
precisely in Section~\ref{sec:ominus-lcaRel}.

\smallskip
\begin{description}
    \item[\textnormal{\textit{(S1)}}] $H$ remains a DAG with leaf set $L(G)$. \smallskip
    \item[\textnormal{\textit{(S2)}}] $V(H) \subseteq V(G)$, meaning no new vertices are introduced. \smallskip
    \item[\textnormal{\textit{(S3)}}] $H$ preserves the ancestor relationships defined by $G$ among the vertices in $H$. \smallskip
    \item[\textnormal{\textit{(S4 \& S5)}}] The set of (unique) least common ancestors for (un)specified subsets $A\subseteq L(G)$ in $G$ and $H$ coincide.     
\end{description}
\smallskip
\noindent
This in turn implies an additional condition \textit{(S0)}: no new clusters in $H$ are introduced.  \smallskip

To transform a given DAG into an $\lca$-\rel or $\LCA$-\rel one, we introduce a simple operator
$\ominus$ that acts on the vertices and edges of $G$ \cite{SCHS:24}. Specifically, we denote with
$G\ominus v$ the DAG obtained from $G$ by removing vertex $v$ and its incident edges and connecting
each parent of $v$ with each child of $v$. Using this method, vertices $v$ can be removed stepwise
from $G$, resulting in an $\lca$-\rel or $\LCA$-\rel DAG $H$ that satisfies the properties
(S0) -- (S5). In particular, we provide conditions under which the set $W$ of vertices such that $G
\ominus W$ is $\lca$-\rel or $\LCA$-\rel is uniquely determined and of minimum size. Furthermore,
polynomial-time algorithms are given to transform any DAG into an $\lca$-\rel or $\LCA$-\rel one.
The established algorithms are implemented in the Python package \texttt{SimpliDAG}
\cite{github-AL}.

Following \cite{Heiss2024}, we discuss in Section~\ref{sec:simplify} a general framework for any
transformation $\varphi(G)$ that ``simplifies'' a DAG $G$, formalized through three axioms (P1) --
(P3). Given a suitable notion of restriction of DAGs, we show that the $\ominus$-operator can be
used to derive simplifications $\varphi(G)$ that satisfy axioms (P1) -- (P3). We exemplify different
types of simplification steps on a biological network with reticulation events that is based on a
study of the \emph{Viola} genus from Marcussen et al.~\cite{MHB+14}.

While we have provided polynomial-time algorithms to verify whether a given DAG $G$ is $\lca$-\rel
or $\LCA$-\rel and to transform $G$ into an $\lca$-\rel or $\LCA$-\rel DAG if it is not, an open
question remained so-far: can it be decided in polynomial-time if a vertex $v$ is a $k$-$\lca$
or $k$-$\LCA$ vertex for a given $k$? Although the answer is affirmative for $k=1$ or $k=2$ (cf.\
e.g.\ \cite[Thm.~4.19]{nowak2009algorithmic} or Observation~\ref{obs:12-rel}), we show in
Section~\ref{sec:complexity-hard} that this problem is NP-complete in general. However, it becomes
polynomial-time solvable for DAGs with the (N3O) property, i.e., DAGs that do not contain three
pairwise overlapping clusters. Such DAGs are of particular interest, as they include important
subclasses of phylogenetic networks, such as rooted phylogenetic trees and galled-trees. 

We close this paper with Section~\ref{sec:sum}, where we summarize the main results and provide open
problems for future work.


\section{Basics}
\label{sec:basics}

\paragraph{Sets and Set Systems.}
All sets considered here are assumed to be finite. Here, $2^X$ denotes the powerset of a set $X$. 
For  $\mathscr{I}\subseteq \{1,\dots, |X|\}$,  we write $X(\mathscr{I}) \subseteq 2^X$ for the set of all
subsets $A$ of $X$ with $|A|\in \mathscr{I}$. 
Two sets $M$ and $M'$ \emph{overlap} if $M \cap M' \notin \{\emptyset, M, M'\}$.

A \emph{set system $\mathfrak{C}$ (on $X$)} is a subset $\mathfrak{C} \subseteq 2^X$. A set system
$\mathfrak{C}$ on $X$ is \emph{grounded} if $\{x\}\in \mathfrak{C} $ for all $x\in X$ and
$\emptyset\notin \mathfrak{C}$, while $\mathfrak{C}$ is a \emph{clustering system} if it is grounded
and satisfies  $X\in \mathfrak{C}$. Furthermore, a set
system \emph{satisfies (N3O)} if it does not contain three distinct pairwise overlapping clusters.

\paragraph{Directed Graphs, DAGs and Networks.}
A \emph{directed graph} $G=(V,E)$ is an ordered pair consisting of a nonempty set $V(G)\coloneqq V$ of
\emph{vertices} and a set $E(G)\coloneqq E \subseteq\left(V\times V\right)\setminus\{(v,v) \mid v\in
V\}$ of \emph{edges}.  
For directed graphs $G=(V_G,E_G)$ and $H=(V_H, E_H)$, an
\emph{isomorphism between $G$ and $H$} is a bijective map $\varphi\colon V_G\to V_H$ such that 
$(u,v)\in E_G$ if and only if $(\varphi(u),\varphi(v))\in E_H$. 
If such a map exist, then $G$ and $H$ are \emph{isomorphic}, in symbols $G\simeq H$. 

A \emph{$v_1v_n$-path} $P = (V,E)$ has an ordered vertex set $V = \{v_1,v_2,\ldots,v_n\}$ and the edges in $E$
are precisely of one of the form $(v_i,v_{i+1})$ or $(v_{i+1},v_i)$, $i=1,2,\ldots, n-1$. The
\emph{length} of $P$ is the number $|E|=n-1$ of its edges. A directed
graph $G$ is \emph{connected} if there exists an $xy$-path between any pair of vertices $x$ and $y$.
If all edges in $P$ are precisely of the form $(v_i,v_{i+1})$ for each $i=1,2,\ldots, n-1$, then $P$
is called \emph{directed}. 

For a directed graph $G=(V,E)$, we define $\indeg_G(v)\coloneqq \left|\left\{u\in V \colon (u,v)\in
E\right\}\right|$ and $\outdeg_G(v)\coloneqq \left|\left\{u \in V\colon (v,u)\in E\right\}\right|$
for each $v\in V$ as the \emph{in-degree} respectively \emph{out-degree} of $v$ in $G$. A directed
graph $G$ is \emph{phylogenetic}, if it does not contain a vertex $v$ with $\outdeg_G(v)=1$ and
$\indeg_G(v)\leq 1$.
	
A directed graph $G$ is \emph{acyclic} if there exist no \emph{directed cycle}, that is, no sequence
of $k\geq 2$ distinct vertices $v_1,v_2,\ldots, v_k\in V$ such that
$(v_1,v_2),(v_2,v_3),\ldots,(v_{k-1},v_{k}),(v_k,v_1)\in E$. A directed acyclic graph is called \emph{DAG}.
An edge $e=(u,w)$ in a DAG $G$ is a
\emph{shortcut} if there is a directed $uw$-path that does not contain the edge $e$
\cite{linz2020caterpillars, DOCKER2019129}. A DAG without shortcuts is \emph{shortcut-free}.

Let $G$ be a DAG with an associated partial order $\preceq_G$ on its vertex set $V(G)$ defined by
$v\preceq_G w$ if and only if there is a directed path (possibly of length zero) from $w$ to $v$. In
this case, we say that $w$ is an ancestor of $v$ and $v$ is a descendant of $w$. If $v\preceq_G w$
and $v\neq w$, we write $v\prec_G w$. Two vertices $u,v\in V(G)$ are
\emph{$\preceq_{G}$-incomparable} if neither $u\preceq_G v$ nor $v\preceq_G u$ is true. We denote by
$L(G)\subseteq V(G)$ the $\preceq_G$-minimal vertices of $G$ and we call $x\in L(G)$ a leaf of $G$.
Note that $\outdeg_G(x)=0$ for all $x\in L(G)$. It easy to verify that $\preceq_G$-minimal vertices
must exist in any DAG $G$: take a longest directed $uv$-path $P$ in $G$, i.e., $P$ has a maximum number of
edges among all paths in $G$. In this case $v$ must be a leaf as, otherwise, there is an edge
$(v,w)$ such that either $w\notin P$ in which case $P$ was not a longest directed path or $w\in P$ in which
case $G$ contains a cycle; both cases leading to a contradiction. Thus, $L(G)\neq \emptyset$ for all
DAGs $G$. A vertex of $G$ that is not contained in $L(G)$ is called an \emph{inner vertex}.
Moreover, if $(u,v)$ is an edge of $G$, then $u$ is a \emph{parent} of $v$, while $v$ is a
\emph{child} of $u$. We let $\child_G(v)$ denote the set of all children of a vertex $v$.

\begin{figure}
	\centering
	\includegraphics[width=1.\textwidth]{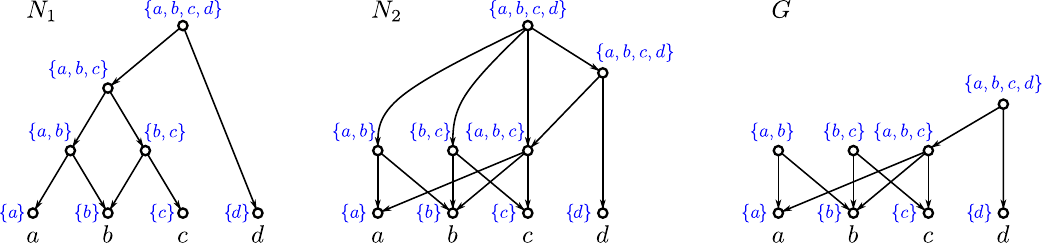}	
	\caption{
	Shown are two phylogenetic networks $N_1$ and $N_2$ and a phylogenetic DAG $G$ such that
	$\mathfrak{C}_{N_1} = \mathfrak{C}_{N_2} = \mathfrak{C}_G$. The clusters $\CC(v)$ are drawn next
	to each individual vertex $v$ and highlighted by blue text. Out of the shown DAGs, only $N_1$
	satisfies (PCC), is regular and has the strong-(CL) property (i.e., $v =
	\lca_{N_1}(\CC_{N_1}(v))$ for all $v$ in $N_1$; cf.\ Def.~\ref{def:CLstrongCL}). Moreover, only
	$N_1$ is $\lca$-\rel and $\LCA$-\rel (cf.\ Def.~\ref{def:LCAlcarel}). Here, $G$ is $\LCA$-\rel
	but $N_2$ is not.
	}		
\label{fig:cluster-N-DAG}		
\end{figure}

If $L(G)=X$, then $G$ is a \emph{DAG on $X$}. A vertex $v\in V(G)$ of a DAG $G$ that is
$\preceq_G$-maximal is called a \emph{root} and the set of roots of $G$ is denoted by $R(G)$. Note
that $\indeg_G(r)=0$ for all $r\in R(G)$. By similar arguments as for leaves, $R(G)\neq \emptyset$
for all DAGs $G$. For every $v\in V(G)$ in a DAG $G$, the set of its descendant leaves
\begin{equation*}
  \CC_G(v)\coloneqq\{ x\in L(G)\mid x \preceq_G v\}
\end{equation*}
is a \emph{cluster} of $G$. We write $\mathfrak{C}_G\coloneqq\{\CC_G(v)\mid v\in V(G)\}$ for the set
of all clusters in $G$. By construction, $\CC_G(x)=\{x\}$ for all $x\in L(G)$. Moreover, for all
$v\in V(G)$, there are vertices $x$ with $x \preceq_G v$ that are $\preceq_G$-minimal and, thus
contained in $L(G)$. Hence, $\CC_G(v)\neq\emptyset$ for all $v\in V(G)$, in particular
$\emptyset\notin\mathfrak{C}_G$ holds. Therefore, $\mathfrak{C}_G$ is a grounded set system on
$L(G)$ for every DAG $G$. For later reference we provide
\begin{lemma}[{\cite[Lem.~1]{S+24}}]\label{lem:prec-subset}
	For all DAGs $G$ and all $u,v\in V(G)$ it holds that $u\preceq_G v$ implies
	$\CC_G(u)\subseteq\CC_G(v)$.
\end{lemma}
The converse of Lemma~\ref{lem:prec-subset} is, in general, not satisfied. 
By way of example, consider the DAG $G$ in Figure~\ref{fig:cluster-N-DAG}, 
where the three roots $r_1$, $r_2$ and $r_3$, read from left to right,
are pairwisely $\preceq_G$-incomparable but  satisfy $\CC_G(r_1), \CC_G(r_2)\subsetneq\CC_G(r_3)$.

The following simple result shows that the cluster associated with a vertex
can be expressed as the union of the clusters associated to its children.

\begin{lemma}\label{lem:union-A}
	Let $G=(V,E)$ be a DAG on $X$ and $A\subseteq X$ nonempty. 		
	For all inner vertices $v\in V$, it holds that 
	\[\CC_G(v) = \bigcup_{u\in\child_G(v)} \CC_G(u) \quad\text{and}\quad A\setminus\CC_G(v)=\bigcap_{u\in\child_G(v)}\left(A\setminus\CC_G(u)\right).\]
\end{lemma}
\begin{proof}
	Let $v$ be an inner vertex of $G$.
	Lemma~\ref{lem:prec-subset} implies that $\cup_{u\in\child_G(v)} \CC_G(u)\subseteq \CC_G(v)$. 
	Now, let $x\in \CC_G(v)$. Since  $x\prec v$, there
	is a directed path from $v$ to $x$ and consequently a child $u$ of $v$ with $x\preceq u$
		and thus, $x\in \CC_G(u)$. Hence, $\CC_G(v) = \cup_{u\in\child_G(v)} \CC_G(u)$. Therefore,
	$A\setminus\CC_G(v)=A\setminus (\cup_{u\in\child_G(v)}\CC_G(u) )=\cap_{u\in\child_G(v)} (A\setminus\CC_G(u) )$.
\end{proof}

A \emph{(rooted) network} $N$ is a DAG for which $|R(N)|=1$, i.e., $N$
has a unique root $\rho\in V(N)$. In a network $N$, we have $v\preceq_N \rho$ for all $v\in V(N)$ and, 
thus, in particular, $\CC_N(\rho)=X$, i.e., $X\in \mathfrak{C}_N$.
Hence, $\mathfrak{C}_N$ is a clustering system (cf.\ \cite[Lem.~14]{Hellmuth2023}).
The converse, however, is in general not satisfied, see Figure~\ref{fig:cluster-N-DAG} for an example
where $G$ is not a network but  $ \mathfrak{C}_G$ is a clustering system.
A network $N$ is a \emph{tree}, if there is no vertex $v$ with $\indeg(v)>1$.

Lemma~\ref{lem:prec-subset} shows that if two vertices are $\preceq_G$-comparable, then their respective clusters are comparable with respect to inclusion. 
The following property ensures the converse, namely, $\preceq_G$-comparability
of vertices $u$ and $v$ based on subset-relations between the underlying clusters $\CC_G(u)$
and $\CC_G(v)$. 

\begin{definition}
	A DAG $G$ has the path-cluster-comparability (PCC) property if it satisfies, for all $u, v \in V(G)$:
	$u$ and $v$ are $\preceq_G$-comparable if and only if $\CC_G(u) \subseteq  \CC_G(v)$ or  $\CC_G(v) \subseteq  \CC_G(u)$.
\end{definition}

By \cite[Lem.~24]{Hellmuth2023}, for every clustering system $\mathfrak{C}$ there is a network 
$N$ with $\mathfrak{C}_N = \mathfrak{C}$ that satisfies (PCC). 
This result builds on the concepts of Hasse diagrams and regular networks. Before delving into the
properties of these specific types of DAGs, we first demonstrate that the $\preceq_G$-ancestor
relationship and, consequently (PCC), is preserved under the removal of shortcuts.
\begin{definition}
	We denote with $G^-$ the DAG obtained from the DAG $G$ by removal	of all shortcuts. 
\end{definition}

\begin{lemma}\label{lem:shortcutfree}
	Let $G = (V,E)$ be a DAG on $X$ with $\ell>0$ shortcuts and let $e$ be a shortcut in
	$G$. Then, $G'\coloneqq (V,E\setminus \{e\})$ is a DAG on $X$ with $\ell-1$ shortcuts. 
	In particular, $G^-$ is uniquely determined for all DAGs $G$.
	Moreover,  for all $u,v\in V$, it holds
	that $u\prec_G v$ if and only if $u\prec_{G'} v$ and, for all $v\in V$, it holds that
	$\CC_G(v)= \CC_{G'}(v)$. Furthermore, $G$ satisfies (PCC) if and only if $G'$ satisfies (PCC). 
\end{lemma}
\begin{proof}
	Let $G = (V,E)$ be a DAG on $X$ with $\ell>0$ shortcuts and let $e$ be a shortcut in $G$. Put $G'
	\coloneqq (V,E\setminus \{e\})$. Reusing exactly the same argument as used in the proof of
	\cite[Lem.\ 1]{Hellmuth2023}, where an analogous result was provided for networks, shows that
	$G'$ is a DAG on $X$ such that, for all $u,v\in V$, it holds that $v\prec_G u$ if and only if
	$v\prec_{G'} u$ and, for all $v\in V$, it holds that $\CC_G(v)= \CC_{G'}(v)$. Since $e =
	(a,b)$ is a shortcut, there is a directed $ab$-path $P_{ab}$ that does not contain $e$.
	Consider now an arbitrary edge $f=(u,v)\neq e$ of $G$. If $f$ is a shortcut of $G$, there is a
	directed $uv$-path $P_{uv}$ in $G$ that does not contain $f$. There are two cases: $e$
	is not an edge in $P_{uv}$, or it is. In the first case put $P\coloneqq P_{uv}$. In the latter
	case, replace the edge $e$ in $P_{uv}$ by the path $P_{ab}$ and denote the resulting subgraph by
	$P$. Since $G$ is a DAG, the edge $f$ is not contained in $P_{ab}$ and
	thus $P$ is a directed path that does not contain $f$. Clearly, $P$ remains a directed $uv$-path in $G'$ that does not
	contain $f$, so $f$ is a shortcut of $G'$. If, instead, $f$ is not a shortcut of $G$, then
	any $uv$-path in $G$ must
	coincide with the edge $f$. Clearly, any $uv$-path must coincide with the edge $f$ in $G'$. In
	summary, an edge distinct from $e$ is a shortcut of $G$ if and only if
	it is a shortcut of $G'$. Consequently, $G'$ has $\ell-1$ shortcuts.
	The latter arguments directly imply that $G^-$ is uniquely determined.
	
	Finally, suppose that $G$ satisfies (PCC). Hence, $\CC_G(u)\subseteq
	\CC_G(v)$ precisely if $u$ and $v$ are $\prec_G$-comparable. Since $v\prec_G u$ if and only if
	$v\prec_{G'} u$ for all $u,v\in V$ and $\CC_G(w)= \CC_{G'}(w)$ for all $w\in V$, it immediately
	follows that $G'$ satisfies (PCC). By similar arguments, if $G'$ satisfies (PCC), then $G$
	satisfies (PCC).
\end{proof}

\paragraph{Hasse Diagrams and Regular DAGs.}

The \emph{Hasse diagram} $\Hasse(\mathfrak{C})$ of a set system $\mathfrak{C}\subseteq 2^X$ is the
DAG with vertex set $\mathfrak{C}$ and directed edges from $A\in\mathfrak{C}$ to $B\in\mathfrak{C}$
if (i) $B\subsetneq A$ and (ii) there is no $C\in\mathfrak{C}$ with $B\subsetneq C\subsetneq A$. We
note that $\Hasse(\mathfrak{C})$ is also known as the \emph{cover digraph} of $\mathfrak{C}$
\cite{Baroni:05}. The Hasse diagram $\Hasse(\mathfrak{C})$ is not necessarily phylogenetic. By way
of example, for $\mathfrak{C} = \{\{x,y\},\{x\}\}$, $\Hasse(\mathfrak{C})$ is a non-phylogenetic
network since its unique root $\{x,y\}$ has out-degree $1$ and in-degree $0$. Nevertheless, if
$\mathfrak{C}$ is a grounded set system, then the underlying Hasse diagram is phylogenetic, as we
will show in Lemma~\ref{lem:hasse-phylo}.

In general, we are interested in DAGs $G$ with certain properties and that satisfy
$\mathfrak{C}_G=\mathfrak{C}$ for a given grounded set system $\mathfrak{C}$. Structural properties
of $\Hasse(\mathfrak{C})$ are, in this context, often helpful. However,
$\mathfrak{C}_{\Hasse(\mathfrak{C})}\neq \mathfrak{C}$ holds as the leaves of $\Hasse(\mathfrak{C})$
are labeled with the inclusion-minimal elements in $\mathfrak{C}$, i.e., as sets. To circumvent
this, we write \[G\doteq \Hasse(\mathfrak{C})\] for the directed graph that is obtained from
$\Hasse(\mathfrak{C})$ by relabeling all vertices $\{x\}$ in $\Hasse(\mathfrak{C})$ by $x$. Thus,
for $G\doteq \Hasse(\mathfrak{C})$ it holds that $\mathfrak{C}_G=\mathfrak{C}$ provided that
$\mathfrak{C}$ is a grounded set system on $X$.

\begin{definition}[{\cite{Baroni:05}}]
  \label{def:regular-N}
  A DAG $G=(V,E)$ is \emph{regular} if the map
  $\varphi\colon V\to V(\Hasse[\mathfrak{C}_G])$ defined by  $v\mapsto \CC_G(v)$ is an
  isomorphism between $G$ and $\Hasse(\mathfrak{C}_G)$.  
\end{definition}

We emphasize that not every DAG $G\doteq \Hasse(\mathfrak{C})$ is regular. By way of example,
consider the set system $\mathfrak{C} = \{\{x\},\{x,y\}\}$ where $G\doteq \Hasse(\mathfrak{C})$
consists of a single edge and where each $v\in V(G)$ satisfies $\CC_{G}(v) = \{x\}$, i.e.,
$\varphi\colon G\to V(\Hasse(\mathfrak{C}_G))$ via $v\mapsto \CC_{G}(v)$ will map both of the
vertices of $G$ to $\{x\}$ and thus, does not yield an isomorphism between $G$ and
$\Hasse(\mathfrak{C})$. However, as we will see in Lemma~\ref{lem:hasse-phylo},
$\hasse(\mathfrak{C})$ is regular whenever $\mathfrak{C}$ is grounded.


\section{Least Common Ancestors and lca- \& LCA-Relevant DAGs}
\label{sec:LCA-rel}

For a given a DAG $G$ and a subset $A\subseteq L(G)$, a vertex $v\in V(G)$ is a \emph{common
ancestor of $A$} if $v$ is an ancestor of every vertex in $A$. Moreover, $v$ is a \emph{least common
ancestor} (LCA) of $A$ if $v$ is a $\preceq_G$-minimal vertex that is an ancestor of all vertices in
$A$. The set $\LCA_G(A)$ comprises all LCAs of $A$ in $G$. In general, not every set $A\subseteq
L(G)$ has a least common ancestor in a DAG: consider the DAG with three leaves $\{x,y,z\}$ and two
$\prec_G$-maximal vertices $p,q$ such that $\CC_G(p)=\{x,y\}$ respectively $\CC_G(q)=\{x,z\}$, in
which case $x$ and $y$ have no common ancestor at all and, therefore, $\LCA_G(\{y,z\})=\emptyset$.
In a network $N$, the unique root is a common ancestor for all $A\subseteq L(N)$ and, therefore,
$\LCA_N(A)\neq\emptyset$. We now  provide a simple characterization of vertices that belong to
$\LCA(A)$, generalizing \cite[Lem.~35, Obs.~11]{Hellmuth2023} from networks to all DAGs.

\begin{lemma}\label{lem:alg_LCA}
	For all DAGs $G=(V,E)$ on $X$, all nonempty subsets $A\subseteq X$  and vertices $v\in V$
	the following statements are equivalent.
	\smallskip
	\begin{enumerate}[label=(\arabic*)]
		\item  $v\in\LCA_G(A)$.
		\item  $A\subseteq \CC_G(v)$ and $A\not\subseteq \CC_G(u)$ for all $u\in\child_G(v)$. 
		\item  $A\subseteq \CC_G(v)$ and $A\not\subseteq \CC_G(u)$ for all $u\in V$ with $u\prec_G v$. 
	\end{enumerate}
	\smallskip
	In particular, if $v\in\LCA_G(A)$ for some $\emptyset\neq A\subseteq X$, 
	then $\CC_G(u)\neq\CC_G(v)$ for all $u\in\child_G(v)$.
\end{lemma}
\begin{proof}
	Let $G$ be a DAG on $X$ and $A\subseteq X$ nonempty. By definition, if $v\in\LCA_G(A)$, then $v$
	is an ancestor of every vertex in $A$ i.e. $A\subseteq \CC_G(v)$ and no descendant of $v$ is an
	ancestor of every vertex in $A$. Hence, for all $u\in V$ with $u\prec_G v$ at least one vertex
	in $A$ is not contained in $\CC_G(u)$, which implies that $A\not\subseteq \CC_G(u)$. Hence, (1)
	implies (3). Trivially, (3) implies (2). Now, suppose that Statement (2) is satisfied. Since $A$
	is not empty and $A\subseteq\CC_G(v)$, $v$ is a common ancestor of every vertex in $A$.
	Moreover, $A\not\subseteq \CC_G(u)$ for all $u\in\child_G(v)$ implies together with Lemma
	\ref{lem:prec-subset} that $A\not\subseteq \CC_G(w)$ for all descendants $w$ of $v$. Hence, $v$
	is a least common ancestor of the vertices in $A$, i.e., $v\in \LCA_G(A)$. Thus, (2) implies
	(1).

	Suppose now that $v\in\LCA_G(A)$ for some non-empty  $A\subseteq X$. 
	Thus, $A\subseteq \CC_G(v)$.	
	If $v$ has no children, then the statement is vacuously true. Hence, assume that 
	$v$ is an inner vertex. By statement (2), $A\not\subseteq \CC_G(u)$ and, therefore,
	$\CC_G(u)\neq\CC_G(v)$ for all $u\in\child_G(v)$.
\end{proof}

We will, in particular, be interested in situations where the LCA of
certain sets of leaves is uniquely defined. More precisely, we are
interested in DAGs where $|\LCA_G(A)|=1$ holds for certain subsets
$A\subseteq X$. For simplicity, we will write
$\lca_G(A)=v$ in case that $\LCA_G(A)=\{v\}$  and say that
\emph{$\lca_G(A)$ is well-defined}; otherwise, we leave $\lca_G(A)$
\emph{undefined}. 

\begin{definition}[$\kLCA$ and $\klca$ vertices]\label{def:Ilca}
	Let $G$ be a DAG on $X$, $k\geq 1$ be an integer and $v\in V(G)$. 
	\smallskip
	\begin{enumerate}
		\item	The vertex $v$ is a \emph{$\kLCA$ vertex} if $v\in \LCA_G(A)$ for some subset $A\subseteq X$ of size $|A|=k$.

		\item	The vertex	$v$ is a \emph{$\klca$ vertex} if $v = \lca_G(A)$  for some subset $A\subseteq X$ of size $|A|=k$.
	\end{enumerate}		
	\smallskip
	For a subset $\mathscr{I}\subseteq \{1,\dots,|X|\}$, the vertex
	$v$ is an \emph{$\mathscr{I}$-$\LCA$ vertex} (resp., $\mathscr{I}$-$\lca$ vertex) if it
		is a $\kLCA$ vertex (resp., $\klca$ vertex) for some $k\in \mathscr{I}$. 
\end{definition}

\begin{figure}
	\centering
	\includegraphics[width=1.\textwidth]{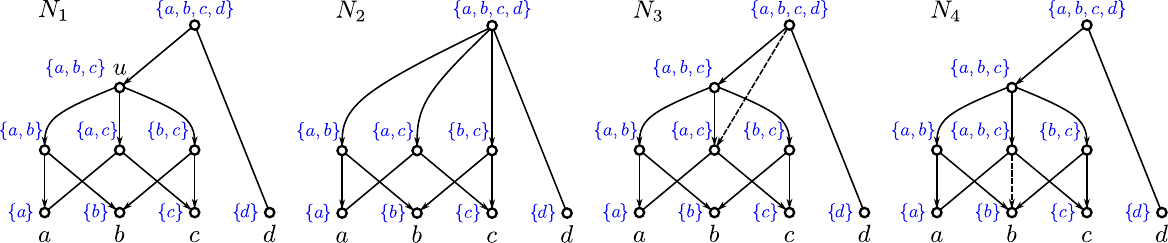}	
	\caption{Shown are four phylogenetic networks $N_1$, $N_2$, $N_3$ and $N_4$ with the same set of leaves.
				Here, $N_1$ and $N_2 =N_1\ominus u$ are regular networks. The networks $N_3$ and $N_4$ only differ 
				from $N_1$ by one edge each, as highlighted by dashed lines.
				Each inner vertex $v$ of these networks with $\CC_{N_i}(v)\neq \{a,b,c\}$ is a $2$-$\lca$-vertex.
				In $N_1$, the vertex $u$ with cluster 
				$\CC_{N_1}(u)=\{a,b,c\}$ is not a $2$-$\lca$ vertex, but a $3$-$\lca$ vertex.
				Consequently, $N_1$ is a $\{1,2,3\}$-\lca-\rel network but not $\{1,2\}$-\lca-\rel. One
				may also verify that the same holds for the network $N_3$ but that $N_2$ is
				$\{1,2\}$-\lca-\rel. For $N_4$ we can apply Lemma~\ref{lem:not_kLCA}
				to the edge ($u,u')$
				connecting the vertices $u$ and $u'$
				for which $\CC_{N_4}(u)=\{a,b,c\}=\CC_{N_4}(u')$ holds and conclude
				that $N_4$ is not \LCA-\rel and, therefore, not \lca-\rel. 
				In particular, the vertex $u$ in $N_4$ is not the LCA of any subset of leaves.
				}
	\label{fig:lca-relevant-expls}
\end{figure}

By Lemma~\ref{lem:alg_LCA}, a vertex $v$ can neither be a $k$-$\LCA$ nor a $k$-$\lca$ vertex
whenever $k>|\CC(v)|$. A less obvious relationship between $k$-$\LCA$ and $\ell$-$\LCA$ vertices
(respectively $k$-$\lca$ and $\ell$-$\lca$ vertices) is captured by the following lemma.

\begin{lemma}\label{lem:k>lca}
	If $v$ is a $k$-$\LCA$ vertex of a DAG $G$ for some $k\geq1$, then $v$ is an $\ell$-$\LCA$ vertex of $G$ for all $\ell$ with $k\leq \ell\leq |\CC_G(v)|$.
	If $v$ is a $k$-$\lca$ vertex of a DAG $G$ for some $k\geq1$, then $v$ is an $\ell$-$\lca$ vertex of $G$ for all  $\ell$ with $k\leq \ell\leq |\CC_G(v)|$.
\end{lemma}
\begin{proof}
	Let $G$ be a DAG on $X$. It is an easy task to verify that a vertex is a $1$-$\lca$ vertex if
	and only if it is a leaf which, in turn, happens if and only if it is a $1$-$\LCA$ vertex.
	Since, for a leaf $x\in X$ we have $\CC_G(x)=\{x\}$, the two statements are trivial for the case
	when $k=1$.

	Suppose now that $v$ is a $k$-$\LCA$ vertex of $G$ for some $k\geq 2$. Hence, there is some set
	$A\subseteq X$ of size $|A|=k\geq 2$ such that $v\in \LCA_G(A)$. By Lemma~\ref{lem:alg_LCA},
	$A\subseteq \CC_G(v)$ and $A\not\subseteq \CC_G(u)$ for all children $u$ of $v$. Clearly, the latter
	property remains for all subsets $A'\subseteq \CC_G(v)$ with $A\subseteq A'$. By Lemma
	\ref{lem:alg_LCA}, $v\in \LCA_G(A')$ for all such $A'\subseteq \CC_G(v)$ with $A\subseteq A'$.
	Therefore, $v$ is a $\ell$-$\LCA$ vertex of $G$, $k\leq \ell\leq |\CC_G(v)|$.
	
	Suppose now that $v$ is a $k$-$\lca$ vertex. Hence, there is some some set $A\subseteq X$ of
	size $|A|=k$ such that $v = \lca_G(A)$ and, therefore, $\LCA_G(A)=\{v\}$. Let $A'\subseteq X$ be
	such that $A\subseteq A'\subseteq \CC_G(v)$. We show that $\LCA_G(A')=\{v\}$. Since $A'\subseteq
	\CC_G(v)$, $v$ is a common ancestor of $A'$. Hence, there is a vertex $w\preceq_G v$ that is a
	least common ancestor of $A'$. Since $A\subseteq A'$, this vertex $w$ is also a common ancestor
	of $A$. But this implies that $w\prec_G v$ is not possible since $v=\lca_G(A)$. Hence, $w=v$ must
	hold, i.e., $v\in\LCA_G(A')$. Assume, for contradiction, that there exists some $u\in\LCA_G(A')$
	such that $u\neq v$. Note that $u$ and $v$ must be $\preceq_G$-incomparable. Since $A\subseteq
	A'$ and $u\in\LCA_G(A')$, the vertex $u$ is in particular a common ancestor of the vertices in
	$A$. This, together with $\lca_G(A)=v$, means $v\preceq_G u$; a contradiction. Consequently,
	$\LCA_G(A') =\{v\}$. Hence, $v=\lca_G(A')$ must hold for all $A\subseteq A'\subseteq \CC_G(v)$. In
	summary, $v$ is a $\ell$-$\lca$ vertex of $G$, $k\leq \ell\leq |\CC_G(v)|$. 
\end{proof}

Generally, the converse of Lemma~\ref{lem:k>lca} is not satisfied. Consider, for example, 
the vertex $u$ of the network $N_1$ that satisfies $\CC_{N_1}(u)=\{a,b,c\}$ in Figure~\ref{fig:lca-relevant-expls}, which is a $3$-$\lca$ vertex (thus, in
particular, a $3$-$\LCA$ vertex), but neither a $2$-$\lca$ vertex nor a $2$-$\LCA$ vertex. 

The following two results provide a characterization of vertices that are not $\{1,\dots,|X|\}$-$\LCA$, 
resp., not $\{1,\dots,|X|\}$-$\lca$ vertices. These result will be employed in Section~\ref{sec:ominus-lcaRel}
to efficiently transform a given DAG $G$ into a DAG $G'$
in which all vertices are $\kLCA$ or $\klca$ vertices for at least one $k\in \{1,\dots,|X|\}$.

\begin{lemma}\label{lem:not_kLCA}
	For a  DAG $G$ on $X$ and a vertex $v\in V(G)$, the following statements are equivalent. 
	\smallskip
	\begin{enumerate}[label=(\arabic*)]
	\item $v$ is not a $\kLCA$ vertex for any $k\in \{1,\dots,|X|\}$. 
	\item there is a child $u$ of $v$ in $G$ such that $\CC_G(u)=\CC_G(v)$. 
	\item $v\notin \LCA_G(\CC_G(v))$
	\end{enumerate}
	\smallskip
\end{lemma}
\begin{proof}
	Let $G$ be a DAG on $X$,  $v\in V(G)$ and put $C\coloneqq \CC_G(v)$.
	If $v$ is not a $\kLCA$ vertex for any $k\in \{1,\dots,|X|\}$, then 
	in particular, $v\notin \LCA_G(C)$.
	By Lemma~\ref{lem:alg_LCA}, there must be a child $u$ of $v$ such that $C\subseteq \CC_G(u)$. 
	Since $u\prec_G v$, Lemma~\ref{lem:prec-subset} implies $\CC_G(u)\subseteq \CC_G(v)=C$ and thus, 
   $\CC_G(u)=C$. Hence, (1) implies (2). 
    If there is a child $u$ of $v$ in $G$ such that $\CC_G(u)=C$, then   
    $u\prec_G v$ implies that $v\notin \LCA_G(C)$, i.e.,  (2) implies (3). 
	Assume now that  $v\notin \LCA_G(C)$ and put $\ell\coloneqq |C|$. 
	Clearly, $v$ is not a $\kLCA$ vertex for any $A\subseteq X$ of size $|A|>\ell$
	as, in this case, $A\not\subseteq C$.
	Moreover, since $v$ is not an $\ell$-$\LCA$ vertex,
	contraposition of Lemma~\ref{lem:k>lca} implies that $v$ is not a $\klca$ vertex for any $k\in \{1,\dots,\ell\}$.
	Thus, (3) implies (1). 
\end{proof}

\begin{lemma}\label{lem:not_klca}
	For	 a  DAG $G$ on $X$ and a vertex $v\in V(G)$, 
	the following statements are equivalent.
	\smallskip
	\begin{enumerate}[label=(\arabic*)]
	\item $v$ is not a $\klca$ vertex for any $k\in \{1,\dots,|X|\}$. 
	\item there is a child $u$ of $v$ in $G$ such that $\CC_G(u)=\CC_G(v)$ or $|\LCA_G(\CC_G(v))|\geq2$. 
	\item $v\neq \lca_G(\CC_G(v))$.
	\end{enumerate}
	\smallskip
\end{lemma}
\begin{proof}
	Let $G$ be a DAG on $X$, $v\in V(G)$ and put $C\coloneqq \CC_G(v)$. 
	We start with showing that (1) implies (2). Suppose that $v$ is not a
	$\klca$ vertex for any $k\in \{1,\dots,|X|\}$, then in particular, $ v\neq \lca_G(C)$. Thus, if
	$v\in\LCA_G(C)$, then $|\LCA_G(C)|\geq2$. If
	$v\not\in\LCA_G(C)$, then Lemma~\ref{lem:alg_LCA} implies that there must be a child $u$ of $v$ such
	that $C\subseteq \CC_G(u)$. Since $u\prec_G v$, Lemma~\ref{lem:prec-subset} implies
	$\CC_G(u)\subseteq C$ and thus, $\CC_G(u)=C$. Thus, (1) implies (2). Assume now that statement (2)
	holds. If $|\LCA_G(C)|\geq2$, then in particular $v\neq \lca_G(C)$. If there is a child $u$ of
	$v$ in $G$ such that $\CC_G(u)=C$, then Lemma~\ref{lem:not_kLCA} implies that $v\notin \LCA_G(C)$
	and thus, $v\neq \lca_G(C)$. Hence, (2) implies (3). Finally, suppose that $v\neq \lca_G(C)$.
	Hence, $v$ is not an $\ell$-$\lca$ vertex for $\ell=|C|$. Contraposition of Lemma~\ref{lem:k>lca}
	implies that $v$ is not a $\klca$ vertex for any $k\in \{1,\dots,\ell\}$. Clearly, $v$ is not an
	$\klca$ vertex for any $A\subseteq X$ of size $|A|>\ell$ as, in this case, $A\not\subseteq C$.
	Consequently, (3) implies (1).
\end{proof}

Lemma~\ref{lem:not_kLCA} and \ref{lem:not_klca} will be useful in both Section~\ref{sec:regular-CL}
and Section~\ref{sec:ominus-lcaRel}. The contrapositive of statements (1) and (3) in these lemmas
together with the fact that $v$ can only be a $\klca$ or $\kLCA$ vertex if $k\leq |\CC_G(v)|$ imply
\begin{corollary}\label{cor:klca<=>v=lca(C(v))}
	Let $G$ be a DAG on $X$,  $v\in V(G)$. Then,  
	$v$  is a  $\kLCA$ vertex in $G$ for some $k$ if and only if $v\in \LCA_G(\CC_G(v))$.
	Moreover, $v$  is a $\klca$ vertex in $G$ for some $k$ if and only if $v=\lca_G(\CC_G(v))$.
	In both cases, $k\leq |\CC_G(v)|$.
\end{corollary}

In what follows, we consider DAGs for which each vertex $v$ satisfies $v\in \LCA(A)$ or $v =
\lca(A)$ for some set $A$ whose size $k=|A|$ is contained in a specified set $\mathscr{I}$ of
integers. 

\begin{definition} \label{def:LCAlcarel}
	Let $G$ be a DAG on $X$, $v\in V(G)$ and $\mathscr{I}$ be a set of integers. 
		\smallskip
	\begin{enumerate}
		\item  
		$G$ is \emph{\OlcaTMP-relevant} (in short \OlcarelTMP)	if  all vertices in $V(G)$ are \OlcaTMP vertices.
			
					DAGs that are $\{1,2,\dots,|X|\}$-$\lca$-\rel 	are simply called $\lca$-\rel.
		\smallskip
		
		\item 
		$G$ is \emph{\OLCATMP-relevant} (in short \OLCArelTMP) if  all vertices in $V(G)$ are \OLCATMP vertices. 
		
					DAGs that are $\{1,2,\dots,|X|\}$-$\LCA$-\rel are simply called $\LCA$-\rel.
	\end{enumerate}
		\smallskip
\end{definition}

Thus, $G$ is $\LCA$-\rel if each vertex $v$ in $G$ is a least common ancestor for at least some set
$A\subseteq X$. Similarity, $G$ is $\lca$-\rel if, for all $v\in V(G)$, there is some set
$A\subseteq X$ such that $v= \lca_G(A)$. 
\begin{observation}\label{obs:Ilca-rel}
Every \OlcarelTMP DAG is $\lca$-\rel and $\LCA$-\rel. Every \OLCArelTMP DAG is $\LCA$-\rel.
\end{observation}

There are DAGs and even networks that are not $\mathscr{I}$-$\lca$-\rel for any
$\mathscr{I}\subseteq \{1,\dots, |X|\}$, see the network $N_4$ in
Figure~\ref{fig:lca-relevant-expls}. In contrast, for every set $\mathscr{I} \subseteq
\{1,\dots,n\}$ with $1\in \mathscr{I}$, there is an \OlcarelTMP DAG $G$. To see this, let
$\mathscr{I} = \{i_1,\dots, i_\ell\}$ with $1=i_1<i_2<\dots<i_\ell$ and take the Hasse diagram of
the clustering system $\{\{j\}\mid 1\leq j\leq i_\ell\} \cup \{\{1,\dots,j\} \mid j\in
\{i_2,\dots,i_\ell\}\}$ which is \OlcarelTMP.

The requirement that $1\in \mathscr{I}$ is indispensable for \OlcarelTMP DAGs. To see this,
observe that every leaf $x \in  L(G)$ of a DAG  $G$  satisfies $\LCA_G(\{x\}) = \{x\}$
and that $x$ cannot be an ancestor of any vertex $y\neq x$. Hence, a leaf $x \in  L(G)$ is
always a $1$-$\lca$ vertex but never a $k$-$\lca$ vertex for $k>1$.
Since $L(G)\neq \emptyset$ for all DAGs $G$, 
every DAG contains  at least one $1$-$\lca$ vertex and thus, at least one $1$-$\LCA$ vertex.
Hence, if \emph{all} vertices  are $\mathscr{I}$-$\LCA$ or $\mathscr{I}$-$\lca$ vertices, 
then  $1\in \mathscr{I}$ must hold. The latter is captured by the following

\begin{definition}\label{def:1inI}
For a given DAG $G$, the set $\One$ always denotes a subset of $\{1,\dots, |L(G)|\}$ that satisfies $1\in \One$.
\end{definition}

A useful structural property of $\LCA$-$\rel$ and $\lca$-$\rel$ DAGs is provided next.
\begin{lemma}\label{lem:LCA-REL=>phylo} 
An $\LCA$-$\rel$ or $\lca$-$\rel$ DAG does not contain vertices $w$ with $\outdeg_G(w)=1$ and is, thus, phylogenetic.
\end{lemma}
\begin{proof}
Let $G$ be an $\LCA$-$\rel$ DAG. Contraposition of Lemma~\ref{lem:not_kLCA} shows that $G$ cannot
contain vertices $w$ that have a child $u$ such that $\CC_G(w) = \CC_G(u)$. Hence, $G$ can, in
particular, not have any vertex with a single child, i.e., $\outdeg_G(w)\neq 1$ for all $w\in V(G)$.
Thus, $G$ is phylogenetic. Since every $\lca$-$\rel$ DAG is, in particular, $\LCA$-$\rel$ the
statement holds for $\lca$-$\rel$ DAGs as well.
\end{proof}

We finally show that the set $\LCA_G(A)$ in a DAG $G = (V, E)$ can be determined in $O((|V|+|E|)|A|)$ time.
Assuming that the size of $A$ is treated as constant, i.e., $|A|\in O(1)$, this
result implies that $\LCA_G(A)$ can be determined in linear time. To achieve this goal, we use a
\emph{topological order} $\ll$ on $V$, i.e., a total order on the vertices in $G$ such that $(u,v) \in E$
implies $v\ll u$. Since we consider DAGs, such an order always exists \cite{cormen2022introduction}.
The pseudocode of the underlying algorithm is provided in Algorithm~\ref{alg:LCA}. The main idea of this
algorithm is as follows: we determine the set $A\setminus\CC_G(v_i)$, which is stored in $C[i]$, for
each $v_i\in V$. We then employ Lemma~\ref{lem:alg_LCA} which states that $v_i\in \LCA_G(A)$ if and
only if $A\subseteq \CC_G(v_i)$ and $A\not\subseteq \CC_G(v_j)$ for all $v_j\in\child_G(v_i)$ which, in
turn, is precisely if $C[i] = \emptyset$ and $C[j] \neq \emptyset$ for all $j$ such that
$v_j\in\child_G(v_i)$ (see Line~\ref{algLCA:checkA} in Algorithm~\ref{alg:LCA}).

\begin{algorithm}[t]\small
	\caption{\texttt{Find\_\LCA\_of\_set\_$A$}}\label{alg:LCA}
	\begin{algorithmic}[1]
		\Require A DAG $G=(V,E)$ on $X$ and a nonempty set $A\subseteq X$
		\Ensure The set $\LCA_G(A)$
		\If{$|A|=1$}
			\Return $A$ \label{algLCA:A=1}
		\EndIf
		
		\State Compute topological order $ \ll $ on the elements in $V$ resulting in the order $v_1  \ll  v_2  \ll  \dots  \ll  v_n$  \label{algLCA:topo}
		
		\State Sort $A$ w.r.t.\  $\ll$ \label{algLCA:sortA}
		\State Initialize the array $C$ of size $n = |V|$ whose entries are empty sets \Comment{$C[i]$ will store $A\setminus\CC_G(v_i)$} \label{algLCA:initC}
		\State Initialize the empty set $\mathsf{LCA}$ \label{algLCA:initLCA}

		\For{$i=n, \dots, 1$ (in this order)} \label{algLCA:forStart}
			\If{$v_i$ is a leaf of $G$} \label{algLCA:leafCheck}
				\State $C[i]\gets A\setminus\{v_i\}$, kept sorted w.r.t.\ $\ll$ \label{algLCA:leafC}
			\Else
				\State $C[i]\gets \bigcap_{v_j\in\child_G(v_i)} C[j]$, kept sorted w.r.t.\ $\ll$  \label{algLCA:otherC}
				\If{$C[i]=\emptyset$ and $C[j]\neq\emptyset$ for every $v_j\in\child_G(v_i)$}
					Add $v_i$ to $\mathsf{LCA}$ \label{algLCA:checkA}
				\EndIf
			\EndIf
		\EndFor \label{algLCA:forEnd}
		\State \Return $\mathsf{LCA}$
	\end{algorithmic}
\end{algorithm}

\begin{proposition}\label{prop:algoLCA}
	For a given DAG $G=(V,E)$ on $X$ and a non-empty set $A\subseteq X$, 
	Algorithm~\ref{alg:LCA} correctly determines $\LCA_G(A)$. Moreover,  
	Algorithm~\ref{alg:LCA}  can be implemented to run in $O((|V|+|E|)|A|)$ time. 
\end{proposition}
\begin{proof}
	Let the DAG $G=(V,E)$ on $X$ and the non-empty set $A\subseteq X$ serve as an input for Algorithm
	\ref{alg:LCA}. For simplicity, put $\LCA(\cdot) \coloneqq \LCA_G(\cdot)$.
	We start with proving the correctness of Algorithm~\ref{alg:LCA}. In Line
	\ref{algLCA:A=1}, we first check if $|A|=1$ and, in the affirmative case, $\LCA(A)=A$ must hold
	and the algorithm correctly returns $A$. Otherwise, if $|A|>1$, the algorithm continues as
	follows. In Line~\ref{algLCA:topo}, the vertices in $V$ are topologically ordered resulting in
	$v_1 \ll v_2 \ll \dots \ll v_n$, where $n\coloneqq|V|$. Sorting $A$ in Line~\ref{algLCA:sortA} and maintaining
	sorted elements in Lines~\ref{algLCA:leafC} and \ref{algLCA:otherC} are primarily used to
	establish the runtime but do not influence the correctness proof. Thus, we can treat the set $A$
	and the array $C$ as unordered for now. The array $C$ and
	the empty set $\mathsf{LCA}$ is initialized in Line~\ref{algLCA:initC} and \ref{algLCA:initLCA},
	respectively. The entry $C[i]$ will store the elements of $A\setminus\CC_G(v_i)$. The main idea
	of this algorithm is based on Lemma~\ref{lem:alg_LCA} which states that $v_i\in \LCA(A)$
	precisely if $A\subseteq \CC_G(v_i)$ and $A\not\subseteq \CC_G(v_j)$ for all $v_j\in\child_G(v_i)$. The
	latter is precisely if $C[i] = \emptyset$ and $C[j] \neq \emptyset$ for all $j$ that correspond
	to indices of the children $v_j$ of $v_i$. It thus suffices to show that $C[i] $ is correctly
	determined for all $i\in \{1,\dots, n\}$. In the \emph{for}-loop in Line~\ref{algLCA:forStart}
	the vertices are processed in the order $v_n, v_{n-1},\dots v_1$. Based on the topological
	order, this ensures that, whenever $v_i$ is processed, all its descendants have been processed as
	they must be located in $v_{i+1},\dots,v_n$. If $v_i$ is a leaf of $G$, then $\CC_G(v_i) =\{v_i\}$
	holds. Hence, $C[i]=A\setminus\CC_G(v_i)=A\setminus\{v_i\}$ is correctly determined in
	Line~\ref{algLCA:leafC}. Otherwise, i.e. if $v_i$ is an inner vertex, then we put $C[i] =
	\cap_{v_j\in\child_G(v_i)} C[j]$ in Line~\ref{algLCA:otherC}. By the latter arguments and
	induction, for each such $v_j$ the set $C[j]$ has already been correctly determined. This and
	Lemma~\ref{lem:union-A} implies that
	$C[i]=\cap_{v_j\in\child_G(v_i)}A\setminus\CC_G(v_j)=A\setminus\CC_G(v_i)$ is correctly determined. In
	Line~\ref{algLCA:checkA}, we simply verify if the conditions of Lemma~\ref{lem:alg_LCA} are
	satisfied and, in the affirmative case, $v_i$ is correctly added to $\mathsf{LCA}$. In summary, $\LCA(A)$
	is correctly determined. 
	
	Let us now consider the runtime of Algorithm~\ref{alg:LCA}. To this end, we assume that $n=|V|$, 
	$m=|E|$ and $k=|A|$. Line~\ref{algLCA:A=1} takes constant time. Determining the topological order
	$\ll$ of $V$ can be done in $O(n+m)$ time and sorting $A$ in Line~\ref{algLCA:sortA} can be done
	in $O(k\log{k})$ time \cite{cormen2022introduction}. The tasks in Line
	\ref{algLCA:initC}-\ref{algLCA:initLCA} can be accomplished in $O(n)$ time. We may assume that
	the DAG $G$ is represented as an adjacency list $L$. In this case, we can traverse all entries of
	$L$ and check whether the entry $L[v_i]$ is empty (resp.\ non-empty) in which case $v_i$ is a
	leaf (resp.\ inner vertex). As a pre-processing step this takes $O(n)$ time and we can,
	afterwards, check in constant time as whether $v_i$ is a leaf or not. Thus, the
	\emph{if}-condition in Line~\ref{algLCA:leafCheck} can be evaluated in constant time. In
	Line~\ref{algLCA:leafC}, we can traverse the sorted set $A$, adding all elements except
	$v_i$ to the ordered set $C[i]$ in $O(k)$ time, keeping the order of the elements. Since
	this is repeated for the $|X|$ leaves of $G$, Line~\ref{algLCA:leafC} contributes with $O(k|X|)$
	over all iterations.
	In Line~\ref{algLCA:otherC}, we compute the intersection of ordered sets and keep the order. The
	intersection of two sorted sets $S$ and $S'$ resulting in a sorted set can be done in
	$O(|S|+|S'|)$ time \cite{AHU:83}. Each set $C[j]$ has $O(k)$ elements, as they are subsets of
	$A$. Thus, computing the $\ll$-sorted set $C[i]$ as the intersection $\bigcap_{v_j\in\child_G(v_i)}
	C[j]$ can be done $O(|\child_G(v_i)|k)$ time. As the latter task is repeated for all inner vertices
	of $G$, the total runtime, for Line~\ref{algLCA:otherC} is $O(\sum_{i=1}^n (|\child_G(v_i)| k)) =
	O(mk)$ time. In Line~\ref{algLCA:checkA}, we simply make $|\child_G(v_i)|+1$ constant time look-ups
	in the array $C$ to determine whether the ordered sets are empty or not. Once again summing over
	all inner vertices, this contributes with $O(m)$ to the total runtime.

	In summary, the total runtime of Algorithm~\ref{alg:LCA} is in $O(n+m+k\log{k}+k|X|+mk)$ time.
	Since $\log{k}<k\leq|X|$ and $|X|\leq n$, the
	terms $k|X|$ and $k\log{k}$ are both dominated by the term $nk$. Thus, the total runtime
	simplifies to $O(nk+mk)$.
\end{proof}

	Since $\LCA_G(A)$ can be determined for $G=(V,E)$ in $O((|V|+|E|)|A|)$ time and since 
	$|\LCA_G(A)| \in O(|V(G))|$, the additional costs for checking if  $v\in \LCA_G(A)$ or
	$v=\lca_G(A)$ add $O(|V|)$ to the cost of computing  $\LCA_G(A)$. 
	Moreover, by Corollary~\ref{cor:klca<=>v=lca(C(v))}, 
		a DAG is $\LCA$-\rel, resp., $\lca$-\rel if and only if, for all of its vertices
		$v$, it holds that $v\in \LCA_G(\CC_G(v))$, resp., $v=\lca_G(\CC_G(v))$.
		Summarizing the latter arguments, we obtain
\begin{corollary}\label{cor:check-LCA/lca-fast}
	For a given DAG $G=(V,E)$ on $X$, a non-empty set $A\subseteq X$ and a vertex $v\in V$,
	it can be determined in  $O((|V|+|E|)|A|)$ time if $v\in \LCA_G(A)$ and if  $v=\lca_G(A)$.
	Moreover, it can decided in polynomial time if $G$ is $\LCA$-\rel (resp., $\lca$-\rel) or not.
\end{corollary}


\section{Characterization of  lca- \& LCA-Relevant DAGs and Regular DAGs }
\label{sec:regular-CL}

By  Corollary~\ref{cor:klca<=>v=lca(C(v))},
 every $\klca$ vertex $v$ and therefore, every vertex $v$ in an \Olca-\rel DAG,
satisfies $v = \lca_G(\CC_G(v))$. To cover such type of DAGs we provide

\begin{definition}\label{def:CLstrongCL}
A DAG $G$ satisfies the \emph{cluster-lca (CL)} property if 
$\lca_G(\CC_G(v))$  is well-defined for all  $v\in V(G)$. 
A DAG $G$ has the \emph{strong cluster-lca (strong-(CL))} property if
$v = \lca_G(\CC_G(v))$ for all $v\in V(G)$. 
\end{definition}

\begin{figure}
	\centering
	\includegraphics[width=0.5\textwidth]{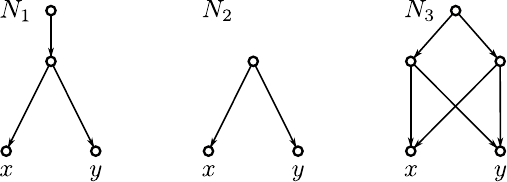}	
	\caption{Shown are three networks $N_1$, $N_2$ and $N_3$ having the same
				clustering system $\mathfrak{C} = \{\{x\}, \{y\}, \{x,y\}\}$. 
				The network $N_1$ has the (CL) but not the strong-(CL) property. 
				The network $N_2$ has the strong-(CL) and, thus, also the (CL) property. 
				The network $N_3$ has neither the strong-(CL) nor the (CL) property.
	} 
	\label{fig:CL}
\end{figure}

By definition,  strong-(CL) implies (CL). However,
there are DAGs with the (CL) property but without the strong-(CL) property
and DAGs without (CL) property, see Figure~\ref{fig:CL} for an example.

\begin{lemma}\label{lem:PCC=>CL}
If a DAG $G$  satisfies (PCC) then it satisfies (CL). If $G$ satisfies (CL),  then $\lca_G( \CC_G (v)) \preceq_G v$ and
$\CC ( \lca_G (\CC_G (v))) = \CC_G (v)$  for all $v\in V(G)$.
\end{lemma}
\begin{proof}
We emphasize first that these results have been proven for the case that $G$ is a network, 
cf.\ \cite[Lem.~36 \& 38]{Hellmuth2023}. Suppose that $G$ is DAG on $X$ that is not a network and thus, $|R(G)|>1$. 
Let $N$ be the network obtained from $G$ by adding a new root $\rho$ to $G$ and edges $(\rho,r)$
for all $r\in R(G)$. By construction $V(G) = V(N)\setminus \{\rho\}$, $\CC_N(\rho)=X$ and 
$\CC_N(v) = \CC_G(v)$ for all $v\in V(G)$. Assume first that $G$ that satisfies (PCC).
It is straightforward to verify that $N$ satisfies (PCC). Therefore, $N$ satisfies (CL).
One easily observes that, for all $v\in V(G)$, we have $\lca_N (\CC_N (v))=\lca_G (\CC_G (v))$.
Since $N$ satisfies (CL),  $G$ satisfies (CL). Moreover, since 
the statements are true for $N$ and since $\lca_N (\CC_N (v))=\lca_G (\CC_G (v))$, 
we can conclude that the second statement is satisfied for the DAG $G$.
\end{proof}

We provide now a simple characterization of DAGs with (CL) property.

\begin{proposition}
A DAG $G = (V,E)$ has the (CL) property if and only if, for every vertex $v\in V$,
$v = \lca_G(\CC_G(v))$ or $v$ has a child $u$ such that $\CC_G(v)=\CC_G(u)$. 
\end{proposition}
\begin{proof}
	 Let $G = (V,E)$ be DAG and $v\in V$. Suppose that
    $G$ has the (CL) property. If $v = \lca_G(\CC_G(v))$, then we are done. Hence, assume that
    $v \neq \lca_G(\CC_G(v))$. Then, Lemma~\ref{lem:not_klca} implies that there either
    is a child $u$ of $v$ in $G$ with $\CC_G(v)=\CC_G(u)$, or $|\LCA_G(\CC_G(v))|\geq2$. However, 
    the latter cannot hold since $G$ has the (CL) property, which establishes the \emph{only if}-direction.
	
	Conversely, assume that every vertex $v\in V$ satisfies: (a) 
	$v = \lca_G(\CC_G(v))$ or (b) $v$ has a child $u$ such that $\CC_G(v)=\CC_G(u)$. 
	If $v = \lca_G(\CC_G(v))$, then $\lca_G(\CC_G(v))$ is well-defined.
	Suppose that $v$ has a child $u$ such that
	$\CC_G(v)=\CC_G(u)$. We can now take a $\preceq_G$-minimal vertex $w$ that satisfies $w\preceq_G
	u$ and $\CC_G(w)=\CC_G(v)$. If $w$ is a leaf, then $\CC_G(w)=\{w\}$ and we have $w=\lca_G(\CC_G(w))$
	which implies that $\lca_G(\CC_G(w)) = \lca_G(\CC_G(v))$ is well-defined. Otherwise, $w$ is an
	inner vertex. By choice of $w$, all children $u'$ of $w$ must satisfy $\CC_G(u')\neq \CC_G(w)$,
	i.e., $w$ does not satisfy (b) and must therefore, satisfy (a) 
	i.e. that $w =\lca_G(\CC_G(w))$. This together with $\CC_G(w)=\CC_G(v)$
	implies that $\lca_G(\CC_G(v))$ is well-defined. In summary, $G$ satisfies (CL).
\end{proof}

As we shall see later, there is a close relationship between regular DAGs, DAGs that
are $\lca$-$\rel$ and DAGs with the strong-(CL) property. 
Before considering $\lca$-$\rel$ DAGs, we provide a characterization of $\LCA$-$\rel$ DAGs
that is an immediate consequence of Lemma~\ref{lem:not_kLCA} applied to all vertices.

\begin{theorem}\label{thm:LCArel-char}
	A DAG $G$ is $\LCA$-$\rel$ if and only if there are no adjacent vertices $u$ and $v$ in $G$ that satisfy $\CC_G(u)=\CC_G(v)$. 
\end{theorem}

The more specific property of being $\lca$-$\rel$ imposes more structural constraints on the DAG in question
which, in turn,	 allows us to provide the following characterization.

\begin{theorem}\label{thm:strongCLChar}
	The following statements are equivalent for every DAG $G$. 
	\smallskip
	\begin{enumerate}[label=(\arabic*)]
		\item $G$ is $\lca$-\rel
		\item $G$ has the strong-(CL) property.
		\item $G$ has the (CL) property and is $\LCA$-\rel.
		\item $G$ satisfies (PCC) and is $\LCA$-\rel.
		\item $G$ satisfies (PCC) and $u\neq v$ implies $\CC_G(u)\neq \CC_G(v)$ for all $u,v\in V(G)$. 
		\item $\CC_G(u)\subseteq \CC_G(v)$ if and only if $u\preceq_G v$ for all $u,v\in V(G)$.
	\end{enumerate}
	\smallskip
\end{theorem}
\begin{proof}
	By Corollary~\ref{cor:klca<=>v=lca(C(v))}, Statements (1) and (2) are equivalent. Now, assume
	that Statement (3) holds. Let $v$ be a vertex of $G$. Corollary~\ref{cor:klca<=>v=lca(C(v))}
	together with the fact that $v$ is a $k$-$\LCA$ vertex for some $k$ implies that $v\in\LCA_G(\CC_G(v))$. 
	Since $G$ has the (CL) property, we have $|\LCA_G(\CC_G(v))|=1$. The latter two arguments
	imply that $v=\lca_G(\CC_G(v))$. Since $v$ was chosen arbitrarily, $G$ is an $\lca$-\rel DAG,
	i.e., Statement (1) holds. Since the two equivalent Statements (1) and (2) together
	immediately imply Statement (3), we conclude that Statements (1), (2) and (3) are equivalent.
	
	Therefore, it suffices to show that the following implications 
	(2) $\Rightarrow$ (4) $\Rightarrow$ (5) $\Rightarrow$ (6) $\Rightarrow$ (2) hold.
	Assume that Condition (2) holds. Hence, $G$ is a DAG that has the strong-(CL) property.
	To show (PCC),
	observe that, by Lemma~\ref{lem:prec-subset}, $u\preceq_G v$ implies $\CC_G(u)\subseteq \CC_G(v)$
	for all $u,v\in V(G)$. Suppose now that $u,v\in V(G)$ are such that $\CC_G(u)\subseteq \CC_G(v)$.
	Since $G$ satisfies strong-(CL), $u=\lca_G(\CC_G(u))$. This together with $\CC_G(u)\subseteq
	\CC_G(v)$ and the definition of $\lca$s implies that $u=\lca_G(\CC_G(u)) \preceq_G v$. Hence,
	$G$ satisfies (PCC). In addition,
	$G$ is $\LCA$-\rel since (2)  and (3) are equivalent. In summary, (2) implies (4).	
	
	Assume that Condition (4) holds. Hence, $G$ satisfies (PCC)  and is $\LCA$-\rel. 
	Since $G$ satisfies (PCC), there cannot be any incomparable vertices $u,v$ in $G$
    with $\CC_G(v) = \CC_G(u)$. Thus, any two incomparable vertices have distinct clusters. 
	 By Theorem~\ref{thm:LCArel-char}, 
 	 no adjacent vertices $u$ and $v$ in $G$ can satisfy $\CC_G(u)=\CC_G(v)$. 
   This together with Lemma~\ref{lem:prec-subset} implies that for any two vertices $u,v$ in $G$ with 
   $u\prec_G v$ it holds that $\CC_G(u)\subsetneq \CC_G(v)$. In summary, 
   $u\neq v$ implies $\CC_G(u)\neq \CC_G(v)$ for all $u,v\in V(G)$. 
	  Hence, (4) implies (5).
 
	Assume that Condition (5) holds.
	Hence, $G$ is a DAG that satisfies (PCC) and where $u\neq v$ implies $\CC_G(u)\neq
	\CC_G(v)$ for all $u,v\in V(G)$. Let $u,v\in V(G)$ be chosen arbitrarily. If $u\preceq_G v$, then
	Lemma~\ref{lem:prec-subset} implies that $\CC_G(u)\subseteq \CC_G(v)$. Suppose now that
	$\CC_G(u)\subseteq \CC_G(v)$. Since $G$ satisfies (PCC), $u$ and $v$ must be
	$\preceq_G$-comparable. However, the case $v\prec_G u$ cannot occur since then 
	$u\neq v$ and, thus,	$\CC_G(u)\neq
	\CC_G(v)$ which together with Lemma~\ref{lem:prec-subset} implies that $\CC_G(v)\subsetneq \CC_G(u)$;
	a contradiction to $\CC_G(u)\subseteq \CC_G(v)$. Thus, $u\preceq_G v$ holds and (5) implies (6).

	Assume that Condition (6) holds.
	Hence, $G$ is a DAG such that $\CC_G(u)\subseteq \CC_G(v)$ if and only if $u\preceq_G
	v$ for all $u,v\in V(G)$. Thus, $G$ satisfies (PCC). By Lemma~\ref{lem:PCC=>CL}, $G$ satisfies
	(CL). Thus, $w = \lca_G(\CC_G(v))$ is well-defined for all $v\in V(G)$. Again, by Lemma
	\ref{lem:PCC=>CL}, $\CC_G (w) = \CC_G (v)$. Thus, we have $\CC_G (w) \subseteq \CC_G (v)$
	implying $w\preceq_G v$. In addition, $\CC_G (v) \subseteq \CC_G (w)$ implies $v\preceq_G w$.
	Consequently, $w=v$ holds. Therefore, $v = \lca_G(\CC_G(v))$ for all $v\in V(G)$ and $G$ has the
	strong-(CL) property. In summary, (6) implies (2), which completes this proof.
\end{proof}

We continue to establish a few further key results that enable us to show the close connection
between regular DAGs and $\lca$-\rel DAGs in Theorem~\ref{thm:regular-Char}. Regular networks have
very constrained structural properties, as characterized in \cite[Thm.\ 2]{Hellmuth2023}. Here, we
generalize these results to arbitrary DAGs.

\begin{theorem}\label{thm:org-char-regular}
The following statements are equivalent for every DAG $G$.
	\smallskip
	\begin{enumerate}[label=(\arabic*)]
		\item $G$ is regular 
		\item $G$ does not contain vertices with out-degree 1, is
		      shortcut-free and satisfies (PCC).
	\end{enumerate}
	\smallskip
	In particular, a regular DAG is  phylogenetic.
\end{theorem}
\begin{proof}
	If $G$ is a network, then we can use \cite[Thm.\ 2]{Hellmuth2023} which states that (1) and (2)
	are equivalent for networks. If $G$ is a DAG that is not a network, then $|R(G)|>1$. In this
	case, we can obtain a network $N_G$ from $G$ by adding a new root $\rho$ to $G$ and edges
	$(\rho,r)$ for all $r\in R(G)$. It is now a straightforward task -- which we leave to the reader
	-- to verify that $G$ satisfies (1) if and only if $N_G$ does and that $G$ satisfies (2) if and
	only if $N_G$ does. Since (1) and (2) are equivalent for networks, the latter arguments show that
	(1) and (2) are equivalent for DAGs. In particular, since regular DAGs have no vertices of
	out-degree~1, they must be phylogenetic.
	\end{proof}

We note that DAGs that are shortcut-free and satisfy (PCC) are also known as \emph{semi-regular} \cite{Hellmuth2023}. 
The following result generalizes \cite[Lem.~22]{Hellmuth2023} that has been established for networks. 
\begin{lemma}\label{lem:hasse-phylo}
	For every set system $\mathfrak{C}$, the Hasse diagram $\Hasse(\mathfrak{C})$ is a shortcut-free
	DAG that satisfies (PCC). Moreover, if $\mathfrak{C}$ is grounded, then $\Hasse(\mathfrak{C})$
	is regular and phylogenetic. Furthermore, if $\mathfrak{C}$ is a clustering system, then $\Hasse(\mathfrak{C})$
	is a regular network.
\end{lemma}	
\begin{proof}
	The first statement is a direct consequence of the definition of the Hasse diagram. Hence,
	to prove that $\Hasse\coloneqq \Hasse(\mathfrak{C})$ is regular for grounded set systems
	$\mathfrak{C}$, it suffices to show that $\Hasse$ has no vertex of out-degree 1 (cf.\
	Theorem~\ref{thm:org-char-regular}). For contradiction, assume that $\hasse$ has a vertex 
	$C$ such that $\outdeg_{\hasse}(C)=1$. Let $C'$ be the unique child of $C$. 
	Since $C,C'\in \mathfrak{C}$ are distinct clusters with $C'\subsetneq C$ and $C'\neq\emptyset$, there is
	some element $x\in C\setminus C'$. Since $\mathfrak{C}$ is grounded, $\{x\}\in\mathfrak{C}$. 
	But then the definition of $\hasse$ together with $\{x\}\subseteq C$ and 
	$\{x\}\not\subseteq C'$ implies $\{x\}\prec_{\hasse} C$ while $\{x\}$ and $C'$ are $\preceq_{\hasse}$-incomparable.
	One easily verifies that this implies that $C$ must have at least two children; a
	contradiction. Thus, $\Hasse$ is
	regular. By Theorem~\ref{thm:org-char-regular}, $\Hasse$ is phylogenetic.

	Finally, by definition, every clustering system $\mathfrak{C}$ on $X$ is grounded and thus,
    $\Hasse(\mathfrak{C})$ is regular. Since $X$ is the unique
    inclusion-maximal cluster in $\mathfrak{C}$, it follows that $X$ is the unique root of
    $\Hasse(\mathfrak{C})$. Taking the latter arguments together, $\Hasse(\mathfrak{C})$ is a
    regular network for clustering systems $\mathfrak{C}$.
\end{proof}

Next, we show that, roughly speaking, DAGs with the strong-(CL) property differ from regular DAGs
only by the presence of additional shortcuts.

\begin{theorem}\label{thm:Char-strongCL}
 A DAG $G$ has the strong-(CL) property if and only if $G$ is isomorphic to the regular DAG 
 $\Hasse(\mathfrak{C}_G)$ to which $\ell\geq 0$ shortcuts have been added. 
\end{theorem}	
\begin{proof}
	Suppose that $G$ is a DAG on $X$ with the strong-(CL) property. Let $H\doteq
	\Hasse(\mathfrak{C}_G)$ in which also every inner vertex $C$ obtains a new label $v$ for some
	$v\in V(G)$ with $\CC_G(v)=C$. By definition, $H$ is a DAG on $X$. By
	Theorem~\ref{thm:strongCLChar}, $\CC_G(u)\neq \CC_G(v)$ for all distinct $u,v\in V(G)$ and thus,
	$v\in V(G)$ if and only if there is a unique cluster $C\in \mathfrak{C}_G$ such that $C=
	\CC_G(v)$. Hence, the aforementioned relabeling of the inner vertex is well-defined and uniquely
	determined and we have, in particular, $V(H) = V(G)$ and $\CC_H(v) = \CC_G(v)$ for all $v\in
	V(G)$. Note that, since $G$ is a DAG, $\mathfrak{C}_G$ is a grounded set system and
	Lemma~\ref{lem:hasse-phylo} implies that $\hasse{(\mathfrak{C}_G)}$ is regular. Since
	$\hasse{(\mathfrak{C}_G)}\simeq H$, the DAG $H$ is regular. We show now that $H$ is a subgraph
	of $G$. Let $u,v\in V$ be such that $\CC_G(u)\subsetneq \CC_G(v)$ and there is no cluster $C \in
	\mathfrak{C}_G$ such that $\CC_G(u)\subsetneq C \subsetneq \CC_G(v)$. Thus, $u\neq v$ and
	Theorem~\ref{thm:strongCLChar} implies that $u\prec_G v$. Hence, there is a directed $vu$-path
	$P$ in $G$. If there would be vertex $w$ in $P$ such that $u\prec_G w \prec_G v$, then
	Theorem~\ref{thm:strongCLChar} together with $\CC_G(w) \neq \CC_G(v)$ and $\CC_G(w) \neq
	\CC_G(u)$ implies that $\CC_G(u)\subsetneq \CC_G(w) \subsetneq \CC_G(v)$; a contradiction.
	Consequently, $P$ just consists of the single edge $(u,v)\in E(G)$. By definition of regular
	DAGs these type of edges $(u,v)$ are precisely the edges in $H$ and, therefore, $E(H)\subseteq
	E(G)$, i.e., $H$ is a subgraph of $G$ with $V(H) = V(G)$.

	Now, let $(u,v) = e\in E(G)\setminus E(H)$ and thus, $v\prec_G u$. Since $\CC_G(u)\neq \CC_G(v)$,
	Lemma~\ref{lem:prec-subset} implies $\CC_G(v)\subsetneq \CC_G(u)$. As argued above, $u,v\in V(G)
	= V(H)$ and $\CC_H(v) = \CC_G(v)$ and $\CC_H(u) = \CC_G(u)$. Thus, $\CC_H(v)\subsetneq \CC_H(u)$.
	Since $H$ is regular, Theorem~\ref{thm:org-char-regular} implies that $H$ satisfies (PCC). Hence,
	$u\prec_H v$ or $v\prec_H u$ holds. However, $u\prec_H v$ would together with
	Lemma~\ref{lem:prec-subset} imply that $\CC_H(u)\subsetneq \CC_H(v)$; a case that cannot occur.
	Thus, only $v\prec_H u$ is possible. Hence, there is a directed path from $u$ to $v$ in $H$.
	Since $E(H)\subseteq E(G)$, this path exists in $G$ and, in particular, avoids the edge $(u,v)$.
	Hence, $(u,v)$ is a shortcut in $G$. As the latter arguments hold for all edges in $E(G)\setminus
	E(H)$, every edge in $E(G)\setminus E(H)$ is a shortcut in $G$. Thus, $G$ is isomorphic to the
	regular DAG $H\doteq \Hasse(\mathfrak{C}_G)$ to which $\ell\geq 0$ shortcuts have been added.

	Let $H\doteq \Hasse(\mathfrak{C}_G)$ and 
	suppose now that $G$ is isomorphic to $H'$, where $H'$ is is obtained from $H$ by adding 
	$\ell\geq 0$ shortcuts. Since there is a bijection between 
	$V(G)$ and $V(H')$, we can w.l.o.g.\ assume that
	 $V\coloneqq V(G) = V(H') = V(H)$. This together with
	stepwise application of Lemma~\ref{lem:shortcutfree} implies that $\CC_H(v)= \CC_{G}(v)$ for all
	$v\in V$. By definition of $H$, $\CC_H(v)\neq \CC_{H}(u)$ and, therefore, $\CC_G(v)\neq
	\CC_{G}(u)$ for all distinct $u,v \in V$. In addition, Theorem~\ref{thm:org-char-regular} implies
	that $H$ satisfies (PCC) implies that $G$
	satisfies (PCC). This allows us to apply Theorem~\ref{thm:strongCLChar} and to conclude that $G$
	has the strong-(CL) property. 
\end{proof}

By Observation~\ref{obs:Ilca-rel} every \Olcarel DAG is $\lca$-$\rel$ and, by
Theorem~\ref{thm:strongCLChar}, has the strong-(CL) property. This together with
Theorem~\ref{thm:Char-strongCL} implies
\begin{corollary}\label{cor:Onerel=>regular}
	Every \Olcarel DAG $G$ from which all shortcuts have been removed is regular.
\end{corollary}

	The converse of Corollary~\ref{cor:Onerel=>regular} is, in general, not satisfied
	without specifying $\One$, i.e.,
	not every regular DAG is \Olcarel for arbitrary $\One$. By way of example, 
	the regular network $N_1$ in 
	Figure~\ref{fig:lca-relevant-expls} is not $\{1,2\}$-$\lca$-\rel. 
	Nevertheless, we obtain the following new characterization
	of regular DAGs. 	
\begin{theorem}\label{thm:regular-Char}
		For every DAG $G$, 
		the following statements are equivalent.	
		\smallskip
		\begin{enumerate}[label=(\arabic*)]
		\item  $G$ is  regular.
		\item  $G$ is  shortcut-free and has the strong-(CL) property.
		\item  $G$ is  shortcut-free and $\lca$-\rel. 
		\end{enumerate}
		\smallskip
\end{theorem}
\begin{proof}
	The equivalence between (1) and (2) is an immediate consequence of Theorem~\ref{thm:Char-strongCL} and  \ref{thm:org-char-regular}. 
	The equivalence between (2) and (3) follows from Theorem~\ref{thm:strongCLChar}.
\end{proof}

Theorem~\ref{thm:regular-Char} together with Lemma~\ref{lem:hasse-phylo} implies
\begin{corollary}\label{cor:setsytem-phylolcaDAG}
	For every grounded set system $\mathfrak{C}$, there is a phylogenetic $\lca$-\rel, 
	and thus also $\LCA$-\rel, DAG
	$G$ with $\mathfrak{C}_G = \mathfrak{C}$.
\end{corollary}
As argued in the example succeeding Def.~\ref{def:regular-N},
$\Hasse(\mathfrak{C})$ is in general not regular in case $\mathfrak{C}$ is not grounded. 
Moreover, for every DAG $G$, the set system $\mathfrak{C}_G$ is always grounded.
Consequently, the requirement that $\mathfrak{C}$ is grounded cannot be omitted in Corollary~\ref{cor:setsytem-phylolcaDAG}.

In phylogenetic trees, the number of inner vertices and edges is bounded from above by a linear
function on the number of its leaves. In general, phylogenetic DAGs and thus, general DAGs, lack
this property. For \lca-\rel DAGs and thus, also $\One$-$\lca$-\rel DAGs, we nevertheless obtain the
following simple result.
\begin{lemma}\label{lem:bound-V-by-X}
	The number of vertices and edges in $\lca$-\rel DAGs $G$ is asymptotically bounded from above by
	a function depending only on the number of leaves and it holds that $|V(G)| = |\mathfrak{C}_G|$.
\end{lemma}
\begin{proof}
	Let $G=(V,E)$ be an $\lca$-\rel DAGs $G$ on $X$. Hence, by the equivalence between Statements (2)
	and (3) of Theorem~\ref{thm:strongCLChar} it holds that $u\neq v$ implies $\CC_G(u)\neq
	\CC_G(v)$, for all $u,v\in V(G)$. Trivially, $\CC_G(u)\neq \CC_G(v)$ implies $u\neq v$. Taken the
	latter two arguments together, $|V(G)| = |\mathfrak{C}_G|$. Clearly $|\mathfrak{C}_G| \in
	O(2^{|X|})$ and, therefore, the number of vertices in $G$ is asymptotically bounded from above by
	a function on the number of leaves. As the number of edges in any DAG $G$ is always bounded from
	above by a function on the number of vertices in $G$, the number of edges in $G$ is
	asymptotically bounded above by a function on the number of leaves.
\end{proof}

Lemma~\ref{lem:bound-V-by-X} cannot be extended to the case of $\LCA$-\rel DAGs. To see this,
consider, for example, the DAG $G_k$ obtained from any $\LCA$-$\rel$ network $N$ on $X$ by adding
$k$ additional roots $r_1$, ..., $r_k$ connected to the leaves in $X$ by edges $(r_i,x)$ for each
$x\in X$ and $1\leq i\leq |X|$. In this case, $G_k$ remains $\LCA$-$\rel$ and $|V(G_k)|=|V(N)|+k$.
Since $k$ does not depend on $|X|$ and can be chosen arbitrarily, no upper bound on 
$|V(G_k)|$ depending on $|X|$ can be found.


\section{The $\mathbf{\ominus}$-Operator and Computation of lca- \& LCA-Relevant DAGs}
\label{sec:ominus-lcaRel}

Not all DAGs are $\One$-$\lca$-\rel or $\One$-$\LCA$-\rel. This raises the question of whether it is
possible to ``transform'' a non-$\One$-$\lca$-\rel resp., non-$\One$-$\LCA$-\rel DAG $G$ into an
$\One$-$\lca$-\rel, resp., $\One$-$\LCA$-\rel DAG $H$ while preserving as many structural properties
of $G$ as possible. To clarify, for a given DAG $G$ on $X$, we aim to maintain the following
structural properties in $H$:
\smallskip
\begin{enumerate}[label=\emph{(S\arabic*)}]
    \item $H$ remains a DAG on $X$. 
    \item $V(H) \subseteq V(G)$, meaning no new vertices are introduced.
    \item $H$ preserves the  ancestor relationship $\prec_G$, i.e., $u \prec_G w$ if and only if $u \prec_{H} w$ for all $u, w \in V(H)$.
    \item $H$ is \emph{$\One$-$\lca_G$-preserving}, i.e., $\lca_{H}(A) = \lca_G(A)$ for all $A\in X(\One)$ for which $\lca_G(A)$ is well-defined.
\end{enumerate}
\smallskip
In case we are interested in \OLCA-\rel DAG, we  strengthen (S4) to
\smallskip
\begin{enumerate}[resume, label=\emph{(S\arabic*)}]
	\item $H$ is \emph{$\One$-$\LCA_G$-preserving}, i.e., $\LCA_{H}(A) = \LCA_G(A)$ for all $A\in X(\One)$.
\end{enumerate}
\smallskip
Note that (S5) implies (S4). Moreover, Property (S4), resp., (S5) implies that that $\One$-$\lca$,
resp., $\One$-$\LCA$ vertices in $G$ remain $\One$-$\lca$, resp., $\One$-$\LCA$ vertices in $H$.
This together with (S2) implies that no new vertices that violate the property of being
$\One$-$\lca$, resp., $\One$-$\LCA$ vertices are introduced.

Moreover, Property (S1) states that $L(H)=L(G)=X$. In addition, for all $C\in\mathfrak{C}_H$, there
is a vertex $v\in V(H)$ such that $\CC_H(v)=C$. By (S2), it holds that $v\in V(G)$ and by (S3) we
have $\CC_H(v)=\CC_G(v)$. Consequently, each $C\in\mathfrak{C}_H$ is contained in $\mathfrak{C}_G$.
We summarize the latter into 
\begin{observation}\label{obs:newS} 
If $H$ satisfies (S1), (S2) and (S3) w.r.t.\ to a DAG $G$, then the following property is satisfied: 
	\smallskip
\begin{enumerate}[label=\emph{(S\arabic*)}]
    \item[(S0)] $\mathfrak{C}_{H} \subseteq \mathfrak{C}_G$, meaning no new clusters are introduced.
\end{enumerate}
	\smallskip
\end{observation}

A powerful tool in this context is the following $\ominus$-operator.

\begin{definition}[{\cite{SCHS:24}}]\label{def:ominus}
  Let $G=(V,E)$ be a DAG and $v\in V$. Then $G\ominus v=(V',E')$ is the directed graph with vertex
 set $V'=V\setminus\{v\}$ and edges $(p,q)\in E'$ precisely if $v\ne p$, $v\ne q$ and $(p,q)\in E$,
 or if $(p,v)\in E$ and $(v,q)\in E$. For a non-empty subset $W = \{w_1,\dots,w_\ell\} \subsetneq
 V$, define $G\ominus W \coloneqq (\dots ((G \ominus w_1) \ominus w_2) \dots)\ominus w_\ell$.
\end{definition} 
  In simple words, the directed graph $G\ominus v$ is obtained from $G=(V,E)$ by removing $v$ and
  its incident edges and connecting each parent $p$ of $v$ with each child $q$ of $v$. In case $v$
  is a leaf or a root in $G$, then $v$ and its incident edges are simply deleted. The
  $\ominus$-operator was formally introduced in \cite{SCHS:24} and is also known as
  \texttt{collapse} in the Biopython package \cite{Biopython,Talevich2012}, or as ``suppression''
  when the vertex in question has both in-degree and out-degree one \cite{Huson:11}. However, the
  properties of the $\ominus$-operator seem not to have been studied in the literature so far.

  By construction, $G\ominus v$ remains a DAG such that $p\preceq_G q$ if and only if
 $p\preceq_{G\ominus v} q$ for all $p,q\ne v$. This preservation of the partial order implies that
 also the clusters remain unchanged as long as the deleted vertex is not a leaf. Moreover, if $v\in
 V\setminus X$, then the latter arguments imply that leaves of $G$ remain leaves in $G\ominus v$,
 i.e., $G\ominus v$ is a DAG on $X$. Finally, it is an easy task to verify that for distinct $u,v\in
 V$ it holds that $(G\ominus u)\ominus v=(G\ominus v)\ominus u$. Thus, $G\ominus W$ is well-defined.
 We summarize the latter into

\begin{observation}\label{obs:ominus} 
Let $G$ be a DAG on $X$ and $W\subseteq V(G)\setminus X$ be a non-empty subset. Then, $G\ominus W$
is a DAG on $X$ that satisfies (S1), (S2), (S3) and, by Observation~\ref{obs:newS}, also (S0). In
particular, $\CC_G(u)=\CC_{G\ominus W}(u)$ for all $u\in V(G\ominus W)$.
\end{observation}

We provide now a sufficient condition under which the $\ominus$-operator preserves connectivity.
\begin{lemma}\label{lem:ominusConnected}
	Let $G=(V,E)$ be a connected DAG on $X$ and $v\in V$. If $v$ is not a $\kLCA$ or $\klca$ vertex
	of $G$ for any $k\in\{1,\ldots,|X|\}$, then $G\ominus v$ is connected.
\end{lemma}
\begin{proof}
	Let $G=(V,E)$ be a connected DAG on $X$ and $v\in V$ a vertex that is not a $\kLCA$ or $\klca$
	vertex of $G$ for any $k\in\{1,\ldots,|X|\}$. Assume, for contradiction, that $G\ominus v$ is not
	connected. Hence, there are distinct vertices $u,w$ in $G\ominus v$ for which there is no
	$uw$-path in $G\ominus v$. Since $G$ is connected and $u,w\in V(G\ominus v)\subseteq V(G)$, there
	is an $uw$-path $P_{uw}$ in $G$ which, by assumption, is not contained in $G\ominus v$. Hence,
	$P_{uw}$ must contain the vertex $v$ and $u,v,w$ are pairwise distinct. To recall, paths do not
	contain ``repeated'' vertices, that is, $v$ is contained in exactly two edges contained in
	$P_{uw}$. Let $u'$, resp., $w'$ be the neighbor of $v$ along the subpath of $P_{uw}$ from $v$ to
	$u$, resp., $v$ to $w$. Note that $u'=u$ or $w'=w$ is possible. Hence $P_{uw}$ consists of an
	$uu'$-path $P_{uu'}$, an $u'w'$-path $P_{u'w'}$ that contains $v$ and consist of exactly two
	edges and a $w'w$-path $P_{w'w}$.

	Note that any $u'w'$-path $P'$ in $G\ominus v$ would imply the existence of an undirected
	$uw$-path in $G\ominus v$, by combining $P_{uu'}$, $P'$ and $P_{w'w}$. Since there is no
	$uw$-path in $G\ominus v$, it follows that there is no $u'w'$-path in $G\ominus v$. Note that in
	$G$ the vertices $u'$ and $w'$ must be children or parents of $v$. Hence, we consider the
	following three possible cases that can appear in $G$: $\{w',u'\}$ contains (i) one parent and
	one child of $v$, (ii) two parents of $v$ and (iii) two children of $v$.

	In Case (i), we can assume without loss of generality that $w'$ is a child of $v$ and $u'$ a
	parent of $v$ in $G$. By construction, we have in $G\ominus v$ the edge $(u',w')$ and, therefore,
	an $u'w'$-path in $G\ominus v$; a contradiction.

	In Case (ii), both $u'$ and $w'$ are parents of $v$. Since every leaf is a $1$-$\lca$ vertex of
	$G$ and $v$ is, in particular, not a $1$-$\lca$ vertex, $v$ must have some child $v'$ in $G$. By
	construction, we have in $G\ominus v$ the edges $(u',v')$ and $(w',v')$ which results in an
	$u'w'$-path in $G\ominus v$; a contradiction.
	
	In Case (iii), both $u'$ and $w'$ are children of $v$. Consider the set
	$A\coloneqq\CC_{G}(u')\cup \CC_{G}(w')$. Since $u',w'\prec_G v$, Lemma~\ref{lem:prec-subset}
	ensures that $A\subseteq \CC_G(v)$ and hence, $v$ is a common ancestor of the elements of $A$.
	Therefore, $\LCA_G(A)\neq\emptyset$. If $v$ is not a $\kLCA$ vertex, then $v\notin \LCA_G(A)$. If
	$v$ is not a $\klca$ vertex, then $v\notin \LCA_G(A)$ or $v\in\LCA_G(A)$ and $|\LCA_G(A)|>1$. In
	either case, we may choose $z\in \LCA_G(A)$ such that $z\neq v$. Thus $z\in V(G\ominus v)$. By
	Observation~\ref{obs:ominus} we have $\CC_{G\ominus v}(u')=\CC_G(u')$, $\CC_{G\ominus
	v}(w')=\CC_G(w')$ and $\CC_{G\ominus v}(z)=\CC_G(z)$. Clearly there are paths in $G\ominus v$
	from $u'$ to every element in $\CC_{G\ominus v}(u')$ respectively from $w'$ to every element in
	$\CC_{G\ominus v}(w')$. Moreover, there are paths from $z$ to every element in $\CC_{G\ominus
	v}(z)$. Since $\CC_{G\ominus v}(z)=\CC_G(z)$ contains every element of $A$, there is a
	$u'w'$-path in $G\ominus v$; a contradiction.
	
	In summary, all three possible Cases (i), (ii) and (iii) yield a contradiction. 
	Consequently, $G\ominus v$ must be connected.
\end{proof}

\begin{figure}[t]
	\centering
	\includegraphics[width=0.95\textwidth]{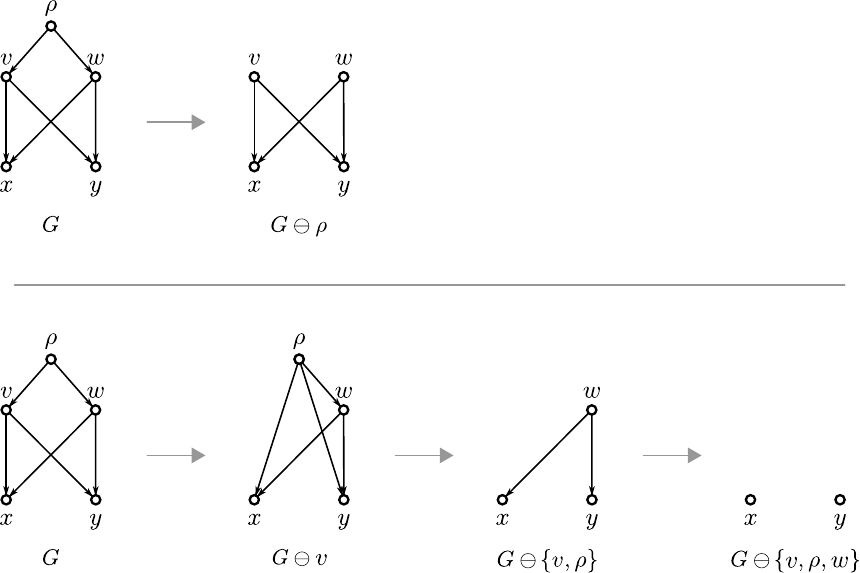}	
	\caption{The network $G$ is neither $\lca$-\rel nor $\LCA$-\rel, since none of the vertices $v$,
	         $w$ and $\rho$ in $G$ are $\{1,2\}$-$\lca$ vertices. Since $\rho$ is the only vertex
	         that is not a $\{1,2\}$-$\LCA$, $G\ominus \rho$ is $\LCA$-\rel. The set $W=\{\rho,v,w\}$
	         is the set of all vertices that are not $\{1,2\}$-$\lca$ vertices. The stepwise
	         computation of $G\ominus v$, $(G\ominus v)\ominus \rho$ and $G\ominus W$ is shown in the
	         lower part and results in a disconnected DAG. However, Algorithm~\ref{alg:lca-relevant}
	         determines whether a vertex is a $\{1,2\}$-$\lca$ vertex in the updated DAG. Hence, if
	         we start with $v$ to obtain $G\ominus v$, there is only one vertex left that is not an
	         $\{1,2\}$-$\lca$ vertex, namely $\rho$. In $(G\ominus v)\ominus \rho$ each vertex is a
	         $\{1,2\}$-$\lca$ vertex and the algorithm terminates.
	} 
	\label{fig:gominusv}
\end{figure}

In contrast to Lemma~\ref{lem:ominusConnected}, $G\ominus v$ may be disconnected if $v$ is a
$\klca$ vertex or a $\kLCA$ vertex, even if $G$ is connected 
The possibly simplest example here is
the DAG $H$ consisting of a single root $w$ with leaf-children $x$ and $y$; in $H$, the root
satisfy $w=\lca_G(\{x,y\})$ and $G\ominus w$ is the disconnected DAG $(\{x,y\},\emptyset)$, see
also Figure~\ref{fig:gominusv}.

\begin{theorem}\label{thm:S4S5}
	Let $G$ be a DAG on $X$ and $W\subseteq V(G)$ be a non-empty subset of
	vertices that are not $\One$-$\lca$ (resp., not $\One$-$\LCA$) vertices in $G$. 
	Then, $G\ominus W$ is a DAG on $X$ that satisfies (S0) -- (S4) (resp., (S0) -- (S5)) w.r.t.\ $G$.
		
	In particular, if $W$ contains every vertex of $G$ that is not an \Olca vertex (resp. not an
	\OLCA vertex) of $G$, then $G\ominus W$ is \Olcarel (resp. \OLCArel).
\end{theorem}
\begin{proof}
	Let $G=(V,E)$ be a DAG on $X$ and $W\subseteq V$ be a non-empty subset of vertices that are not
	$\One$-$\lca$ (resp., not $\One$-$\LCA$) vertices. Observe first that for all $x\in X$ we have, by
	definition, $\LCA_G(x)=\{x\}$ and thus, $W\subseteq V\setminus X$. This together with
	Observation~\ref{obs:ominus} implies that $G\ominus W$ is a DAG on $X$ that satisfies (S0), (S1), (S2) and
	(S3) and $\CC_G(w)=\CC_{G\ominus W}(w)$ for all $w\in V(G\ominus W)$.

	We show first that $u\in \LCA_{G}(A)$ and $u\neq v$ implies that $u\in \LCA_{G\ominus v}(A)$ for
	any vertex $v\in V\setminus X$ and $A\subseteq X$. By (S2), $u\in V(G\ominus v)$. By
	Lemma~\ref{lem:alg_LCA}, $u\in\LCA_G(A)$ only if $A\subseteq\CC_G(u)$ and
	$A\not\subseteq\CC_G(u')$ for all $u'\in V(G)$ with $u'\prec_G u$. Since $\CC_G(u)=\CC_{G\ominus
	v}(u)$, we have $A\subseteq\CC_{G\ominus v}(u)$. Since $G\ominus v$ satisfies (S3) and since
	$\CC_G(w)=\CC_{G\ominus v}(w)$ for all $w\in V(G\ominus v)$, we can conclude that
	$A\not\subseteq\CC_{G\ominus v}(u')$ for all $u'\in V(G\ominus v)$ with $u'\prec_{G\ominus v} u$.
	Application of Lemma~\ref{lem:alg_LCA} now shows that $u\in\LCA_{G\ominus v}(A)$ must hold.

	Suppose now that $W$ is a subset of non-$\One$-$\LCA$ vertices in $G$ and let $v\in W$. We show
	now that $\LCA_{G\ominus v}(A) = \LCA_G(A)$ for all $A\in X(\One)$. Let $A\in X(\One)$ and assume
	first that $\LCA_{G}(A) = \emptyset$. This in particular implies that there is no vertex $w\in V$
	such that $A\subseteq \CC_G(w)$. Since $\CC_G(w)=\CC_{G\ominus v}(w)$ for all $w\in V(G\ominus
	v)$ it follows that there is no vertex $w\in V(G\ominus v)$ such that $A\subseteq \CC_{G\ominus
	v}(w)$. Hence, $\LCA_{G\ominus v}(A) = \emptyset$. Assume now that $\LCA_{G}(A)\neq \emptyset$.
	Hence, there is some vertex $u\in \LCA_{G}(A)$. Since $v$ is not an $\One$-$\LCA$ vertex, $u\neq
	v$ and $v\notin \LCA_G(A)$ must hold. This together with the result of the preceding paragraph 
	implies that $\LCA_G(A)\subseteq\LCA_{G\ominus v}(A)$.
	Conversely, assume that $u\in \LCA_{G\ominus v}(A)$. By Lemma~\ref{lem:alg_LCA},
	$A\subseteq\CC_{G\ominus v}(u)$ and $A\not\subseteq\CC_{G\ominus v}(u')$ for all $u'\in
	V(G\ominus v)$ with $u'\prec_{G\ominus v} u$ must hold. If $v\not \prec_{G} u$, then
	Observation~\ref{obs:ominus} together with (S2) implies that $A\subseteq\CC_{G }(u)$ and
	$A\not\subseteq\CC_{G}(u')$ for all $u'\in V(G)$ with $u'\prec_{G} u$, in which case, $u\in
	\LCA_{G}(A)$. Assume that $v\prec_{G} u$. Then either $A\not\subseteq \CC_{G}(v)$ or $A\subseteq
	\CC_{G}(v)$. In the first case, Lemma~\ref{lem:alg_LCA} implies that $u\in \LCA_{G}(A)$. In the
	last case, Lemma~\ref{lem:alg_LCA} together with the assumption that $v$ is not an $\One$-$\LCA$
	vertex in $G$ implies that there must be a child $w$ of $v$ such that $A\subseteq \CC_{G}(w)$.
	But then, $w\prec_G u$ and $w\in V\setminus \{v\}$ must hold. Again, Observation~\ref{obs:ominus}
	together with (S2) implies that $w\prec_{G\ominus v} u$ and $A\subseteq \CC_{G\ominus v}(w)$
	which together with Lemma~\ref{lem:alg_LCA} implies that $u\notin \LCA_{G\ominus v}(A)$; a
	contradiction. Hence, $u\in \LCA_{G}(A)$ must hold. In summary, $\LCA_{G\ominus v}(A) =
	\LCA_G(A)$ for all $A\in X(\One)$. Thus, $G\ominus v$ satisfies (S5) and, thus, in particular
	(S4). We can now repeat the latter arguments on $G\ominus v$ and an element in $v'\in W\setminus
	\{v\}$ to conclude that $(G\ominus v)\ominus v'$ is a DAG on $X$ that satisfies $\LCA_{(G\ominus
	v)\ominus v'}(A) = \LCA_{G\ominus v}(A)= \LCA_{G}(A)$ for all $A\in X(\One)$ and thus, that
	$(G\ominus v)\ominus v'$ satisfies (S4) and (S5). By induction, $G\ominus W$ is a DAG on $X$ that
	satisfies (S4) and (S5).
	
	Assume now that $v\in W$ is not an $\One$-$\lca$ vertex in $G$. Let $A\in X(\One)$ and suppose
	that $u = \lca_G(A)$ is well-defined. Hence, $u\neq v$ and, since $G\ominus v$ satisfies (S2),
	$u\in V(G\ominus v)$. Moreover, by the arguments in the second paragraph of this proof,
	$u\in\LCA_{G\ominus v}(A)$. Assume, for contradiction, that $u\neq \lca_{G\ominus v}(A)$ and,
	thus, $|\LCA_{G\ominus v}(A)|>1$. Thus, there is a vertex $w\in \LCA_{G\ominus v}(A)$ such that
	$u$ and $w$ are $\preceq_{G\ominus v}$-incomparable. By (S2), $w$ is $\preceq_G$-incomparable to
	$u$. By Observation~\ref{obs:ominus} and Lemma~\ref{lem:alg_LCA}, $A\subseteq\CC_{G\ominus
	v}(w)=\CC_G(w)$. Hence, $w$ is ancestor of all vertices $x\in A$ in $G$. Therefore, there is a
	vertex $w'\preceq_G w$ such that $w'\in \LCA_G(A)$. Since by assumption $\LCA_G(A) = \{u\}$ it
	follows that $w'=u$ must hold. But then $w$ and $u$ are not $\preceq_G$-incomparable; a
	contradiction. Thus, $u = \lca_{G\ominus v}(A)$ must hold and $G\ominus v$ satisfies (S4). Again,
	by induction, $G\ominus W$ satisfies (S4).

	For the last statement, note that if $W$ contains every vertex of $G$ that is not an \Olca vertex
	(resp., not an \OLCA vertex), then $v\in V(G\ominus W)$ if and only if $v$ is an \Olca vertex
	(resp., an \OLCA vertex) of $G$. Hence, $G\ominus W$ contains precisely all \Olca vertices
	(resp., \OLCA vertices) of $G$. As $G\ominus W$ satisfies (S4) (resp., (S5)) it follows that
	every vertex in $G\ominus W$ is an \Olca (resp., \OLCA) vertex. Thus, $G\ominus W$ is \Olca-\rel
	(resp., \OLCA-\rel).
\end{proof}

As we shall see in Section~\ref{sec:complexity-hard}, it is in NP-hard to determine as whether a
given DAG is $\One$-$\lca$-\rel or $\One$-$\LCA$-\rel for general $\One$. However, for the special
case that $\One = \{1,2,\dots,|X|\}$, we deal with $\lca$-\rel or $\LCA$-\rel DAGs. In fact, simplifying
$G$ into an $\lca$-\rel or $\LCA$-\rel DAG using the $\ominus$-operator is tractable and we provide 
here polynomial-time algorithms to achieve these transformations.

We start with Algorithm~\ref{alg:LCA-relevant} to compute an $\LCA$-\rel version $H$ of an
input DAG $G$ that satisfies (S0) -- (S5). To recall, $G$ is not $\LCA$-\rel if there is a vertex
$v$ in $G$ such that $v$ is not a least common ancestor for any non-empty $A\subseteq X$. 
By Lemma~\ref{lem:not_kLCA}, the latter is precisely the case if $v\notin \LCA_G(\CC_G(v))$
which is the only condition that needs to be checked in Algorithm~\ref{alg:LCA-relevant} 
(Line~\ref{alg:LCA-relevant:check}). 

 \begin{table}
   	\begin{center}
	    \begin{minipage}[t]{1\textwidth}
		\begin{algorithm}[H]\small
			\caption{\texttt{$\LCA$-\rel}} \label{alg:LCA-relevant}
			\begin{algorithmic}[1]
				\Require A DAG $G=(V,E)$ on $X$
				\Ensure  An $\LCA$-\rel DAG $G\ominus W$ satisfying (S0) -- (S5) w.r.t.\ $G$
				\State $W\gets\emptyset$		 
				\For{all vertices $v\in V$}
					\If{ $v\notin \LCA_G(\CC_G(v))$}\label{alg:LCA-relevant:check}
						\State Add $v$ to $W$ \label{alg:LCA-add2W}
					\EndIf
				\EndFor		
				\State \Return $G\ominus W$. 
			\end{algorithmic}
		\end{algorithm}
    \end{minipage}\\
    \begin{minipage}[t]{1\textwidth}
		\begin{algorithm}[H]\small
			\caption{\texttt{$\lca$-\rel}}\label{alg:lca-relevant}
			\begin{algorithmic}[1]
				\Require A DAG $G=(V,E)$ on $X$
				\Ensure  An $\lca$-\rel DAG on $X$ satisfying  (S0) – (S4) w.r.t.\ $G$.
				\For{all vertices $v\in V(G)$}
					\If{$v\neq\lca_G(\CC_G(v))$} 
						\State $G\gets G\ominus v$ \label{alg:lca-rel-ominus}
					\EndIf
				\EndFor
				
				\State \Return $G$. 
			\end{algorithmic}
		\end{algorithm}
    \end{minipage}
  \end{center}
\end{table}

\begin{proposition}\label{prop:alg_kLCA}
	Let $G$ be a DAG on $X$ and $W$ be the set of all vertices that are not $\{1,\ldots,|X|\}$-$\LCA$
	vertices of $G$. Then, Algorithm~\ref{alg:LCA-relevant} with input $G$ returns the DAG $G\ominus
	W$ on $X$ that is $\LCA$-$\rel$, phylogenetic and satisfies Properties (S0) -- (S5) w.r.t.\ $G$.
	In particular, $W$ is the unique and, therefore, smallest subset of $V(G)$ such that $G\ominus W$
	is $\LCA$-$\rel$ and satisfies (S0) -- (S5) w.r.t.\ $G$. Moreover, it holds that
	$\mathfrak{C}_{G\ominus W}=\mathfrak{C}_G$ and, if $G$ is connected, then $G\ominus W$ is
	connected. Finally, Algorithm~\ref{alg:LCA-relevant} can be implemented to run in polynomial
	time.
\end{proposition}
\begin{proof}
	Let $G=(V,E)$ be a DAG on $X$ that serves as input for Algorithm~\ref{alg:LCA-relevant}. 
	By construction in Line~\ref{alg:LCA-relevant:check} and \ref{alg:LCA-add2W},
	the set $W$ contains a vertex $w$ if and only
	if $w\notin\LCA_G(\CC_G(w))$. By Lemma~\ref{lem:not_kLCA}, the latter is precisely if
	$w$ is not a $\{1,\ldots,|X|\}$-$\LCA$ vertex of $G$. In other words, after the last iteration
	of the \emph{for}-loop of Algorithm~\ref{alg:LCA-relevant} the set $W$ comprises all
	vertices of $G$ that are not $\{1,\ldots,|X|\}$-$\LCA$ vertices of $G$. By Theorem~\ref{thm:S4S5},
	the output DAG $H\coloneqq G\ominus W$ thus satisfies Properties (S0) -- (S5) w.r.t.\ $G$ and $H$ is
	$\LCA$-$\rel$. Moreover, $H$ is phylogenetic due to Lemma~\ref{lem:LCA-REL=>phylo}.

	We continue with showing that $W$ is the unique subset of $V$ such that $H = G\ominus W$ is
	$\LCA$-$\rel$ and satisfies (S0) -- (S5). Let $W^*$ be some subset of $V$ such that $H^*
	\coloneqq G\ominus W^*$ is $\LCA$-$\rel$ and satisfies (S0) -- (S5). Observe first that, since
	$H$ and $H^*$ satisfy (S5), $\LCA_G(A) = \LCA_H(A) = \LCA_{H^*}(A)$ for all $A\subseteq X$. This
	together with $H$ and $H^*$ being $\LCA$-$\rel$ implies that each vertex of $H$ and $H^*$ is
	contained in $\LCA_G(A)$ for some $A\subseteq X$ and, in particular, $V(H^*) = \cup_{A\subseteq
	X} \LCA_G(A) = V(H)$. Consequently, $W^* = V\setminus V(H^*) = V\setminus V(H) = W$.
	
	We show now that $\mathfrak{C}_G =\mathfrak{C}_H$.
	Since $H$ satisfies Property (S0) w.r.t.\ $G$, it holds that
	$\mathfrak{C}_H\subseteq\mathfrak{C}_G$. To show that
	$\mathfrak{C}_G\subseteq\mathfrak{C}_H$, let $C\in\mathfrak{C}_G$ be a cluster of $G$ and $v$ a
	vertex of $G$ such that $\CC_G(v)=C$. Since $v$ is a common ancestor of the vertices in $C$,
	there is some $u\in\LCA_G(C)$ such that $u\preceq_G v$. Note that $u\notin W$. Since
	$u\in\LCA_G(C)$, we have $C\subseteq \CC_G(u)$. By Lemma~\ref{lem:prec-subset},
	$\CC_G(u)\subseteq \CC_G(v)=C$. Taken the latter two arguments together, $C=\CC_G(u)$ must hold.
	Since $u\notin W$ and $V(H)=V(G\ominus W)$,
	$u$ is a vertex of $H$. This together with Observation~\ref{obs:ominus} implies that $C = \CC_G(u) = \CC_H(u)\in
	\mathfrak{C}_H$. In summary, $\mathfrak{C}_G =\mathfrak{C}_H$.

	Assume now that $G$ is connected. If $W =\emptyset$, then $G = G\ominus W$ and there is nothing
	to show. Hence suppose that $v\in W$. If $W = \{v\}$, then Lemma~\ref{lem:ominusConnected}
	implies that $H$ is connected. Suppose that there is some $u\in W\setminus \{v\}$. We show that
	$u$ cannot be a $\kLCA$ vertex in $G\ominus v$. To see this, observe first that, since $G\ominus
	v$ satisfies (S5), $\kLCA$ vertices of $G$ remain $\kLCA$ vertices in $G\ominus v$. Hence, the
	set $U$ of all vertices that are not $\{1,\ldots,|X|\}$-$\LCA$ vertices of $G\ominus v$ is a
	subset of $W\setminus \{v\}$. Assume, for contradiction, that $u$ is a $\kLCA$ vertex in
	$G\ominus v$ and thus, $U\subsetneq W\setminus \{v\}$. By Theorem~\ref{thm:S4S5}, $(G\ominus
	v)\ominus U=G\ominus(U\cup\{v\})$ is $\LCA$-\rel. However, $U\subsetneq W\setminus \{v\}$
	implies $U\cup\{v\}\subsetneq W$; the latter two statements yield a contradiction to the
	uniqueness of $W$. Thus, $u$ is not a $\kLCA$ in $G\ominus v$. Now we can apply
	Lemma~\ref{lem:ominusConnected} to conclude that $(G\ominus v)\ominus u$ is connected. Repeating
	the latter arguments until all vertices in $W$ have been processed shows that $G\ominus W$ is
	connected.

	Finally, consider the runtime of Algorithm~\ref{alg:LCA-relevant}. The \emph{if}-condition of the
	algorithm can be implemented to run in polynomial time, since the cluster $\CC_G(v)$ can be
	computed by a simple post-order traversal of $G$ and due to Corollary~\ref{cor:check-LCA/lca-fast}.
	Note furthermore that with an adjacency list representation of $G$, computation of $G\ominus v$
	can be implemented in polynomial time for a given vertex $v$, as it amounts to adding at most
	$|\child_G(v)|$ entries to each list associated to the respective parent of $v$ in $G$ (and there
	are at most $|V(G)|-1$ parents of $v$). In extension, $G\ominus W$ can be computed in polynomial
	time. Since the remaining tasks are clearly possible to perform in constant time, we conclude the
	overall runtime to be polynomial.
\end{proof}

\begin{figure}
	\centering
	\includegraphics[width=0.7\textwidth]{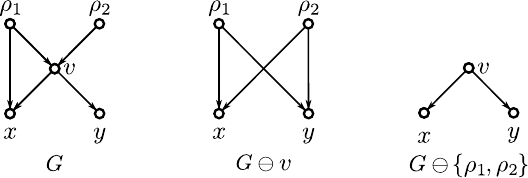}	
	\caption{Shown are DAGs $G$, $G\ominus \{v\}$ and $G\ominus \{\rho_1,\rho_2\}$. 
					 The vertices $\rho_1$ and $\rho_2$ are not LCAs of any subset of leaves in $G$, 
					 while $v = \lca_G(\{x,y\})$. 
					According to Proposition~\ref{prop:alg_kLCA}, $W = \{\rho_1,\rho_2\}$ is the unique and smallest set of vertices 
					such that $G\ominus W$ is $\LCA$-\rel and satisfies (S0) -- (S5). Nevertheless, the set
					$W'=\{v\}$ is the smallest set of vertices 
					such that $G\ominus W'$ is $\LCA$-\rel. However, $G\ominus W'$ violates (S5)
					 since $\LCA_G(\{x,y\})=\{v\} \neq 	\LCA_{G\ominus W'}(\{x,y\})=\{\rho_1,\rho_2\}$. }
	\label{fig:not-minimum-notS5}		
\end{figure}

	Although the set $W$ of all non-$\{1,\ldots,|X|\}$-$\LCA$ vertices of $G$
	(as chosen in Algorithm~\ref{alg:LCA-relevant}) is the unique minimum-sized
	set such that $G\ominus W$ is $\LCA$-\rel \emph{and} satisfies (S0) -- (S5),
	it is not necessarily the smallest set transforming $G$ to an $\LCA$-\rel DAG, 
	see Figure~\ref{fig:not-minimum-notS5} for an example.

\begin{figure}
	\centering
	\includegraphics[width=0.4\textwidth]{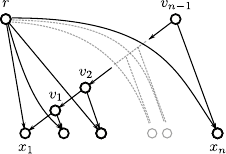}	
	\caption{Consider the clustering system $\mathfrak{C}=\{\{x_1,x_2\},\{x_1,x_2,x_3\},\ldots,X\}
	         \cup \{\{x\} \mid x\in X\}$ on $X=\{x_1,\ldots,x_n\}$, $n>1$.
	         The DAG $G$ on $X$ as shown in the figure is obtained from
	         $H\doteq\hasse(\mathfrak{C})$ by adding a second root $r$ that is adjacent to each leaf
	         $x_i\in X$. Here, for every non-empty $A\subseteq X$, we
	         have $\LCA_G(A)=\{r,v_{i-1}\}$ where $i$ is the maximal index such that $x_i\in A$.
	         Consequently, $G$ is $\LCA$-\rel but not $\lca$-$\rel$. 
	         The set $W$ of all non-$\lca$ vertices of $G$ is the set 
	         $V(G)\setminus X$ of all inner vertices of $G$. Here, $G\ominus W$ would
	         be the disconnected DAG with vertex set $X$ and no edges.
	         The output of Algorithm~\ref{alg:lca-relevant} applied on $G$ is always
	         a connected DAG but
	         heavily  depends on the order in which the vertices have
	         been considered. If $r$ is processed
	         first, i.e., before any of the $v_i$, then the
	         output will be the $\lca$-\rel DAG $G\ominus r\simeq H$. In contrast, if $r$ is processed
	         last, i.e., after each $v_i$, then the $\lca$-\rel DAG $G\ominus\{v_1,\ldots,v_{n-1}\}$ is
	         returned. Here, $\mathfrak{C}_{G\ominus r}=\mathfrak{C}_G=\mathfrak{C}$ while
	         $\mathfrak{C}_{G\ominus\{v_1,\ldots,v_{n-1}\}}=\{X,\{x_1\},\ldots,\{x_n\}\} \subsetneq
	         \mathfrak{C}$.}	
	\label{fig:ominus-diff-no-of-vertices}
\end{figure}

We next show that one can simplify a given DAG $G$ to an $\lca$-\rel DAG $H$ satisfying (S0) -- (S4)
in polynomial time.
Recall that $H$ is $\lca$-\rel if every vertex $v$ in $H$ is the unique least common ancestor for
at least some set $A \subseteq X$. Unsurprisingly, a similar approach to that used in
Algorithm~\ref{alg:LCA-relevant} can be applied in the context of $\lca$-\rel DAGs as well. The
reader may verify that by modifying the \emph{if}-condition in Algorithm~\ref{alg:LCA-relevant} to
``\emph{check if $v \neq \lca_G(\CC_G(v))$}'', one obtains an algorithm that, due to
Lemma~\ref{lem:not_klca} and Theorem~\ref{thm:S4S5}, outputs an $\lca$-\rel DAG that satisfies
Properties (S1) -- (S4). However, the output of this algorithm may be a disconnected DAG even if the
initial input was connected, see Figure~\ref{fig:gominusv} for an example. In particular, the set
$W$ of all non-$\lca$ vertices in $G$ is not necessarily of minimum-size, that is, there are cases
where $G\ominus W'$ is $\lca$-\rel for $W'\subsetneq W$, see
Figure~\ref{fig:ominus-diff-no-of-vertices} for an example. Informally,
the approach in Algorithm~\ref{alg:LCA-relevant} can be overly destructive: it removes \emph{all}
non-$\{1,\ldots,|X|\}$-$\lca$ vertices, including those that are $\{1,\ldots,|X|\}$-$\LCA$ vertices. To
address this issue, we propose Algorithm~\ref{alg:lca-relevant} that, instead of taking
\emph{all} non-$\{1,\ldots,|X|\}$-$\lca$,  repeats the process of removing vertices only until
we end up with an $\lca$-\rel DAG.

\begin{proposition}\label{prop:alg_klca}
	For a given input DAG $G$ on $X$, Algorithm~\ref{alg:lca-relevant} returns a DAG $H$ on $X$ that
	is $\lca$-$\rel$, phylogenetic and satisfies Properties (S0) -- (S4) w.r.t.\ $G$. Moreover, if
	$G$ is connected, then $H$ is connected. If $G$ satisfies (PCC) or (CL), then
	$\mathfrak{C}_{H}=\mathfrak{C}_G$. In addition, Algorithm~\ref{alg:lca-relevant} can be
	implemented to run in polynomial time.
\end{proposition}
\begin{proof}
	To keep track of the original DAG in this proof, we put $G_{\mathrm{orig}}\coloneqq G$ for the
	DAG $G=(V,E)$ on $X$ that serves as input for Algorithm~\ref{alg:lca-relevant}. We show, by
	induction on the number of calls of Line~\ref{alg:lca-rel-ominus}, that each updated DAG $G$
	satisfies (S0) – (S4) w.r.t.\ $G_{\mathrm{orig}}$. As base case, if no calls appear, $G$
	trivially satisfies (S0) – (S4) w.r.t.\ $G_{\mathrm{orig}}$. Suppose the statement is true prior
	to the current call of Line~\ref{alg:lca-rel-ominus}. Since Line~\ref{alg:lca-rel-ominus} is
	called, $v\neq \lca_G(\CC_G(v))$ and Lemma~\ref{lem:not_klca} implies that $v$ is
	not an $\One$-$\lca$ vertex in $G$ for any $\One$. By Theorem~\ref{thm:S4S5}, $G\ominus v$
	satisfies (S0) -- (S4) w.r.t.\ $G_{\mathrm{orig}}$. The algorithm terminates after
	all vertices in $V$ have been processed. Let $H$ denote the final DAG that is returned by
	Algorithm~\ref{alg:lca-relevant}. By the latter arguments and induction, 
	$H$ satisfies (S0) – (S4) w.r.t.\ $G_{\mathrm{orig}}$. 

	We show now that $H$ is	$\One$-$\lca$-$\rel$ for $\One=\{1,2,\ldots,|X|\}$ and hence, that $H$ is $\lca$-$\rel$.
	Let $v$ be the last
	vertex in the \emph{for}-loop for which Line~\ref{alg:lca-rel-ominus} is called and let $W$ be
	the subset of all vertices in $V\setminus \{v\}$ for which Line~\ref{alg:lca-rel-ominus} was called.
	By definition, $H = (G_{\mathrm{orig}}\ominus W)\ominus v$. Assume, for
	contradiction, that $H$ contains a vertex $u$ that is not an $\One$-$\lca$ vertex.
	By Theorem~\ref{thm:S4S5}, $H$ satisfies (S4) w.r.t.\ $G_{\mathrm{orig}}\ominus W$ and thus, if 
	$w$ is an  $\One$-$\lca$ vertex in $G_{\mathrm{orig}}\ominus W$ so it is in $H$. 
	Contraposition of the latter statement implies that $u$ is 
	not an $\One$-$\lca$ vertex in $G_{\mathrm{orig}}\ominus W$. 	Even more,	
	$u$ is not an
	$\One$-$\lca$ vertex in $G_{\mathrm{orig}}\ominus W'$ for any subset $W'\subseteq W$. Hence, if
	$u$ comes before $v$ in the \emph{for}-loop, it would have resulted in a call of Line
	\ref{alg:lca-rel-ominus} and so, $u\in W$; a contradiction. Therefore, $u$ must come after $v$
	in the \emph{for}-loop, i.e., $v$ is not the last vertex for which Line~\ref{alg:lca-rel-ominus}
	is called; also a contradiction. Therefore, all vertices in $G_{\mathrm{orig}}\ominus (W\cup
	\{v\}) =H$ are $\{1,2,\ldots,|X|\}$-$\lca$ vertices of $H$ and, thus, $H$ is $\lca$-\rel. 
	
	By Lemma~\ref{lem:LCA-REL=>phylo}, $H$ is phylogenetic. Furthermore, 
	if $G_{\mathrm{orig}}$ is connected, then induction on the number of calls of
	Line~\ref{alg:lca-rel-ominus} together with Lemma~\ref{lem:ominusConnected} implies that the
	output DAG $H$ is connected.

	Suppose now that $G_{\mathrm{orig}}$ is a DAG that satisfies (CL). Since $H$ satisfies (S0)
	w.r.t $G_{\mathrm{orig}}$, we have $\mathfrak{C}_H\subseteq \mathfrak{C}_{G_{\mathrm{orig}}}$.
	To see that $\mathfrak{C}_{G_{\mathrm{orig}}}\subseteq \mathfrak{C}_H$, let
	$C\in\mathfrak{C}_{G_{\mathrm{orig}}}$. Since $G_{\mathrm{orig}}$
	satisfies (CL), $\lca_{G_{\mathrm{orig}}}(C)$ is well-defined, i.e.
	$u=\lca_{G_{\mathrm{orig}}}(C)$ for some $u\in V(G_{\mathrm{orig}})$. In particular, $u$ is a
	$|C|$-$\lca$ vertex of $G_{\mathrm{orig}}$. Since $H$ satisfy (S4) w.r.t.\ $G_{\mathrm{orig}}$,
	$u$ is thus also a vertex of $H$.  By Observation~\ref{obs:ominus},
	$\CC_H(u)=\CC_{G_{\mathrm{orig}}}(u)$ and by Lemma~\ref{lem:PCC=>CL} we have
	$\CC_{G_{\mathrm{orig}}}(u)=\CC_{G_{\mathrm{orig}}}(\lca_{G_{\mathrm{orig}}}(C))=C$. Thus
	$\CC_H(u)=C$ and $C\in\mathfrak{C}_H$. In conclusion,
	$\mathfrak{C}_{G_{\mathrm{orig}}}=\mathfrak{C}_H$ must hold. By Lemma~\ref{lem:PCC=>CL},
	(PCC) implies (CL). This together with the latter arguments implies that
	$\mathfrak{C}_{G_{\mathrm{orig}}}=\mathfrak{C}_H$ in case that $G_{\mathrm{orig}}$ is a DAG that
	satisfies (PCC).

	For the runtime of Algorithm~\ref{alg:lca-relevant}, note that with an adjacency list
	representation of $G$, computation of $G\ominus v$ can be implemented in polynomial time for a
	given vertex $v$, as it amounts to adding at most $|\child_G(v)|$ entries to each list
	associated to the respective parent of $v$ in $G$ (and there are at most $|V(G)|-1$ parents of
	$v$). Moreover, the \emph{if}-condition of the algorithm can be implemented to run in
	polynomial time, since the cluster $\CC_G(v)$ can be computed by a simple post-order
	traversal of $G$ and due to Corollary~\ref{cor:check-LCA/lca-fast}. Hence every step of the
	\emph{for}-loop of Algorithm~\ref{alg:lca-relevant} takes polynomial time, concluding the
	overall runtime to be polynomial.
\end{proof}

While the set $W$ of vertices used to transform a DAG $G$ to an $\lca$-\rel DAG $G\ominus W$ is, in
general, not uniquely determined (cf.\ Figure~\ref{fig:gominusv} and
\ref{fig:ominus-diff-no-of-vertices}), this situation changes whenever $G$ satisfies (CL) or (PCC).

\begin{theorem}\label{thm:lca-relALGonPCC}
	Let $G$ be a DAG that satisfies (PCC) or (CL) and $W\subseteq V(G)$ be the set of all vertices
	that are not $\{1,\dots,|X|\}$-$\lca$ vertices of $G$. Then, $W$ is precisely the set of all
	vertices that are not $\{1,\dots,|X|\}$-$\LCA$ vertices of $G$. Moreover, $W$ is the unique and,
	therefore, smallest subset of $V(G)$ such that $H\coloneqq G\ominus W$ is $\lca$-$\rel$ and
	satisfies (S0) -- (S4) w.r.t.\ $G$.
  
   Furthermore, it holds that $\mathfrak{C}_G=\mathfrak{C}_{H}$ and
  $H^-\simeq\hasse(\mathfrak{C}_G)$. In particular, $H$ coincides with the DAG returned by
  Algorithm~\ref{alg:lca-relevant} with input $G$. 
\end{theorem}
\begin{proof}
	Let $G$ be a DAG on $X$ that satisfies (PCC) or (CL) and $W\subseteq V(G)$ be the set of all
	vertices that are not $\{1,\dots,|X|\}$-$\lca$ vertices of $G$. By Theorem~\ref{thm:S4S5},
	$H\coloneqq G\ominus W$ is $\lca$-$\rel$ and satisfies (S0) -- (S4). 
  By Theorem~\ref{thm:strongCLChar}, $H$ satisfies strong-(CL) and, thus, (CL). 
  Hence, we
  can apply the same arguments as used in the proof of Proposition~\ref{prop:alg_klca} to show that
  ``$\mathfrak{C}_{G_{\mathrm{orig}}}=\mathfrak{C}_H$'' to conclude that
  $\mathfrak{C}_G=\mathfrak{C}_{H}$ holds. Since $H$ satisfies strong-(CL) and
  $\mathfrak{C}_G=\mathfrak{C}_{H}$, we can apply 
    Theorem~\ref{thm:Char-strongCL} which implies that $H^-\simeq\hasse(\mathfrak{C}_G)$.
    
	 Now, let $W'$ be the set of all vertices that are not
    $\{1,\dots,|X|\}$-$\LCA$ vertices of $G$. We show that $W=W'$. By definition, $W'\subseteq W$.
    Assume, for contradiction, that there is a vertex $w\in W\setminus W'$. Hence, $w$ is a
    $k$-$\LCA$ vertex for some $k$ and Lemma~\ref{lem:not_kLCA} together with
    Lemma~\ref{lem:prec-subset} implies $\CC_G(v)\subsetneq \CC_G(w)$ for all children $v$ of $w$.
    By Lemma~\ref{lem:alg_LCA}, $w\in \LCA_G(\CC_G(w))$. By assumption, $G$ satisfies (CL) or (PCC), where in the
    latter case, Lemma~\ref{lem:PCC=>CL} implies that $G$ satisfies (CL). Thus, $|\LCA_G(\CC_G(w))|=1$
    and, therefore, $w = \lca_G(\CC_G(w))$; a contradiction to $w\in W$. Consequently, $W'= W$ holds.
    
    We continue with showing that $W$ is the uniquely determined set such that $H$ is $\lca$-$\rel$
    and satisfies (S0)~--~(S4). 
    To this end, assume that there is some set $W''\subseteq V(G)$ such that
    $G\ominus W''$ is $\lca$-\rel and satisfies (S0)~--~(S4). Since $G\ominus W''$ satisfies (S4) w.r.t.\ $G$, 
    the set $W''$ cannot contain any vertex that is $k$-$\lca$ vertex in $G$ for some $k$, 
    that is, $W''\subseteq W$. Assume, for contradiction, that $W''\subsetneq W$.
	 Since $W''\subsetneq W=W'$, the set
    $W''$ is also a proper subset of vertices that are not $\{1,\dots,|X|\}$-$\LCA$ vertices of $G$. By
    Theorem~\ref{thm:S4S5}, $G\ominus W''$ satisfies (S0) -- (S5). Moreover, since $G\ominus W''$ is
    $\lca$-\rel, it is, in particular, $\LCA$-\rel. However, this contradicts
    Proposition~\ref{prop:alg_kLCA} which states that $W=W'$ is the unique and minimum-sized set such
    that $G\ominus W'$ is $\LCA$-\rel and satisfies (S0) -- (S5), enforcing $W''=W$. 
    Hence, $W$ is the unique and, therefore, smallest subset of $V(G)$ such that $H$ is
    $\lca$-$\rel$ and satisfies (S0) -- (S4). It is now straightforward to verify that $H$ coincides
    with the DAG returned by Algorithm~\ref{alg:lca-relevant} with input $G$. 
\end{proof}

	Although the set $W$ of all non-$\{1,\ldots,|X|\}$-$\lca$ vertices of DAGs $G$
	with (PCC) or (CL) property  is the unique and minimum-sized
	set such that $G\ominus W$ is $\lca$-\rel \emph{and} satisfies (S0) -- (S4),
	it is not necessarily a unique set transforming $G$ to an $\lca$-\rel DAG. 
	By way of example, consider the DAG $G$ in Figure~\ref{fig:not-minimum-notS5}. 
	Here, $G\ominus\{r_1,v\} \simeq G\ominus\{r_1,r_2\}$ is $\lca$-\rel, 
	but since $v = \lca_G(\{x,y\})\neq \lca_{G\ominus\{r_1,v\}}(\{x,y\})$, 
	the DAG $G\ominus\{r_1,v\}$ does not satisfy (S4).

By  Proposition~\ref{prop:alg_kLCA},  Algorithm~\ref{alg:LCA-relevant} 
always outputs a DAG $H$  with the same set system as the
input DAG $G$, i.e., $\mathfrak{C}_G=\mathfrak{C}_{H}$. 
By Theorem~\ref{thm:lca-relALGonPCC}, this property is also 
guaranteed whenever $G$ satisfies  (PCC) or (CL) when using 
Algorithm~\ref{alg:lca-relevant}. 
In general, however, Algorithm~\ref{alg:lca-relevant}
may return a DAG $H$  with $\mathfrak{C}_H\subsetneq \mathfrak{C}_{G}$
depending on the order in which the vertices are traversed; see Figure~\ref{fig:ominus-diff-no-of-vertices} for an
illustrative example.


\section{The $\ominus$-Operator as Transformation to Simplify Networks}
\label{sec:simplify}

In a recent work, Heiss, Huson and Steel \cite{Heiss2024} proposed a general framework that every
transformation $\varphi(N)$ that ``simplifies'' a network $N$ should satisfy, stated as three
axioms. To be more precise, let $\mathbb{N}(X)$ be the set of all networks on $X$ and
$\mathbb{N}'(X)\subseteq \mathbb{N}(X)$ be some subset of networks that is closed under permuting
the leaves, i.e., if $N\in \mathbb{N}'(X)$ then $N^\sigma \in \mathbb{N}'(X)$, where $N^\sigma$ is
the network obtained from $N$ by relabeling the leaves in $X$ according to some permutation
$\sigma\in \Sigma^X$ in the group $\Sigma^X$ of permutations on $X$. Let $N|Y$ be a restriction of
$N$ to a subset of leaves $Y\subseteq X$ that can be defined in different ways
\cite{10.1371/journal.pcbi.1004135,Heiss2024,FRANCIS2021107215} (we will come to this point later
again). A transformation is then a map \[\varphi\colon \mathbb{N}(X) \to \mathbb{N}'(X)\subseteq
\mathbb{N}(X)\] that assigns to each $N\in \mathbb{N}(X)$ a network $\varphi(N)\in \mathbb{N}'(X)$.
Following Dress et al.\ \cite{Dress2010}, Heiss et al.\ \cite{Heiss2024} proposed three axioms that
are desirable for such transformations, namely
\smallskip
\begin{enumerate}
\item[(P1)] $N\in \mathbb{N}'(X)$ $\implies$ $\varphi(N)=N$, and 
\item[(P2)] $\sigma\in \Sigma^X$, $N\in \mathbb{N}(X)$ $\implies$ $\varphi(N^\sigma)\simeq\varphi(N)^\sigma$, and 
\item[(P3)] $\emptyset\neq Y\subseteq X$, $N\in \mathbb{N}(X)$ $\implies$ $\varphi(N|Y)\simeq\varphi(N)|Y$. 
\end{enumerate}
\smallskip
Property (P1) ensures that any transformation applied on $N\in \mathbb{N}'(X)$ always yields $N$
unchanged. This is justified by the fact that one usually wants to transform or simplify a network
to some network with specific properties encoded by the subclass $\mathbb{N}'(X)$. If our network
$N$ is contained in $\mathbb{N}'(X)$, then it has the required properties and thus, no further
transformation is required. Property (P2) ensures that transformations are invariant under
permutation of leaf labels: transforming a network with permuted leaf labels results in the same
network as when one transforms the original network first and then relabel the leaves. In other
words, the transformation is not dependent on the leaf labels. Finally, Property (P3) ensures that
transformations are invariant under restrictions: taking a restricted network $N|Y$ on a subset of
leaves $Y$ and transforming it results in the same network as that obtained by applying the
transformation $\varphi(N)$ first on $N$ and then taking the restriction $\varphi(N)|Y$. The latter
two properties are mathematically sound but are also motivated from a biological point of view, see
\cite{Dress2010,Heiss2024} for further details. What we have not yet defined is the concept of the
restriction $N|Y$. Due to the lack of an axiomatic framework for ``restriction'', several approaches
to defining $N|Y$ are possible. In \cite{Heiss2024}, Heiss et al.\ defined one such restriction in
terms of subnetworks induced by so-called $\mathrm{LSA}$ vertices of $N$ and their descendants.
Using this definition, they demonstrated that the transformation $\varphi_{\mathrm{LSA}}$ of
phylogenetic networks to a specific tree, called the ${\mathrm{LSA}}$-tree, satisfies properties
(P1), (P2) and (P3). In particular, this type of restriction enforces (P3), ensuring that if
additional species are added to a phylogenetic network (without otherwise altering the original
network), transforming the enlarged network into an ${\mathrm{LSA}}$-tree induces the same
${\mathrm{LSA}}$-tree on the original species set as transforming the original network. However,
there are examples that show that such ${\mathrm{LSA}}$-trees lack our desired property {(S3). 
To see this observe first that  ${\mathrm{LSA}}$-trees do, in general, not satisfy (S0), 
 that is, the ${\mathrm{LSA}}$-tree may contain clusters that are not contained in the original network
$N$. As an example, the ${\mathrm{LSA}}$-tree
$\varphi_{\mathrm{LSA}}(N)$ for the network $N$, shown in Figure~\ref{fig:plantsN}, includes a
cluster containing \textit{Rubellium}, \textit{Chilenium} and \textit{Erpetion} but not
\textit{Tridens}. This suggests that the first three taxa are more closely related evolutionarily
compared to \textit{Tridens}. However, such a cluster does not appear in $\mathfrak{C}_N$; instead,
$\mathfrak{C}_N$ includes the cluster $\{\textit{Chilenium}, \textit{Erpetion}, \textit{Tridens}\}$.
As a consequence of Observation~\ref{obs:newS} and since ${\mathrm{LSA}}$-trees satisfy (S1)
and (S2), ${\mathrm{LSA}}$-trees violate, in general, Condition (S3).

The ${\mathrm{LSA}}$ of a subset of leaves has different properties than the $\LCA$ or $\lca$ as
defined here. For example the ${\mathrm{LSA}}$ of a leaf $x\in X$ with in-degree one is its parent
\cite{HRS:10}, whereas $\lca_G(x) = x$ and $\LCA_G(x) = \{x\}$. This makes their type of restriction
not applicable to our developed methods. In particular, we want to show that the transformation of a
network $N$ with set $W$ of non-$\LCA$ vertices into the network $N\ominus W$ from which all
shortcuts have been removed has all three desired properties. However, it can be shown that this
transformation does not satisfy (P3) when using a restriction defined by $\mathrm{LSA}$s.

\begin{figure}
	\centering
	\includegraphics[width=0.95\textwidth]{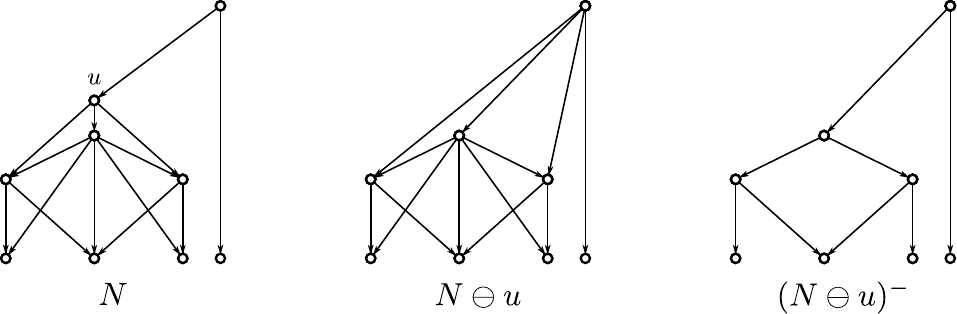}
	\caption{Shown are three phylogenetic networks $N$, $N\ominus u$ and $(N\ominus u)^-$ having the
	         same clustering system $\mathfrak{C}_{N}=\mathfrak{C}_{N\ominus u} =
	         \mathfrak{C}_{(N\ominus u)^-}\eqqcolon \mathfrak{C}$. Here $(N\ominus u)^-\simeq N_1$ with $N_1$
	         being the network as shown in Figure~\ref{fig:cluster-N-DAG}.
	         In $N$, the vertex $u$ is neither
	         a $k$-$\lca$ nor a $k$-$\LCA$\ vertex, for any $k$. In particular, $u$ is the only
	         vertex in $N$ with this property. According to Proposition~\ref{prop:alg_kLCA},
	         $N\ominus u$ is $\LCA$-\rel and satisfies (S0) -- (S5). In addition, Proposition~\ref{prop:alg_klca}
	         implies that      $N\ominus u$ is $\lca$-\rel.
	         Removal of all shortcuts in
	         $N\ominus u$ yields $(N\ominus u)^-$ which is, by
	         Corollary~\ref{cor:setsytem-phylolcaDAG}, regular and thus, isomorphic to the Hasse
	         diagram $\hasse(\mathfrak{C})$. Still, $(N\ominus u)^-$ is $\lca$-\rel and satisfies 
	         (S0) -- (S4). Note that, in this example, $\varphi_{\LCA}(N) = \varphi_{\lca}(N) = (N\ominus
	         u)^- \simeq \Hasse(\mathfrak{C}_{N})$, all satisfying (P1), (P2) and (P3).}
	\label{fig:further-ominus}
\end{figure}

Hence, we will consider a different type of restriction that is solely defined in terms of clusters
of the DAGs under investigation. To be more precise, we define for 
a given DAG $G$ on $X$ and a subset $Y\subseteq X$ the \emph{cluster-restriction} 
\[G \wr Y\coloneqq \hasse(\mathfrak{C}_G\cap Y),
\text{ where } \mathfrak{C}_G\cap Y\coloneqq \{C\cap Y\mid C\in \mathfrak{C}_G, C\cap Y\neq \emptyset\}.\]
In other words, $G\wr Y$ is the restriction of $G$ to the Hasse diagram of
all clusters $C\cap Y$ where $C$ has at least one vertex in $Y$. 
From a phylogenetic point of view, $G \wr Y$ does not make further assumption on the structure 
than what is provided by the clusters in $\mathfrak{C}_G\cap Y$.

In what follows, let $\mathbb{G}(X)$ be the set of all DAGs on $X$ and $\mathbb{R}(X)$ be 
the set of all shortcut-free $\LCA$-\rel DAGs  on $X$. 
Note that all $\lca$-\rel DAGs are $\LCA$-\rel and, therefore, Theorem~\ref{thm:regular-Char} 
implies that the class of regular DAGs is a proper subset of $\mathbb{R}(X)$. 
Denote with $W(G)$ the set of all non-$\LCA$
vertices in the DAG $G$, i.e., the set of all vertices $v\in V(G)$ with $v\notin \LCA_G(A)$ for all $A\subseteq X$. 
To recall, $G^-$ denotes the DAG obtained from $G\in \mathbb{G}(X)$ by removal
of all shortcuts. We will show that the map
\[\varphi_{\LCA} \colon  \mathbb{G}(X) \to \mathbb{R}(X) \text{ defined by } \varphi_{\LCA}(G) = (G\ominus W(G))^-\] 
satisfies (P1), (P2) and (P3) when considering the cluster-restriction $G\wr Y$.
Using the cluster-restriction is motivated by the fact that 
		\[\mathfrak{C}_G = \mathfrak{C}_{(G \ominus W(G))^-} \text{ and } \varphi_{\LCA}(G) \simeq \hasse(\mathfrak{C}_G)\]
for all DAGs satisfying (PCC) or (CL), cf.\ Theorem~\ref{thm:lca-relALGonPCC}. Note that if $G \in
\mathbb{R}(X)$, then $G$ is a shortcut-free DAG on $X$ for which $W(G) = \emptyset$. Hence, each $G
\in \mathbb{R}(X)$ is already in its ``simplified form'' and thus satisfies $\varphi_{\LCA}(G) = G$.
In particular, if $G$ is regular, then $G = G \wr X$. In general, however, it may happen that $G
\not\simeq G \wr X$. This occurs for example if $G$ is not shortcut-free or if $G$ contains non-LCA
vertices, and thus allows for further simplifications.

\begin{figure}[htbp]
	\centering
		\includegraphics[width=1.\textwidth]{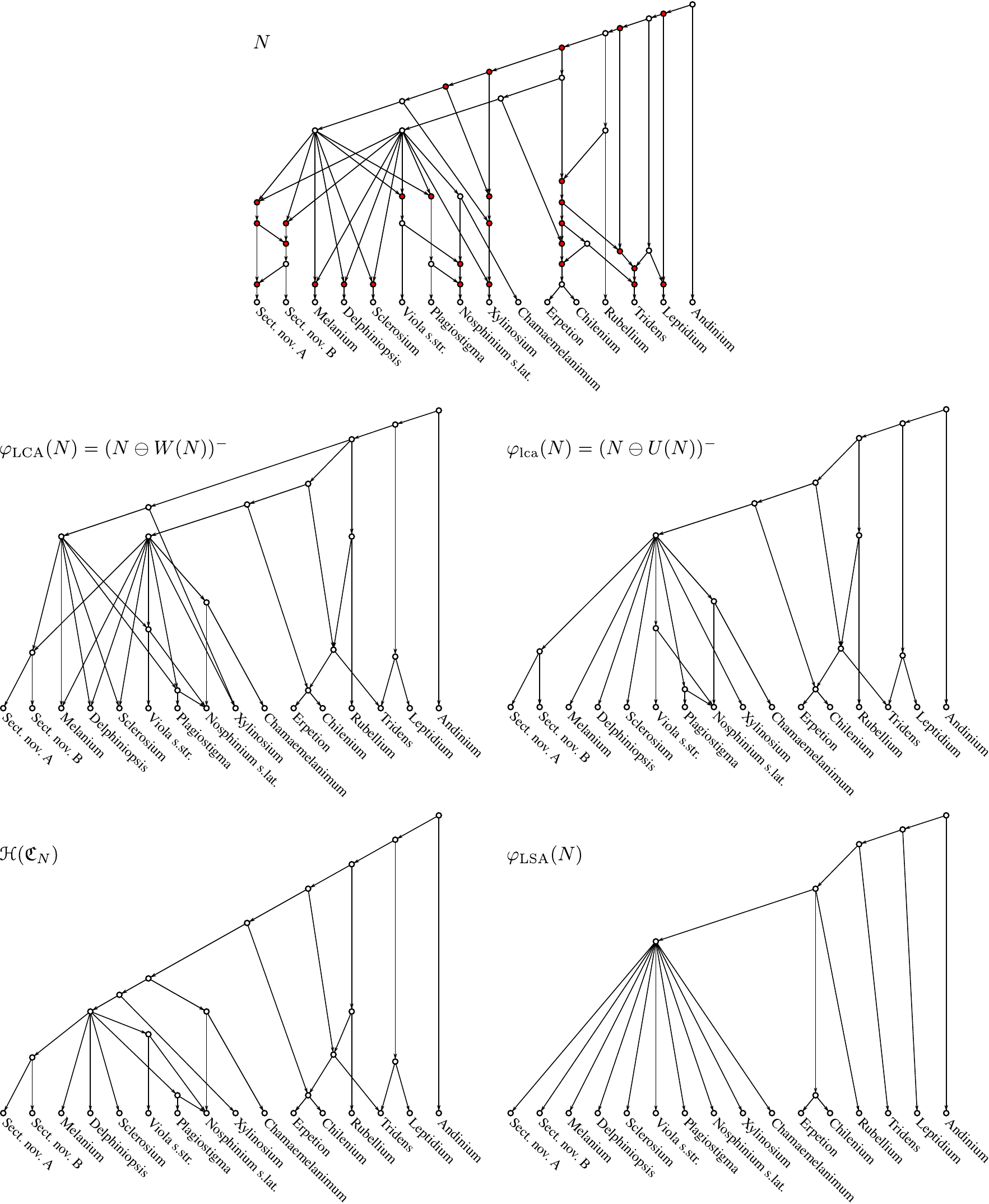}
	\caption{
				Shown is a network $N$ based on a study from Marcussen et al.~\cite{MHB+14} and adapted
				from \cite{JvI:18,HRS:10} together with several simplified versions:
				$\varphi_{\LCA}(N)$, $\varphi_{\lca}(N)$, the Hasse diagram $\hasse(\mathfrak{C}_N)$ and
				the ${\mathrm{LSA}}$-tree $\varphi_{\mathrm{LSA}}(N)$. The ${\mathrm{LSA}}$-tree is
				adapted from \cite[Fig.~3.1]{HRS:10}. Non-$\LCA$ vertices in $N$ are highlighted in red
				and are comprised in the set $W(N)$. The set $U(N)$ comprises all non-$\lca$ vertices in
				$N$. Here, $\varphi_{\LCA}(N)$ satisfies (P1), (P2) and (P3). Since $N$ does not satisfy
				(CL), $\varphi_{\lca}(N)$ satisfies only (P1) and (P2) and we have, therefore,
				$\varphi_{\lca}(N) \neq \hasse(\mathfrak{C}_N)$. 
				} 
	\label{fig:plantsN}	
\end{figure}

\begin{proposition}\label{prop:LCA-P123}
The transformation $\varphi_{\LCA}$ satisfies (P1), (P2) and (P3) under the cluster-restriction for all DAGs. 
To be more precise, it holds that 
\smallskip
\begin{enumerate}[]
\item[(P1)] $G\in \mathbb{R}(X)$ $\implies$ $\varphi_{\LCA}(G)=G$, and
\item[(P2)] $\sigma\in \Sigma^X$, $G\in \mathbb{G}(X)$ $\implies$ $\varphi_{\LCA}(G^\sigma)\simeq\varphi_{\LCA}(G)^\sigma$, and 
\item[(P3)] $\emptyset\neq Y\subseteq X$, $G\in \mathbb{G}(X)$ $\implies$ $\varphi_{\LCA}(G\wr Y)\simeq\varphi_{\LCA}(G)\wr Y$. 
\end{enumerate}
\smallskip
\end{proposition}
\begin{proof}
	If $G\in \mathbb{R}(X)$, then $G$ is shortcut-free
	and $\LCA$-$\rel$ and thus, $W(G) = \emptyset$. Hence,
	$\varphi_{\LCA}(G)=(G\ominus\emptyset)^-=G^- = G$ and (P1) holds. It is a straightforward but tedious task to verify
	that also (P2) is satisfied, which we leave to the reader. We continue with showing (P3). Let $G\in
	\mathbb{G}(X)$ and $\emptyset\neq Y\subseteq X$. Since $\mathfrak{C}_G$ is grounded and $Y$ is nonempty, $\mathfrak{C}_G\cap Y$ is
	grounded. Lemma~\ref{lem:hasse-phylo} implies that $G\wr Y =\hasse(\mathfrak{C}_G\cap Y)$ is
	regular. Theorem~\ref{thm:regular-Char} implies that $G\wr Y$ is shortcut-free 
	and that $G\wr Y$ is $\lca$-rel and thus, in particular, $\LCA$-\rel which implies that $W(G\wr Y) = \emptyset$. 
	Hence, 
	\[\varphi_{\LCA}(G\wr Y) = (G\wr Y \ominus W(G\wr Y))^- = (G\wr Y \ominus \emptyset)^- = (G\wr Y)^- = G\wr Y.\] 
	By repeated application of Lemma~\ref{lem:shortcutfree} to all shortcuts of $H$, it follows that
	$\mathfrak{C}_H = \mathfrak{C}_{H^-}$ for every DAG $H$. This and Proposition~\ref{prop:alg_kLCA}
	implies that $\mathfrak{C}_G = \mathfrak{C}_{G\ominus W(G)} = \mathfrak{C}_{(G\ominus W(G))^-}$
	and, therefore, $\mathfrak{C}_{(G\ominus W(G))^-}\cap Y = \mathfrak{C}_G \cap Y$. Consequently,
	\[\varphi_{\LCA}(G)\wr Y = (G\ominus W(G))^-\wr Y =\hasse(\mathfrak{C}_{(G\ominus W(G))^-}\cap Y)
	= \hasse(\mathfrak{C}_G \cap Y) = G\wr Y = \varphi_{\LCA}(G\wr Y).\] Thus, $\varphi_{\LCA}$
	satisfies (P3).
\end{proof}

We now propose a second transformation which, considering the cluster-restriction, also satisfy (P1)
-- (P3). To this end, let $\mathbb{G}^*(X)\subseteq \mathbb{G}(X)$ be the set of all DAGs that
satisfy (CL). Note that $\mathbb{G}^*(X)$ contains, in particular, all DAGs with (PCC) (cf.\
Lemma~\ref{lem:PCC=>CL}). For $G\in\mathbb{G}^*(X)$, let $U(G)$ be the set of all non-$\lca$ vertices in $G$. 
Consider now the map
\[\varphi_{\lca} \colon  \mathbb{G}^*(X) \to \mathbb{R}(X) \text{ defined by } \varphi_{\lca}(G) = (G\ominus U(G))^-.\] 
Since all DAGs in $\mathbb{G}^*(X)$ satisfy (CL), Theorem~\ref{thm:lca-relALGonPCC} implies that 
$U(G) = W(G)$, i.e., $\varphi_{\lca}(G) = \varphi_{\LCA}(G)$. Hence, we obtain

\begin{proposition}\label{prop:lca-P123}
The transformation $\varphi_{\lca}$ satisfies (P1), (P2) and (P3) under the cluster-restriction, for all DAGs with (CL) or (PCC) property. 
To be more precise it holds that 
\smallskip
\begin{enumerate}[]
\item[(P1)] $G\in \mathbb{R}(X)$ $\implies$ $\varphi_{\lca}(G)=G$, and
\item[(P2)] $\sigma\in \Sigma^X$, $G\in \mathbb{G}^*(X)$ $\implies$ $\varphi_{\lca}(G^\sigma)\simeq\varphi_{\lca}(G)^\sigma$, and 
\item[(P3)] $\emptyset\neq Y\subseteq X$, $G\in \mathbb{G}^*(X)$ $\implies$ $\varphi_{\lca}(G\wr Y)\simeq\varphi_{\lca}(G)\wr Y$.
\end{enumerate}
\smallskip
\end{proposition}

The transformations $\varphi_{\LCA}$ and $\varphi_{\lca}$, in particular, behave very well for DAGs
in $\mathbb{G}^*(X)$ whose set of clusters exhibit a ``tree-like'' structure. 
\begin{corollary}\label{cor:tree-phi}
	Suppose that $G\in \mathbb{G}^*(X)$ is DAG on $X$ for which $\mathfrak{C}_G$ is a hierarchy, i.e., 
	$\mathfrak{C}_G$ is a clustering system that does not contain any overlapping clusters. 
	Then, both $\varphi_{\LCA}(G)$ and $\varphi_{\lca}(G)$ are isomorphic to the phylogenetic tree $T$ on $X$
	with clustering system is $\mathfrak{C}_T = \mathfrak{C}_G$. 
	In particular, the DAG $H$ returned by Algorithm~\ref{alg:lca-relevant} satisfies $H^-\simeq T$.
\end{corollary}
\begin{proof}
	Let  $G\in \mathbb{G}^*(X)$ be a DAG on $X$ for which $\mathfrak{C}_G$ is a hierarchy.
	Put $H\coloneqq \varphi_{\lca}(G)$.	
	Since $G$ satisfies (CL), Theorem~\ref{thm:lca-relALGonPCC} implies that $H$ is an $\lca$-$\rel$ DAG on $X$, 
	that $\mathfrak{C}_H = \mathfrak{C}_G$ and that $H\simeq\hasse(\mathfrak{C}_H)$. Even more, $H = \varphi_{\LCA}(G)$ holds.
	Since $G$ is $\lca$-$\rel$,
  	Lemma~\ref{lem:LCA-REL=>phylo} implies that $H$ is phylogenetic. 
  	It is well-known that $T$ is phylogenetic tree if and only
  	if $\mathfrak{C}_T$ is a hierarchy and 	$T\doteq\Hasse(\mathfrak{C}_T)$ \cite{sem-ste-03a}. 
  	Taking the latter arguments together, $H$ is a phylogenetic tree on $X$
	with clustering system is $\mathfrak{C}_H = \mathfrak{C}_G$. 
	Theorem~\ref{thm:lca-relALGonPCC} implies that the DAG $H$ returned by Algorithm~\ref{alg:lca-relevant} satisfies $H^-\simeq T$.
\end{proof}

We conjecture that similar results to Corollary~\ref{cor:tree-phi} apply to other classes of
phylogenetic networks that are characterized by  properties of their clustering systems, see \cite{Hellmuth2023} for an overview.
Furthermore, it would be of interest to explore in greater detail which 
classes of DAGs or networks $\mathbb{N}^*(X)$
ensure that $\varphi_{\LCA}(G)$ or $\varphi_{\lca}(G)$
remain  in $\mathbb{N}^*(X)$ whenever $G\in \mathbb{N}^*(X)$.

	A simple example of the application of $\varphi_{\LCA}$ and $\varphi_{\lca}$ to a network is shown in Figure~\ref{fig:further-ominus}.
	While the transformation $\varphi_{\LCA}$ applied to any DAG satisfies (P1), (P2) and (P3), 
	it is ensured that the transformation  $\varphi_{\lca}$ satisfies (P1), (P2) and (P3) for DAGs 
	in $\mathbb{G}^*(X)$ and thus, only for DAGs with the (CL) property. 
	In general, $\varphi_{\lca}$ does not satisfy (P3).
	By way of example, consider the network $G$ on $X=\{x_1,\ldots,x_n\}$ in
	Figure~\ref{fig:ominus-diff-no-of-vertices} for some $n\geq2$, where $U(G)=V(G)\setminus X$.
	Consequently, $G'\coloneqq(G\ominus U(G))^-$ is the DAG $(X,\emptyset)$ with no edges or inner
	vertices. Restricting the DAG $G'$ to, say, $Y=\{x_1,x_2\}$ thus also yield a DAG $G'\wr
	Y=\Hasse(\mathfrak{C}_{G'}\cap Y)=\Hasse(\{\{x_1\},\{x_2\}\})$ without edges. In contrast, we
	have \[\varphi_{\lca}(G\wr Y)=\varphi_{\lca}(\hasse(\mathfrak{C}_G\cap
	Y))=\hasse(\{\{x_1\},\{x_2\},\{x_1,x_2\}\}),\] thus $\varphi_{\lca}(G)\wr
	Y\neq\varphi_{\lca}(G\wr Y)$.
	Nevertheless, the application of $\varphi_{\lca}$ to DAGs in $\mathbb{G}(X)\setminus\mathbb{G}^*(X)$
	can reveal meaningful insights, cf.\  $\varphi_{\lca}(N)$ in Figure~\ref{fig:plantsN}	which is
	distinct from $\varphi_{\lca}(N\wr L(N)) = \hasse(\mathfrak{C}_N)$.


\section{Computational Complexity Results for General and (N3O) DAGs}
\label{sec:complexity-hard}

In Section~\ref{sec:ominus-lcaRel}, we have shown that it is possible to compute $\One$-$\lca$-\rel
and $\One$-$\LCA$-\rel DAGs in polynomial time by stepwise removal of certain vertices using the
$\ominus$-operator, given that $\One = \{1, \dots, |X|\}$. However, this situation becomes more
challenging when $\One\subsetneq \{1, \dots, |X|\}$. As we shall see, determining as whether a
vertex is a $\klca$ or $\kLCA$ or verifying that a DAG or network is $\One$-$\lca$-\rel or
$\One$-$\LCA$-\rel are, in general, NP-hard tasks. Nevertheless we provide polynomial time
algorithms for the latter tasks for DAGs $G$ whose set system $\mathfrak{C}_G$ satisfies (N3O), i.e.,
$\mathfrak{C}_G$ does not contain three distinct pairwise overlapping clusters. 

\subsection{General DAGs}
 
For the NP-hardness proofs, we use reductions from the well-known ``Vertex Cover Problem'', which is
based on undirected graphs $H$ where the edge set consists -- unlike in directed graphs -- of
two-element subsets of $V(H)$.

\begin{problem}[\PROBLEM{Vertex Cover}]\ \\
  \begin{tabular}{ll}
    \emph{Input:}    & An undirected graph $H = (V,E)$ and a positive integer $k\leq |V|$.\\
    \emph{Question:} & Is there a \emph{vertex cover} of size $k$ or less, i.e.,  
                      a subset $W\subseteq V$ such that   $|W|\leq k$ \\ & and,  
                      	for each edge $\{u,v\}\in E$,  at least one of $u$ and $v$ is contained in $W$?
  \end{tabular}
  \label{prob:VC}
\end{problem}

\begin{theorem}[\cite{garey1979computers}]\label{thm:VC-NPc}
	\PROBLEM{Vertex Cover} is NP-complete. 
\end{theorem}

For our NP-hardness proofs below we require that the graph $H=(V,E)$ and the integer $k$ that serve
as input for the problem \PROBLEM{Vertex Cover} satisfies certain constrains. To this end, we provide
the following simple observation which is a direct consequence of the fact that $W$ is a vertex
cover of an instance $(H,k)$ if and only if $W' = W\cup \{v\}$ is a vertex cover of an instance
$(H',k')$ obtained from $H$ by adding new vertices $u,v,w$ and edges $\{v,u\}$ and $\{v,w\}$ and by
putting $k'=k+1 > 1$. 
\begin{observation}\label{obs:VC-restricted} 
	\PROBLEM{Vertex Cover} remains NP-complete if the input is restricted to $k>1$ and
	undirected graphs $H=(V,E)$ such that $|V|\geq 4$, $|E|\geq 2$ and $H$ is not star-graph, i.e., a
	connected graph which contains a unique vertex that is contained in all edges.
\end{observation}

For the upcoming proofs, the following simple result will come in handy.
\begin{lemma}\label{lem:CharVC}
	A subset $W\subseteq V$ is a vertex cover of $H = (V,E)$ if and only if $W\not \subseteq
	V\setminus \{u,v\}$ for all $\{u,v\}\in E$. 
\end{lemma}
\begin{proof}
 If  $W \subseteq V\setminus \{u,v\}$ for some $\{u,v\}\in E$, then it can clearly be no vertex cover. 
 Conversely, if $W\subseteq V$ is not a vertex cover of $H$, then there is some edge $\{u,v\}\in E$
 such that $u,v\notin W$. This together with $W \subseteq V$ implies $W \subseteq V\setminus \{u,v\}$. 
\end{proof}

We now formally state the decision problems whose NP-completeness we intend to prove.

\begin{problem}[\PROBLEMK{k}{lca} (resp., \PROBLEMK{k}{LCA})]\ \\
  \begin{tabular}{ll}
    \emph{Input:}    & A DAG $G = (V,E)$, a vertex $v\in V$ and a positive integer $k$ with $1<k\leq |L(G)|$.\\
    \emph{Question:} & Is $v$ a $k$-lca vertex (resp., $k$-LCA vertex) in $G$?
  \end{tabular}
  \label{prob:klca}
\end{problem}

\begin{problem}[\PROBLEMONE{\One}{lca} (resp., \PROBLEMONE{\One}{LCA})]\ \\
  \begin{tabular}{ll}
    \emph{Input:}    & A DAG $G = (V,E)$ and a set $\One$.\\
    \emph{Question:} & Is $G$ an $\One$-$\lca$-\rel (resp., $\One$-$\LCA$-\rel) DAG? 
  \end{tabular}
  \label{prob:kLCA}
\end{problem}

We start with the three problems \PROBLEMK{k}{lca}, \PROBLEMK{k}{LCA} and \PROBLEMONE{\One}{LCA}.

\begin{theorem}\label{thm:kLCa-NPc}
The problems \PROBLEMK{k}{LCA},  \PROBLEMK{k}{lca} and  \PROBLEMONE{\One}{LCA}
are NP-complete, even if the input DAG $G$ is a regular network and, 
thus satisfies (PCC), strong-(CL) and is $\lca$-\rel and shortcut-free.
\end{theorem}
\begin{proof}
	To see that \PROBLEMK{k}{LCA} and \PROBLEMK{k}{lca} are in $\mathrm{NP}$,
	let $A\subseteq X$ of size $k=|A|$ be a given certificate. Now apply Corollary~\ref{cor:check-LCA/lca-fast}.
	To see that \PROBLEMONE{\One}{LCA} is in $\mathrm{NP}$, we assume that as a certificate, we have for each
	vertex $v \in V(G)$ a subset $A_v \subseteq X$ with $|A_v| \in \One$. Verifying whether $v \in
	\LCA_G(A_v)$ can be done in polynomial time due to Corollary~\ref{cor:check-LCA/lca-fast}.

	To prove NP-hardness, we use a reduction from \PROBLEM{Vertex Cover}. Let $(H,k)$ be an
	arbitrary instance of \PROBLEM{Vertex Cover}. By Observation~\ref{obs:VC-restricted}, we can assume
	that $k>1$, $|V(H)|\geq 4$, $|E(H)|\geq 2$ and that $H$ is not a star-graph. Consider the following set
	system \[\mathfrak{C} \coloneqq \left(\bigcup_{x\in V(H)}\{\{x\}\}\right) \cup
	\left(\bigcup_{e\in E(H)} \{V(H)\setminus e\}\right) \cup \{V(H)\}.\] Since $|V(H)|\geq 4$, we have 
	$|V(H)\setminus e|\geq 2$ for all $e\in E(H)$. Thus, $V(H)\setminus e$ appears
	as a non-singleton cluster in $\mathfrak{C}$.
	It is now
	easy to verify that $\mathfrak{C}$ is a well-defined clustering system. Let
	$G\doteq\Hasse(\mathfrak{C})$ be the DAG obtained from the Hasse diagram $\hasse(\mathfrak{C})$
	by relabeling all vertices $\{x\}$ by $x$. Hence, $L(G) = V(H)$ and $\mathfrak{C}_G =
	\mathfrak{C}$. Since $\mathfrak{C}$ is a clustering system and $G\simeq \hasse(\mathfrak{C}_G)$,
	Lemma~\ref{lem:hasse-phylo} implies that $G$ is a regular network. By
	Theorem~\ref{thm:regular-Char}, $G$ is shortcut-free, $\lca$-$\rel$ and satisfies the
	strong-(CL) property. An example of the constructed DAG $G$ is
	shown in Figure~\ref{fig:NPC1-gadgets}.

	\begin{figure}[h!]
		\centering
		\includegraphics[width=0.6\textwidth]{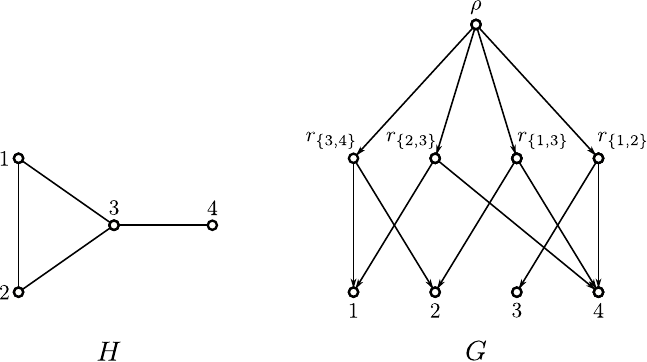}\\
		\caption{Reduction used in the proof of Theorem~\ref{thm:kLCa-NPc}.  In this example, $H$
					and $k=2$ serve as  input for \PROBLEM{Vertex Cover}. Here, 
					$W=\{1,3\}$ and $W=\{2,3\}$  are the only vertex covers of $H$ size $|W|\leq 2$. 
					Moreover, $\rho$ is a $k$-$\lca$, resp., $k$-$\LCA$  vertex for $k=2$ precisely for the 
					two sets $W=\{1,3\}$ and $W=\{2,3\}$. In particular, $\rho$ is a $k$-$\lca$, resp., $k$-$\LCA$ in $G$
					precisely if $H$ has a vertex cover of size $k$. We note that, in general, the vertices $r_e$ are
					not $2$-$\lca$ vertices but always $2$-$\LCA$ vertices.
					}
		\label{fig:NPC1-gadgets}	
	\end{figure}

	We show first NP-hardness of \PROBLEMK{k}{LCA} and \PROBLEMK{k}{lca}.
	Let us denote with $\rho$ the unique root of $G$ which results in the instance $(G,\rho,k)$
	of \PROBLEMK{k}{lca}, as well as of \PROBLEMK{k}{LCA}. By definition, the vertex set of $G$ consists of the
	unique root $\rho$, the leaves in $L(G)=V(H)$ and a vertex $r_e$ that correspond to the cluster
	$V(H)\setminus e$ for each $e\in E(H)$. Note that $r_e$ does not correspond to a leaf in 
	$G$ since $V(H)\setminus e$ is not a singleton cluster in $\mathfrak{C}$ and, since 
	$|E(H)|\geq 2$, at least two such vertices $r_e$ with $e\in E(H)$ exist. Moreover,
	since $H$ is not a star-graph, for each vertex $v\in V(H)$ there is an edge $e\in E(H)$ such
	$v\notin e$. Hence, each vertex $v\in V(H)$ is contained in at least one set $V(H)\setminus e$
	for some $e\in E(H)$. The latter arguments imply that $G$ has edges $(\rho,r_e)$ for all $e\in
	E(H)$ and edges $(r_e,v)$ for all $e\in E(H)$ and all vertices $v \in V(H)\setminus e$. No
	further edges exists. Thus, without explicit construction of the Hasse diagram based on the
	clustering system $\mathfrak{C}$, we can instead directly construct $G$ by adding $1 + |E(H)| +
	|V(H)|$ vertices to $G$ and the prescribed $|E(H)| + |E(H)|(|V(H)|-2)$ edges to $G$. In summary,
	$G$ and thus, the instance $(G,\rho,k)$ of \PROBLEMK{k}{lca} and \PROBLEMK{k}{LCA}, can be
	constructed in polynomial time. 	
   
	Suppose that $(H,k)$ is a yes-instance of \PROBLEM{Vertex Cover}. Thus, there is, in particular, a vertex cover
	$W$ of $H$ such that $|W|=k>1$. By construction, $\rho$ is a common ancestor of all vertices in
	$L(G) = V(H)$ and thus, of all vertices in $W$. The vertex $r_e$ is a common ancestor of
	$A\subseteq L(G) = V(H)$ precisely if $A\subseteq V(H)\setminus e$ and $|A|> 1$. By Lemma~\ref{lem:CharVC},
	$W\not \subseteq V(H)\setminus e$ and thus, $r_e$ is not a common ancestor of $W$ for any $e\in E(H)$. Thus, $\rho$
	is the unique least common ancestor of $W$ and therefore, $\rho = \lca_G(W)$. Since $|W|=k$, the
	root $\rho$ is a $\klca$ vertex and thus, in particular, a $\kLCA$ vertex.
	
	Suppose that $(G,\rho,k)$ is a yes-instance of \PROBLEMK{k}{lca} (resp., \PROBLEMK{k}{LCA}).
	Hence, $\rho = \lca_G(W)$ (resp., $\rho \in \LCA_G(W)$) for some $W\subseteq L(G) = V(H)$ with
	$|W|=k>1$. If $W\subseteq V(H)\setminus e$ for some edge $e\in E(H)$, then $r_e$ is the unique least
	common ancestor of $W$; contradicting $\rho = \lca_G(W)$ (resp.\ $\rho \in \LCA_G(W)$).
	Thus, it must hold that $W\not \subseteq  V(H)\setminus e$ for all $e\in E(H)$ and Lemma
	\ref{lem:CharVC} implies that $W$ is a vertex cover of $H$.
	In summary, \PROBLEMK{k}{lca} and \PROBLEMK{k}{LCA} are NP-hard and, therefore, NP-complete.
	
	To show NP-hardness of \PROBLEMONE{\One}{LCA}, we use the same DAG $G$ and put $\One = \{1,2,k\}$
	which results in an instance $(G,\One)$ of \PROBLEMONE{\One}{LCA}. By the arguments above, this
	reduction can be achieved in polynomial time. As shown above, $\rho$ is a $\kLCA$ vertex, that is
	$\rho \in \LCA_G(W)$ for some $W\subseteq L(G)=V(H)$ with $W=k$ if and only if $W$ is a vertex
	cover of $H$. Moreover, since distinct $r_e$ and $r_f$ are $\preceq_G$-incomparable and since
	each $r_e$ is adjacent to at least two leaves $x,y\in V(H)\setminus e$, it follows that $r_e \in
	\LCA_G(\{x,y\})$, i.e., $r_e$ is a $2$-$\LCA$ vertex for each $e\in E(H)$. In summary, $(H,k)$ is
	a yes-instance of \PROBLEM{Vertex Cover} if and only if $(G,\One)$ is a yes-instance of
	\PROBLEMONE{\One}{LCA}. Therefore, \PROBLEMONE{\One}{LCA} is NP-hard and thus, NP-complete.
\end{proof}

	The NP-hardness of \PROBLEMK{k}{lca} and \PROBLEMONE{\One}{LCA} does not directly imply that 
	\PROBLEMONE{\One}{lca} is NP-hard as well. In particular, the vertices $r_e$ in 
	the instance $G$ constructed in Theorem~\ref{thm:kLCa-NPc} are, in general, 
	not $2$-$\lca$ vertices. Moreover, verifying whether $G$ is
	$\One$-$\lca$-\rel is equivalent to checking that \emph{all} vertices $v
	\in V(G)$ are $\One$-$\lca$ vertices, which imposes strong
	structural constraints on $G$. Nevertheless, it is not at all surprising that we arrive at the following 

\begin{theorem}\label{thm:kLCa-relevant-NPc}
	The problem \PROBLEMONE{\One}{lca} is NP-complete. 
\end{theorem}
\begin{proof}
	Showing that \PROBLEMONE{\One}{lca} is in $\mathrm{NP}$ is done in the same way as showing that
	\PROBLEMONE{\One}{LCA} is in $\mathrm{NP}$. To prove NP-hardness, we use a reduction from
	\PROBLEM{Vertex Cover}. Let $(H,k)$ be an arbitrary instance of \PROBLEM{Vertex Cover}. By
	Observation~\ref{obs:VC-restricted}, we can assume that $k>1$, $|V(H)|\geq 4$ and $|E(H)|\geq 2$.
	We construct now an instance $(G,\One)$ for \PROBLEMONE{\One}{lca}. First, put $\One =
	\{1,2,k\}$. Now construct a DAG $G$ as follows. First, initialize $G$ as the Hasse diagram
	$G\doteq\Hasse(\mathfrak{C})$ of the set system $\mathfrak{C} \coloneqq (\bigcup_{x\in
	V(H)}\{\{x\}\}) \cup (\cup_{e\in E(H)} \{V(H)\setminus e\})$. Here, $G$ is equivalently obtained
	from the DAG constructed in the proof of Theorem~\ref{thm:kLCa-NPc} by removal of the unique root
	$\rho$. For each $e\in E(H)$, let $r_e$ denote the vertex in $G$ that is adjacent to all $x\in
	V(H)\setminus e$. We now add to $G$, for all $e\in E(H)$, two new leaves $x^e_1, x^e_2$ and edges
	$(r_e,x^e_1)$ and $(r_e,x^e_2)$. Let us denote with $Z$ the set comprising all leaves $x^e_1$ and
	$x^e_2$ for all $e\in E(H)$. Finally add to $G$ a vertex $r^*$ and edges $(r^*,x)$ for all $x\in
	V(H)$. By construction, $L(G) = V(H)\cup Z$. This results in the instance $(G,\One)$ for
	\PROBLEMONE{\One}{lca}, see Figure~\ref{fig:NPC2-gadgets} for an illustrative example. By similar
	arguments as in the proof of Theorem~\ref{thm:kLCa-NPc}, the instance $(G,\One)$ can be
	constructed in polynomial time. 

	\begin{figure}[h!]
		\centering
		\includegraphics[width=0.8\textwidth]{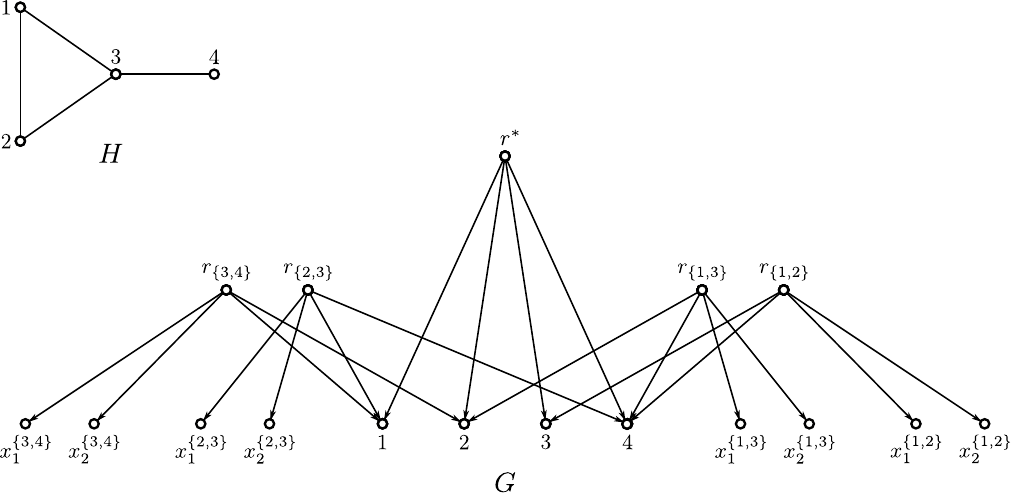}\\
		\caption{Reduction used in the proof of Theorem~\ref{thm:kLCa-relevant-NPc}.  In this example, $H$
					and $k=2$ serve as  input for \PROBLEM{Vertex Cover}. Here, 
					$W=\{1,3\}$ and $W=\{2,3\}$ are the only vertex covers of $H$  size $|W|\leq 2$. 
					Each vertex $v\neq r^*$ in $G$ is, by construction, a $\{1,2\}$-$\lca$ vertex:
					if $v$ is a leaf it is a $1$-$\lca$ vertex and, otherwise, 
					$r_e=\lca_G(\{x^e_1, x^e_2\})$ with $e\in E(H)$ is a $2$-$\lca$ vertex. 
					Moreover, $r^*$ is a $k$-$\lca$ vertex for $k=2$ precisely for the 
					two sets $W=\{1,3\}$ and $W=\{2,3\}$. In particular, $G$ is a $\{1,2,k\}$-$\lca$-\rel
					precisely if $H$ has a vertex cover of size $k$.
					}
		\label{fig:NPC2-gadgets}	
	\end{figure}

	By construction, each $r_e$ in $G$ is precisely adjacent to the leaves in $V(H) \setminus e$ and
	the two leaves $x^e_1$ and $x^e_2$. In particular, $r_e$ is the only ancestor of the leaves
	$x^e_1$ and $x^e_2$. Hence, $r_e=\lca_G(\{x^e_1,x^e_2\})$ which implies that $r_e$ is a
	$2$-$\lca$ vertex for all $e\in E(H)$. Therefore, it is evident that $(G,\One)$ is a yes-instance
	of \PROBLEMONE{\One}{lca} with $\One = \{1,2,k\}$ if and only if $r^*$ is a $\{2,k\}$-$\lca$
	vertex in $G$. Although the DAG $G$ constructed here slightly differs from the one in the proof
	Theorem~\ref{thm:kLCa-NPc} (in particular, $r^*$ is not an ancestor of any $r_e$ with $e\in
	E(H)$), $r^*$ is still an ancestor of all $x\in V(H) = L(G)\setminus Z$ and of no other vertices. 
	
	Suppose that $(H,k)$ is a yes-instance of \PROBLEM{Vertex Cover}. Thus, there is a vertex cover
	$W\subseteq V(H)$ of $H$ such that $|W|=k>1$. By construction, $r^*$ is a common ancestor of all
	vertices in $V(H) = L(G) \setminus Z$ and thus, of all vertices in $W$, but not of $r_e$ for all
	$e\in E(H)$. Since $W \subseteq V(H)$, the vertex $r_e$ is a common ancestor of $W$ precisely if
	$W\subseteq V(H)\setminus e$. However, by Lemma~\ref{lem:CharVC}, for all $e\in E(H)$, $W\not
	\subseteq V(H)\setminus e$ and thus, $r_e$ is not a common ancestor of $W$. Thus, $r^*$ is the
	unique least common ancestor of $W$ and therefore, $r^* = \lca_G(W)$. Since $|W|=k$, the vertex
	$r^*$ is a $\klca$ vertex. In summary, $(G,\One)$ is a yes-instance of \PROBLEMONE{\One}{lca}.
	
	Suppose that $(G,\One)$ is a yes-instance of \PROBLEMONE{\One}{lca}. Hence, $r^* = \lca_G(W)$ for
	some $W\subseteq L(G) = V(H)\cup Z$ with $|W|\in\{2,k\}$. Since $r^*$ is not an ancestor of any
	vertex in $Z$, it follows that $W\subseteq V(H)$. If $W\subseteq V(H)\setminus e$ for some edge
	$e\in E(H)$, then $r_e$ is a common ancestor of $W$. Since $r_e$ and $r^*$ are
	$\preceq_G$-incomparable, this would contradict $r^* = \lca_G(W)$. Thus, it must hold that $W\not
	\subseteq V(H)\setminus e$ for all $e\in E(H)$ and Lemma \ref{lem:CharVC} implies that $W$ is a
	vertex cover of $H$. Hence, $(H,k)$ is a yes-instance of \PROBLEM{Vertex Cover}. Thus,
	\PROBLEMONE{\One}{lca} is NP-hard and, consequently, NP-complete.
\end{proof}

Although the problems \PROBLEMK{k}{lca} and \PROBLEMK{k}{LCA} as well as \PROBLEMONE{\One}{lca} and
\PROBLEMONE{\One}{LCA} are NP-hard, we note in passing that these problems become polynomial-time
solvable whenever $k$ is treated as a constant, i.e., $k\in O(1)$. To test if a vertex $v$ is a
$\klca$ or a $\kLCA$ vertex in a DAG $G=(V,E)$ can then be done by testing all $O(|V|^k)$ subsets $A
\subseteq \CC_G(v)$ with $|A| = k$ using Algorithm~\ref{alg:LCA}. Hence, testing if $v$ is $\klca$
or $\kLCA$ can be done in polynomial time. Similarly, if in a given set $\One$ the maximum integer
$k\in \One$ is treated as fixed and constant, we can repeat the latter arguments for all vertices in
$G$ and all $O(|V|^k)$ subsets of size at most $k$ to obtain a polynomial time approach for
verifying if $G$ is \Olca-\rel or \OLCA-\rel or not. This approach is, in particular, feasible when
we are interested in the special case that all $v\in V$ satisfy $v = \lca(\{x, y\})$ or $v \in
\LCA(\{x, y\})$ for some $x, y \in X$. This discussion together with Proposition~\ref{prop:alg_kLCA}
and \ref{prop:alg_klca} implies
\begin{observation}\label{obs:12-rel}
It can be tested in polynomial time if a vertex $v\in V$ in a DAG $G=(V,E)$
is a $\{1,2\}$-$\lca$ or  $\{1,2\}$-$\LCA$ vertex. Moreover, it can be tested in 
polynomial time if a DAG is $\{1,2\}$-$\lca$-\rel  or  $\{1,2\}$-$\LCA$-\rel. 
Finally, every DAG $G$ can be transformed in polynomial-time into a 
phylogenetic $\{1,2\}$-$\lca$-\rel (resp.,   $\{1,2\}$-$\LCA$-\rel) DAG 
satisfying (S0) -- (S4)  (resp., (S0) -- (S5)) w.r.t.\ $G$
using straightforward modifications of Algorithm~\ref{alg:lca-relevant} (resp., Algorithm~\ref{alg:LCA-relevant}).
\end{observation}

We now summarize the complexity results presented in this paper so far. We have shown that it is
tractable to decide whether a given vertex is the (unique) LCA of a \emph{specified set} $A$ (cf.\
Corollary~\ref{cor:check-LCA/lca-fast}) and whether a given vertex $v$ is a $k$-$\lca$ (resp.\
$k$-$\LCA$) vertex for \emph{some} integer $k$ (cf.\ Corollaries~\ref{cor:check-LCA/lca-fast}
and~\ref{cor:klca<=>v=lca(C(v))}). However, if $k$ is is part of the input and not treated as
constant, it is NP-complete to determine whether $v$ is a $k$-$\lca$ or a $k$-$\LCA$ vertex.
Similarly, deciding whether a DAG is \Olcarel (resp.\ \OLCArel) is tractable when
$\One=\{1,\ldots,|L(G)|\}$ (cf.\ Corollary~\ref{cor:check-LCA/lca-fast}) but NP-complete otherwise.

\subsection{DAGs with (N3O) property}

Although the problems \PROBLEMONE{\One}{lca} and \PROBLEMONE{\One}{LCA}, as well as
\PROBLEMK{k}{LCA} and \PROBLEMK{k}{lca}, are NP-complete, we show that they can be efficiently
solved when the input DAGs \emph{$G$ satisfy (N3O)}, i.e., $\mathfrak{C}_G$ satisfies (N3O) and thus
does not contain three distinct pairwise overlapping clusters. The interest into DAGs that satisfy
(N3O) is two-fold. On the one hand, DAGs with (N3O) property include interesting and non-trivial
classes of networks such as rooted trees or galled-trees \cite{Hellmuth2023}, i,e., networks in
which each connected component $K$ is either a single vertex, an edge or $K$ is composted of exactly
two $uv$-paths that only have $u$ and $v$ in common. An example of a galled-tree is provided by the network $N_1$ in Figure~\ref{fig:cluster-N-DAG}.
In particular, galled-trees form a subclass of
level-1 networks, i.e., networks $N$ in which each biconnected component contains at most one vertex
$v$ with $\indeg_N(v)>1$ \cite{Hellmuth2023}. On the other hand, it can be verified in polynomial
time whether a given set system $\mathfrak{C}$ on $X$ satisfies (N3O) or not, by checking, for all
of the $O(|\mathfrak{C}|^3)$ distinct clusters $C_1,C_2,C_3\in \mathfrak{C}$, if they pairwise
overlap or not in $O(|X|^3)$ time. Since $|\mathfrak{C}_G|\leq |V|$ and since $\mathfrak{C}_G$ can,
together with Lemma~\ref{lem:union-A}, be determined by a simple post-order traversal in polynomial
time for every DAG $G$, this yields a polynomial time approach to test if $G$ satisfies (N3O).
\begin{observation}	
DAGs with (N3O) property can be recognized in polynomial time.
\end{observation}

In what follows, we show that the problems  \PROBLEMONE{\One}{lca} and \PROBLEMONE{\One}{LCA}, as well as
\PROBLEMK{k}{LCA} and \PROBLEMK{k}{lca} can be solved in polynomial time on DAGs with (N3O) property.
To this end, we start with the following results.

\begin{lemma}\label{lem:(N3O)::k-LCA<=>2-LCA}
	Let $G$ be a DAG on $X$ that satisfies (N3O) and let $v\in V(G)$. Then, $v$ is a $\2LCA$ vertex
	in $G$ if and only if (a) $|\CC_G(v)|\geq 2$ and (b) $\CC_G(u)\neq\CC_G(v)$ for all $u\in\child_G(v)$. 
	In particular, the following statements are equivalent. 
	\smallskip
	\begin{enumerate}[label=(\arabic*)]
	\item $v$ is a $\kLCA$ vertex in $G$ for some $k\geq 2$.
	\item $v$ is an $\ell$-$\LCA$ vertex in $G$ for all $\ell\in \{2,3,\dots,|\CC_G(v)|\}$. 
	\end{enumerate}
	\smallskip
	Hence, $G$ is $\One$-$\LCA$-\rel for some $\One$ with $|\One|>1$ if and only if each inner vertex
	of $G$ is a $\2LCA$ vertex. 
\end{lemma}
\begin{proof}
	Let $G$ be a DAG on $X$ that satisfies (N3O) and let $v\in V(G)$. First suppose that $v$ is a
	$\2LCA$ vertex in $G$ and thus that $v\in\LCA_G(A)$ for some $A\subseteq X$ with $|A|=2$. Hence,
	$A\subseteq \CC_G(v)$ and $|\CC_G(v)|\geq 2$ must hold. By Lemma~\ref{lem:alg_LCA},
	$\CC_G(u)\neq\CC_G(v)$ for all $u\in\child_G(v)$. Conversely, assume that Condition (a) and (b)
	are satisfied for $v$. Assume, for contradiction, that $v$ is not a $\2LCA$ vertex. Let
	$\mathfrak{C}^*$ be the set of all inclusion-maximal cluster in the set $\{\CC_G(u)\mid
	u\in\child_G(v)\}$ of the clusters associated with the children of $v$ in $G$. Since
	$|\CC_G(v)|\geq 2$, $v$ must be an inner vertex and Lemma~\ref{lem:union-A} implies that
	\begin{equation}\label{eq:C(u)unionOfCstar}
	\CC_G(v)=\bigcup_{u\in\child_G(v)}\CC_G(u)=\bigcup_{C\in\mathfrak{C}^*}C.\tag{I} \end{equation}
	Since $v$ is not a leaf and not a $\2LCA$ vertex, Lemma~\ref{lem:alg_LCA} implies that, for all
	$A\subseteq \CC_G(v)$ with $|A|=2$, there must be some child $u\in \child_G(v)$ such that
	$A\subseteq \CC_G(u)$. Thus, for every $A\subseteq\CC_G(v)$ of size $|A|=2$ there is, in
	particular, some element $C\in\mathfrak{C}^*$ such that $A\subseteq C$. Note that any two
	clusters in $\mathfrak{C}^*$ are either disjoint or overlap. Condition (b) and Eq.\
	\eqref{eq:C(u)unionOfCstar} imply that $|\mathfrak{C}^*|\geq 2$. Assume, first that the clusters
	in $\mathfrak{C}^*$ are pairwise disjoint. Let $C_1,C_2$ be distinct elements in $\mathfrak{C}^*$
	and $x\in C_1$ and $y\in C_2$. Since $C_1\cap C_2=\emptyset$, we have $x\neq y$. However, as
	argued above, there exists a cluster $C_3\in\mathfrak{C}^*$ such that $\{x,y\}\subseteq C_3$ and,
	thus, $C_1\cap C_3\neq \emptyset$; a contradiction. Hence, the clusters in $\mathfrak{C}^*$
	cannot all be pairwise disjoint. Thus, there are clusters $C_1,C_2 \in \mathfrak{C}^*$ that
	overlap. Therefore, there are $x\in C_1\setminus C_2$ and $y\in C_2\setminus C_1$ and thus,
	$x\neq y$. Again, as argued above, there exists a cluster $C_3\in\mathfrak{C}^*$ such that
	$\{x,y\}\subseteq C_3$ and thus, $C_3\neq C_1,C_2$. Consequently, $C_3$ overlaps with both $C_1$
	and $C_2$, i.e., $\mathfrak{C}^*$ contains the three pairwise overlapping clusters $C_1,C_2$ and
	$C_3$; violating the fact that $G$ satisfies (N3O). Hence, this case cannot occur. In summary,
	$v$ must be a $\2LCA$ vertex.
	
	We show now that Statements (1) and (2) are equivalent. Clearly (2) implies (1). Hence, assume
	that $v$ is a $\kLCA$ vertex in $G$ for some $k\geq 2$. One easily observes that $k\leq |\CC_G(v)|$
	must hold as, otherwise, there is no $A\subseteq X$ of size $|A|=k$ such $A\subseteq \CC_G(v)$ and,
	thus $v\notin \LCA_G(A)$ by Lemma~\ref{lem:alg_LCA}. Moreover, there cannot be a child $u$ of $v$ in $G$ such that
	$\CC_G(u)=\CC_G(v)$ since, otherwise, Lemma~\ref{lem:alg_LCA} would imply that $v$ is not a
	$k$-$\LCA$ vertex. Thus, we have shown that (a) $|\CC_G(v)|\geq k \geq 2$ and (b)
	$\CC_G(u)\neq\CC_G(v)$ for all $u\in\child_G(v)$ must hold. By the previous statement, $v$ is a
	$\2LCA$ vertex. By Lemma~\ref{lem:k>lca}, $v$ is an $\ell$-$\LCA$ vertex in $G$ for all $\ell$
	with $2\leq \ell\leq |\CC_G(v)|$.
	
	It is now an easy task to verify that $G$ is $\One$-$\LCA$-\rel for some $\One$ with
	$|\One|>1$ (which implies that $\ell\in \One$ for some $\ell>1$) if and only if each inner vertex
	of $G$ is a $\2LCA$ vertex.
\end{proof}

\begin{lemma}\label{lem:(N3O)::k-lca<=>2-lca}
	Let $G$ be a DAG on $X$ that satisfies (N3O) and let $v\in V(G)$. 
	Then, the following statements are equivalent. 
	\smallskip
	\begin{enumerate}[label=(\arabic*)]
	\item $v$ is a $\klca$ vertex in $G$ for some $k\geq 2$.
	\item $v$ is an $\ell$-$\lca$ vertex in $G$ for all $\ell\in \{2,3,\dots,|\CC_G(v)|\}$. 
	\end{enumerate}
	\smallskip
	Moreover, the following statements are equivalent.
	\smallskip 
	\begin{enumerate}[resume,label=(\arabic*)]
	\item $G$ is $\One$-$\lca$-\rel for some $\One$ with $|\One|>1$. 
	\item Each inner vertex	of $G$ is a $2$-$\lca$ vertex.
	\item  $G$ has the strong-(CL) property.
	\end{enumerate}
	\smallskip
\end{lemma}
\begin{proof}
	Suppose first that $v$ is a $\klca$ vertex in $G$ for some $k\geq 3$. We start with showing
	that $v$ is a $(k-1)$-$\lca$ vertex in $G$. Since $v$ is a $\klca$ vertex, there is a subset $A
	\subseteq X$ of size $|A|=k$ such that $\lca_G(A)=v$. Assume, for contradiction, that $v$ is not
	a $(k-1)$-$\lca$ vertex in $G$. Since $|A|\geq 3$ we can choose three distinct vertices
	$x_1,x_2,x_3\in A$. Put $A_i\coloneqq A\setminus \{x_i\}$ for all $i\in \{1,2,3\}$. Note that
	$|A_i|=k-1$ and thus, $v\neq \lca_G(A_i)$ for all $i\in \{1,2,3\}$. Since $v$ is a common
	ancestor of every vertex in $A_i$ but
	$v\neq \lca_G(A_i)$ it follows that there is a vertex $v_i\prec_G v$ with $v_i\in \LCA_G(A_i)$
	for all $i\in \{1,2,3\}$ By construction and since $|A|\geq 3$, $A_i\cap A_j\neq \emptyset$ and
	thus, $\CC_G(v_i)\cap \CC_G(v_j)\neq \emptyset$ for all $i,j\in \{1,2,3\}$. Since $v_i
	\prec_G v=\lca_G(A)$ and $A_i\subseteq \CC_G(v_i)$ and $A_i\cup\{x_i\}=A$ it follows that
	$x_i\notin \CC_G(v_i)$ for all $i\in \{1,2,3\}$. By construction, $x_i \in \CC_G(v_j)$ for all
	distinct $i,j\in \{1,2,3\}$. Hence, $\CC_G(v_i)\not\subseteq \CC_G(v_j)$ for all distinct $i,j\in
	\{1,2,3\}$. Consequently, $\CC_G(v_1)$, $\CC_G(v_2)$ and $\CC_G(v_3)$ are three pairwise
	overlapping cluster; contradicting the fact that $G$ satisfied (N3O). Hence, $v$ is a
	$(k-1)$-$\lca$ vertex in $G$.
	
	We are now in the position to prove the equivalence between (1) and (2). Clearly, Statement
	(2) implies (1). Suppose that $v$ is a $\klca$ vertex in $G$ for some $k\geq 2$. If $k=2$, then
	Lemma~\ref{lem:k>lca} implies that $v$ is an $\ell$-$\lca$ vertex in $G$ for all $\ell\in
	\{2,\dots, |\CC_G(v)|\}$. If $k\geq 3$, then repeated application of the latter statement, i.e.,
	$v$ is a $(k-1)$-$\lca$ shows that $v$ is an $\ell$-$\lca$ vertex in $G$ for all $\ell\in
	\{2,3,\dots,k\}$. In addition, Lemma~\ref{lem:k>lca} implies that $v$ is an $\ell$-$\lca$ vertex
	in $G$ for all $\ell\in \{k,\dots, |\CC_G(v)|\}$. In summary, $v$ is an $\ell$-$\lca$ vertex in
	$G$ for all $\ell\in \{2,\dots, |\CC_G(v)|\}$. 
	
	It is now an easy task to verify that $G$ is $\One$-$\lca$-\rel for some $\One$ with
    $|\One|>1$ (which implies that $\ell\in \One$ for some $\ell>1$) if and only if each inner
    vertex of $G$ is a $2$-$\lca$ vertex. Moreover, by the equivalence between (1) and (2), each
    inner vertex of $G$ is a $2$-$\lca$ vertex if and only if each inner vertex of $G$ is a
    $|\CC_G(v)|$-$\lca$ vertex and thus, satisfies $v=\lca_G(\CC_G(v))$. Consequently, each inner
    vertex of $G$ is a $2$-$\lca$ vertex if and only if $G$ has the strong-(CL) property. In
    summary, Statements (3), (4) and (5) are equivalent. 
\end{proof}

Based on the latter results, we derive the following simple characterization of 
$\lca$-\rel trees.
\begin{corollary}\label{cor:tree-k-redFree} 
A tree is $\lca$-\rel if and only if it is phylogenetic. In particular, every inner vertex $v$ in a
phylogenetic tree $G$ is an $\ell$-$\lca$ vertex for each $\ell\in \{2,\dots,|\CC_G(v)|\}$. 
\end{corollary} 
\begin{proof}
By Lemma~\ref{lem:LCA-REL=>phylo}, every $\lca$-\rel DAG is phylogenetic. Suppose that $G$ is a
phylogenetic tree. It is well-known that every phylogenetic tree is isomorphic to the Hasse diagram
of its clustering system \cite{sem-ste-03a}. Hence, $G$ is regular. Theorem~\ref{thm:regular-Char} implies 
that $G$ is $\lca$-\rel. Since $\mathfrak{C}_G$ is a
hierarchy, i.e., a clustering system without overlapping clusters, we can conclude that $G$
satisfies (N3O). By Lemma~\ref{lem:(N3O)::k-lca<=>2-lca}, every inner vertex $v$ of $G$ is an
$\ell$-$\lca$ vertex for each $\ell\in \{2,\dots,|\CC_G(v)|\}$.
\end{proof}

Note that Corollary~\ref{cor:tree-k-redFree} cannot easily be generalized to other phylogenetic DAGs with the
(N3O) property. For example, consider the galled tree $G$ on $X=\{x_1,x_2,x_3\}$, which has a root $\rho$ with two children
$u_1$ and $u_2$, an additional edge $(u_1, u_2)$, and such that $u_1$ has only $x_1$ as child,
while $u_2$ has precisely $x_2$ and $x_3$ as its children. It is
easy to verify that $G$ is phylogenetic, but $\rho \neq \lca_G(A)$ for any $A \subseteq X$, that is,
 $G$ is not $\lca$-$\rel$. Adding a new leaf $x_4$ and the edge $(\rho, x_4)$ to $G$ would
result in an $\lca$-$\rel$ galled tree.

Note that the DAG $G$ in Figure~\ref{fig:gominusv} is a DAG with (N3O) property that is neither
$\One$-$\LCA$-\rel nor $\One$-$\lca$-\rel for any $\One$. However, the results above allow us,
together with the $\ominus$-operator, to transform a DAG $G$ with (N3O) property in polynomial time
into a DAG $G'$ that is $\One$-$\LCA$-\rel or $\One$-$\lca$-\rel for arbitrary $\One$.
To this end, observe first that one can easily verify whether a given DAG with (N3O) property is an 
\Olca-\rel or \OLCA-\rel : Lemma~\ref{lem:(N3O)::k-lca<=>2-lca} implies that we only need
to verify if each inner vertex is a $2$-$\lca$-\rel or $2$-$\LCA$-\rel DAG; a task that can be 
be achieved in polynomial time (cf.\ Observation~\ref{obs:12-rel}). 
Even more, we can transform any DAG $G$ with (N3O) into an $\One$-$\lca$-\rel, resp., $\One$-$\LCA$-\rel DAG 
that satisfies (N3O) and properties (S0) -- (S4), resp., (S0) -- (S5) w.r.t.\ $G$ by utilizing the following

\begin{lemma}\label{lem:(N3O)ominus-closed}
	Let $G$ be a DAG and $v$ be an inner vertex of $G$.
	If $G$ satisfies (N3O), then $G\ominus v$ satisfies (N3O).
\end{lemma}
\begin{proof}
	By contraposition, assume that $G\ominus v$ does not satisfy (N3O) for some inner vertex $v$ of the DAG $G$. 
	Thus, $\mathfrak{C}_{G\ominus v}$ contains three clusters that are pairwise overlapping. 
	By Observation~\ref{obs:ominus} we have $\mathfrak{C}_{G\ominus v}\subseteq\mathfrak{C}_G$. 
	Hence, $\mathfrak{C}_G$ contains three pairwise overlapping clusters and $G$ does not satisfy (N3O).
\end{proof}

Thus, whenever we found a vertex $v$ that is not an $\One$-$\lca$ vertex
resp., an $\One$-$\LCA$ vertex in an (N3O) DAG $G$,  we can compute $G\ominus v$ in polynomial time, to derive a DAG
that,  by Lemma~\ref{lem:(N3O)ominus-closed}, satisfy (N3O) and, by Theorem~\ref{thm:S4S5}, satisfies (S0) -- (S4),
resp., (S0) -- (S5) w.r.t.\ $G$. 
Hence, we can reuse the latter arguments, for checking whether the remaining vertices are  $\One$-$\lca$ 
or  $\One$-$\LCA$ vertices or not and then repeat this process until no such vertices exist 
and always obtain a DAG satisfying (N3O) and (S0) -- (S4), resp., (S0) -- (S5) w.r.t\ $G$. 
We summarize the latter discussion into 	

\begin{theorem}
	For a given DAG $G$ that satisfies (N3O), a vertex $v\in V(G)$ and a set $\mathscr{I} \subseteq \{1,\dots,|X|\}$, 
	it can be verified in polynomial time if $v$ is an $\mathscr{I}$-$\lca$ vertex or an  $\mathscr{I}$-$\LCA$ vertex. 
	In particular, every DAG $G$ that satisfies (N3O) can be transformed in polynomial time into an 
	$\One$-$\lca$-\rel, resp.,  $\One$-$\LCA$-\rel DAG with (N3O)-property and that satisfies (S0) -- (S4) 
	resp., (S0) -- (S5)	w.r.t.\ $G$. 
\end{theorem}

\section{Summary and Outlook}
\label{sec:sum}

In this paper, we introduced \OLCArel and \Olcarel DAGs, with focus on the case when
$\One=\{1,2,\ldots,|L(G)|\}$, resulting in the notion of $\LCA$-\rel and $\lca$-\rel DAGs. In
particular, we have shown that one can efficiently transform any given DAG $G$ into an $\LCA$-\rel
and $\lca$-\rel DAG $H$ by stepwise removal of vertices that are not LCAs, resp., unique LCAs of any
subset of taxa with the help of the $\ominus$-operator. Importantly, the resulting DAG $H$ maintains
significant structural features of the original DAG $G$ specified by the structural properties (S1)
-- (S5) which also imply (S0). The simply defined and, in our opinion, rather inconspicuous
$\ominus$-operator has been a somewhat surprisingly powerful tool in this paper, and may still prove
to be helpful in related contexts. We characterized $\LCA$-$\rel$ and $\lca$-\rel DAGs and showed
their close relationship to regular DAGs. Moreover, we showed that our construction indeed
``simplifies'' a DAG $G$, formalized through three axioms (P1) -- (P3). Although we have provided
polynomial-time algorithms to recognize $\lca$-\rel or $\LCA$-\rel DAGs and to transform DAGs into
$\lca$-\rel or $\LCA$-\rel ones, the problem of determining if a vertex $v$ is a $k$-$\lca$ or
$k$-$\LCA$ vertex for a given $k$ and, recognizing \OLCArel and \Olcarel DAGs for specified sets
$\One$ is an NP-complete task. The latter problems become tractable for DAGs that do not contain
three pairwise overlapping clusters; a class of DAGs which includes rooted phylogenetic trees and
galled-trees.

All questions posed in the introduction have been fully addressed. Question~1 is answered by the
results in Section~\ref{sec:LCA-rel}, where we demonstrate that it is possible to determine in
polynomial time whether a given vertex is the (unique) LCA of a specific subset $A \subseteq L(G)$.
The answer to Question~2 is two-fold. When the size of the unknown set $A \subseteq L(G)$ is
unspecified, Corollary~\ref{cor:klca<=>v=lca(C(v))} provides a characterization, allowing Question~2
to be answered in polynomial time. However, when the size $|A|$ is specified but the set $A$ itself
is unknown, verifying if a vertex is the (unique) LCA of a subset of leaves of size $|A|$ becomes
NP-complete, as shown in Section~\ref{sec:complexity-hard}. Nevertheless, the latter type of problem
becomes tractable for DAGs with N3O property. The answer to Question~3 is provided by
Theorems~\ref{thm:LCArel-char} and~\ref{thm:strongCLChar}, while answers to Question~4 are presented
in Sections~\ref{sec:ominus-lcaRel} and~\ref{sec:complexity-hard}.
	
The attentive reader may have noticed that some of the technical details introduced with the set
$\One$ in the definitions of \Olcarel and \OLCArel DAGs could have been omitted, as we primarily
focused on the special case where $\One = \{1, 2, \ldots, |L(G)|\}$. However, since \OLCArel and
\Olcarel DAGs are, in particular, $\LCA$-\rel and $\lca$-\rel DAGs, respectively, most of the
results also hold for the more general \OLCArel and \Olcarel DAGs. Specifically, Theorem
\ref{thm:S4S5} implies that the structural properties (S0) -- (S5), resp., (S0) -- (S4) are
preserved under the $\ominus$-operator when applied to non-$\One$-$\LCA$, resp., non-$\One$-$\lca$
vertices. Hence, these results naturally generalize those established for so-called $\lca$-networks
\cite{Hellmuth2023} and for the case $\One = \{1, 2, \dots, k-1, k\}$ \cite{S+24}. Moreover, the
established results also allow for a focus on interesting subcases, such as DAGs that are \OLCArel
for $\One = \{1, 2\}$ or $\One = \{1, 3\}$ \cite{Barthelemy:08, nowak2009algorithmic,Huber2018}.
Such DAGs play an important role in reflecting gene relationships, such as so-called orthology
relations \cite{Lafond2016, Hellmuth:13a, Huber2018} or more general combinatorial objects
\cite{BHS:21, ASH:25}. In particular, it is of interest to explore in greater detail which classes
of DAGs and networks allow for a polynomial-time solution to check the properties \Olcarel and
\OLCArel, as well as to transform the underlying DAGs into \Olcarel and \OLCArel ones for specific
sets $\One$. Further questions in this context include: Can we characterize DAGs and networks in
which every inner vertex is a $\klca$ vertex for a specific $k$? If such a $k$ exists, what is the
minimal one? Similarly, if $G$ is $\lca$-\rel or $\LCA$-\rel, what is the smallest integer $k$ in a
subset $\One \subseteq \{1,2,\ldots,|L(G)|\}$ such that $G$ is \Olcarel or \OLCArel? By
Observation~\ref{obs:12-rel}, the latter task can be easily addressed when we consider whether $G$
is $\{1,2\}$-$\lca$-\rel or $\{1,2\}$-$\LCA$-\rel. 

We have shown that the set $W$ of non-$\LCA$ vertices required to transform $G$ into an $\LCA$-rel
DAG $G \ominus W$, satisfying conditions (S0) -- (S5), is uniquely determined. This uniqueness
property is preserved for DAGs that satisfy (CL) or (PCC), i.e., for such DAGs $G$, the set $W$ of
non-$\lca$ vertices in $G$ that ensures $G \ominus W$ is $\lca$-rel while satisfying conditions (S0)
-- (S4) is also unique. In general, however, the set $W$ of non-$\lca$ vertices in $G$ that makes $G
\ominus W$ $\lca$-rel with conditions (S0) -- (S4) is not unique. This raises the question of the
computational complexity involved in finding a minimum-sized set $W$ of non-$\lca$ vertices to
ensure the latter.

By Proposition~\ref{prop:alg_kLCA}, if $G \ominus W$ is the $\LCA$-rel version of $G$, we have
$\mathfrak{C}_{G \ominus W} = \mathfrak{C}_G$. In contrast, if $W$ is the set of all non-$\lca$
vertices of $G$, Observation~\ref{obs:ominus} implies only that $\mathfrak{C}_{G \ominus W}
\subseteq \mathfrak{C}_G$. In fact, the example in Figure~\ref{fig:ominus-diff-no-of-vertices}
demonstrates that $\mathfrak{C}_{G \ominus W} \subsetneq \mathfrak{C}_G$ is possible. This raises
the question of how the set systems $\mathfrak{C}_G$ and $\mathfrak{C}_{G \ominus W}$ are related,
and which clusters, if any, are contained in $\mathfrak{C}_G \setminus \mathfrak{C}_{G \ominus W}$.
Moreover, instead of seeking a minimum-sized set $W$ of non-$\lca$ vertices that ensures $G \ominus
W$ is $\lca$-rel under conditions (S0) -- (S4), one might consider finding a set $W$ that minimizes
the size of the difference $\mathfrak{C}_G \setminus \mathfrak{C}_{G \ominus W}$, thereby preserving
as many clusters in $\mathfrak{C}_G$ as possible in $G \ominus W$.

A further interesting generalization is as follows. For a DAG $G$, define all leaves as
\emph{pertinent}. Recursively, a non-leaf vertex is considered \emph{pertinent} if it serves as a
least common ancestor (LCA) for a subset of pertinent vertices. For example, in the DAG $G$
illustrated in Figure~\ref{fig:gominusv}, the vertices $v$ and $w$ are pertinent because they are
the LCA of the set $\{x,y\}$. Consequently, the non-LCA vertex $\rho$ also becomes pertinent, as it
is the unique LCA of the two pertinent vertices $v$ and $w$. A characterization of networks and DAGs
in which all vertices are pertinent, as well as operations to transform a DAG into such a type of
DAG, might be an interesting avenue for future research.

\section*{Supplementary information}

This work is accompanied by a Python tool, \texttt{SimpliDAG}, which computes 
$\LCA$-\rel or $\lca$-\rel versions for a given DAG and is freely available on GitHub \cite{github-AL}.

\section*{Acknowledgements}

We thank Guillaume Scholz, Nicolas Wieseke and Peter F.\ Stadler for  stimulating discussions on this topic.
Moreover, we thank Daniel Huson for providing the data set of the \emph{Viola} genus network.

\section*{Declarations}

The authors declare no conflict of interest. Both authors contributed equally to this work.
The data used to generate the network $N$ in Figure~\ref{fig:plantsN} is available at \url{https://github.com/husonlab/phylosketch} (Nov 1, 2024).

\bibliographystyle{spbasic}
\bibliography{pc}

\end{document}